\PassOptionsToPackage{table,xcdraw}{xcolor}
\documentclass[manuscript, nonacm]{acmart}
\usepackage[table, xcdraw]{xcolor}
\usepackage{xcolor}
\usepackage{verbatim}
\usepackage{caption}
\usepackage{xspace}
\usepackage{pifont}
\usepackage{wrapfig}
\usepackage{multirow}
\usepackage{graphicx}
\usepackage{listings}
\usepackage{hyperref}
\usepackage[capitalise, nameinlink, noabbrev]{cleveref}
\usepackage[inline]{enumitem}
\usepackage{boxedminipage}
\usepackage{alltt}
\usepackage[justification=centering]{caption}  % center align figure captions

\AtBeginDocument{%
  }

\setcopyright{acmlicensed}
\copyrightyear{2024}
\acmYear{2024}
\acmDOI{XXXXXXX.XXXXXXX}

\acmJournal{TOSEM}
\acmVolume{1}
\acmNumber{1}
\acmArticle{1}
\acmMonth{1}

\newcommand{\Modelizer}{\textsc{Modelizer}\xspace}
\newcommand{\PUT}{~$\textnormal{PUT}$}
\newcommand{\PUTinverse}{~$\textnormal{PUT}^{-1}$}
\newcommand*\rot{\rotatebox{90}}
\newcommand{\cmark}{\ding{51}}
\newcommand{\xmark}{\ding{55}}
\newcommand{\fsquare}{\ding{110}}

\newenvironment{result}{\smallskip\begin{center}\begin{boxedminipage}[t]{14cm}\centering\it\smallskip}{\smallskip\end{boxedminipage}\end{center}\smallskip\par}

\definecolor{lightyellow}{HTML}{FFFDCE}
\definecolor{lightblue}{HTML}{DAE8FC}
\definecolor{lightpurple}{HTML}{C3C2FF}
\definecolor{htmlwhite}{HTML}{FFFFFF}
\definecolor{lightgrey}{HTML}{C0C0C0}

\newcounter{rqcounter}
\Crefname{section}{Section}{Sections}
\Crefname{subsection}{Section}{Sections}
\Crefname{appendix}{Appendix}{Appendices}

\lstdefinestyle{BNF} {
  language=XML,
  basicstyle=\ttfamily\small\linespread{0.9}, 
  keywordstyle=\color{blue}, 
  stringstyle=\color{blue}, 
  commentstyle=\color{black}, 
  numbers=left, 
  numberstyle=\tiny\color{gray}, 
  stepnumber=1, 
  numbersep=10pt, 
  backgroundcolor=\color{white}, 
  showspaces=false, 
  showstringspaces=false, 
  showtabs=false, 
  frame=none, % single 
  rulecolor=\color{black}, 
  tabsize=4, 
  captionpos=b, 
  breaklines=true, 
  breakatwhitespace=false, 
  escapeinside={\%*}{*)}, 
  morekeywords={*,...},
  morestring=[s]{<}{>},
  columns=fullflexible,
}

\makeatletter
\renewcommand\@makefnmark{}
\renewcommand{\footnoterule}{%
  \kern-3\p@
  \noindent\rule{\textwidth}{0.4pt}\kern2.6\p@
}
\makeatother

\begin{document}

\title{Learning Program Behavioral Models from Synthesized Input-Output Pairs}

%% Of note is the shared affiliation of the first two authors and the
\author{Tural Mammadov}
\email{tural.mammadov@cispa.de}
\orcid{0009-0007-8466-3694}
\authornotemark[1]
\affiliation{%
  \institution{CISPA Helmholtz Center for Information Security}
  \city{Saarbr{\"u}cken}
  \country{Germany}
}

\author{Dietrich Klakow}
\email{dietrich.klakow@lsv.uni-saarland.de}
\orcid{0000-0002-4147-9690}
\affiliation{%
  \institution{Saarland University}
  \city{Saarbr{\"u}cken}
  \country{Germany}
}

\author{Alexander Koller}
\email{koller@coli.uni-saarland.de}
\orcid{0000-0002-5317-6689}
\affiliation{%
  \institution{Saarland University}
  \city{Saarbr{\"u}cken}
  \country{Germany}
}

\author{Andreas Zeller}
\email{zeller@cispa.de}
\orcid{0000-0003-4719-8803}
\affiliation{%
  \institution{CISPA Helmholtz Center for Information Security}
  \city{Saarbr{\"u}cken}
  \country{Germany}
}

\renewcommand{\shortauthors}{Mammadov et al.}

\begin{abstract}
We introduce \Modelizer---a novel framework that, given a black-box program, learns a \emph{model from its input/output behavior} using \emph{neural machine translation} algorithms.
The resulting model \emph{mocks} the original program: Given an input, the model predicts the output that would have been produced by the program.
However, the model is also \emph{reversible}---that is, the model can predict the input that would have produced a given output.
Finally, the model is \emph{differentiable} and can be efficiently restricted to predict only a certain aspect of the program behavior.
\Modelizer uses \emph{grammars} to synthesize and inputs and {unsupervised tokenizers} to decompose the resulting outputs, allowing it to learn sequence-to-sequence associations between token streams.
Other than input grammars, \Modelizer only requires the ability to execute the program.
The resulting models are \emph{small,} requiring fewer than 6.3~million parameters for languages such as Markdown or HTML; and they are \emph{accurate,} achieving up to 95.4\% accuracy and a BLEU score of~0.98 with standard error~0.04 in mocking real-world applications.
As it learns from and predicts \emph{executions} rather than code, \Modelizer departs from the LLM-centric research trend, opening new opportunities for \emph{program-specific models} that are fully tuned towards individual programs.
Indeed, we foresee several \emph{applications} of these models, especially as the output of the program can be any aspect of program behavior.
Beyond mocking and predicting program behavior, the models can also synthesize inputs that are likely to produce a particular behavior, such as failures or coverage, thus assisting in program understanding and maintenance.
\end{abstract}

\begin{CCSXML}
  <ccs2012>
     <concept>
         <concept_id>10010147.10010257.10010282.10010290</concept_id>
         <concept_desc>Computing methodologies~Learning from demonstrations</concept_desc>
         <concept_significance>300</concept_significance>
         </concept>
     <concept>
         <concept_id>10011007.10010940.10010971.10010980.10010984</concept_id>
         <concept_desc>Software and its engineering~Model-driven software engineering</concept_desc>
         <concept_significance>100</concept_significance>
         </concept>
     <concept>
         <concept_id>10002978.10003022.10003465</concept_id>
         <concept_desc>Security and privacy~Software reverse engineering</concept_desc>
         <concept_significance>500</concept_significance>
         </concept>
     <concept>
         <concept_id>10011007.10011074.10011092.10010876</concept_id>
         <concept_desc>Software and its engineering~Software prototyping</concept_desc>
         <concept_significance>500</concept_significance>
         </concept>
     <concept>
         <concept_id>10011007.10011074.10011099.10011102.10011103</concept_id>
         <concept_desc>Software and its engineering~Software testing and debugging</concept_desc>
         <concept_significance>300</concept_significance>
         </concept>
   </ccs2012>
\end{CCSXML}

\ccsdesc[300]{Computing methodologies~Learning from demonstrations}
\ccsdesc[100]{Software and its engineering~Model-driven software engineering}
\ccsdesc[500]{Security and privacy~Software reverse engineering}
\ccsdesc[500]{Software and its engineering~Software prototyping}
\ccsdesc[300]{Software and its engineering~Software testing and debugging}

\keywords{Software Testing, Mocking, Deep Learning}

% \received{11 July 2024}
% \received[revised]{2024}
% \received[accepted]{2025}

\maketitle

\section{Introduction}

\begin{wrapfigure}{r}{0.45\textwidth}
  \includegraphics[width=\linewidth]{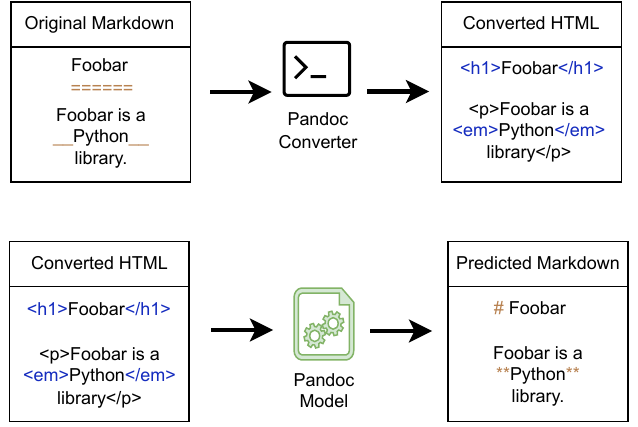}
  \caption[Example]{Example of \Modelizer in action.}
  \label{fig:example}
  \Description{Example of \Modelizer in action.}
\end{wrapfigure}

Understanding and predicting the behavior of a given program is a core challenge of Computer Science. 
Many programs people use are black-box systems that conceal their internal workings, while allowing users to interact with them by providing inputs and observing outputs.
In this work, we present \Modelizer---a novel approach to extract a \emph{behavior model} from a black-box program under test (PUT)---by synthesizing inputs, observing outputs, and learning the relationships between input and output features.
The resulting \emph{neural machine translation model}~$M$ can be used for several purposes.
First, $M$ is able to \emph{mock} the behavior of the PUT: Given an input, $M$~predicts the output that the PUT~would have produced, with high accuracy (\Cref{fig:example}).
$M$ can then replace the PUT in case the PUT can no longer be executed---for instance, because it requires a special environment, or is expensive to run, or because it may be suspected to contain malware (which will not be extracted into~$M$).
As the output can be any aspect of program behavior (such as execution traces or coverage), the model $M$ can also predict such aspects for any given input.

The most interesting point, however, is that the learned model $M$ is \emph{reversible:} it can predict an input that likely produces a given output. Assuming the PUT translates \emph{Markdown} inputs to HTML, then a reversed model~$M^{-1}$ would take HTML inputs and \emph{translate them back} to the original Markdown code, again with high accuracy.
And if the output is, say, a particular execution trace, then the reversed model $M^{-1}$ will produce an input that is likely to produce this very trace.
The learned model~$M$ is \emph{differentiable,} and can thus be efficiently restricted to particular output features.

\Modelizer learns the model from the PUT by executing it again and again, observing how its inputs translate to its outputs. 
However, like all machine learning models, \Modelizer depends on good training. This raises data availability problem, for which we need to provide a solution.
In Software Engineering, the field of \emph{test generation} discusses how to obtain input data for a program.
The test data must be \emph{valid,} such that it gets accepted by the PUT, as well as \emph{diverse,} such that a wide variety of input features is covered. In a black-box setting, obtaining valid and diverse input data requires a dedicated \emph{generator} or \emph{producer,} leveraging knowledge about the input language of the program.

\Modelizer addresses the data availability problem by synthesizing inputs with the help of \emph{grammars} - formal languages for generating strings.
The \emph{input grammar} is used to \emph{produce} syntactically valid and diverse inputs; it can also \emph{parse} existing inputs into derivation trees. In both cases, we know the structure of the input.
The \emph{output grammar} can also be used to \emph{parse} the output, again delivering a tree structure allowing decomposition into tokens;
if the output language specification is unknown, the program output can be decomposed by automatically training a tokenizer from scratch using the SentencePiece~\cite{kudo-richardson-2018-sentencepiece} algorithm.
Using the input grammar as a producer, we can synthesize valid and diverse input data automatically and thus obtain as many input-output pairs from the PUT as we want.
Additionally, \Modelizer distinguishes structure from content in the decomposed sequences and can abstract content by replacing them with placeholders to further improve the learning quality and efficiency.

Developers who want to use \Modelizer thus only need to provide the PUT in an executable form, as well as grammars that describe its input and output languages; such grammars can also be inferred from the PUT or input samples. \Modelizer synthesizes inputs for the PUT, eventually extracting its behavior model~$M$. The model can then be used for a variety of purposes, including:
\begin{itemize}
  \item as a \emph{mock} for the PUT, predicting its behavior---for instance, for testing or development when the PUT is not available for execution;
  \item as a \emph{reverse engineering} tool---for instance, allowing developers to simulate and analyze the behavior of the PUT without requiring its code; or for plain \emph{model stealing,} replicating the behavior of an existing model;
  \item as an \emph{anomaly detector} monitoring the PUT and detecting behavior changes---for instance, flagging situations where the PUT and model deviate;
  \item as a \emph{reverse predictor,} producing likely inputs that result in given outputs---for instance, to produce inputs that trigger behaviors of interest; this is the core challenge of software test generation. 
\end{itemize}
These scenarios only scratch the surface of possible usages.
In this paper, however, we focus on the \emph{mock} and \emph{reverse predictor} scenarios---that is, predicting outputs from inputs and vice versa.
We evaluate Modelizer across a range of challenging PUTs and achieve excellent accuracy.
Notably, the models produced by \Modelizer showed high accuracy on non-trivial programs despite having a small number of neurons (and thus efficient to learn) when compared to large language models (LLMs). For instance, on the \emph{Pandoc} markup converter, \Modelizer achieved over 90\% accuracy in converting Markdown to HTML and vice-versa with a BLEU score of 0.71.

Despite the potential of \Modelizer, we also identify several limitations:
\begin{itemize}
  \item Some \emph{complex programs} remain out of reach for our current methods due to the computational complexity of the underlying learning algorithm which scales quadratically relative to the processed data length. 
  The legnths of tokenized sequences with input-output data can vary significantly depending on the selected data decomposition and abstraction strategies.
  For example, intermediate abstraction layers can significantly shorten sequence lengths but can cause context loss, which will affect the prediction quality. 
  Existing sequence-to-sequence translation models tend to make mistakes when predicting long sequences.
  Thus, lack of prior knowledge about the subject properties and hardware limits makes it hard to pre-specify all tasks where \Modelizer will succeed.
  \item We still require minor \emph{developer effort} in configuring the \Modelizer framework. 
  Our experiments suggest that improving synthetic input generation is crucial for automating behavior extraction.
  Feedback from real program runs can already help to automate grammar tweaks, reconfigure fuzzers, and adjust learners' hyperparameters.
  Continuous growth of the computational power will soon enable complete behavior extraction of any subjects.
\end{itemize}

\noindent
In summary, our contributions are as follows:
\begin{description}
  \item[A novel, generic means to extract behavior models from programs.] To the best of our knowledge, \Modelizer is the first approach to automatically extract execution behavior from arbitrary programs by converting them into sequence-to-sequence translation models.
  \item[Using grammars to produce inputs for training.] \Modelizer uses grammars to produce inputs (and parse given inputs), and, thus, can learn from an unbounded number of training inputs and resulting outputs. 
  %\todo{I decided not to mention the abstract placeholder as the main contribution. It will still be covered in the approach/implementation section -- Tural}\todo{OK -- AZ}
  % \item [Abstract placeholders for tokenizers.] 
  % Tokenizers that implement the suggested interface can automatically replace the real content with abstract placeholders, thus focusing on translating \emph{text structure} rather than text content.
  \item[Predictive behavior models with high accuracy.] The models produced by the \Modelizer framework show high accuracy when predicting outputs for given inputs.
  \item[Reversible behavior models.] The models produced by \Modelizer can be reversed, producing inputs that are likely to result in given outputs. 
\end{description}

The remainder of this paper is organized as follows:
\Cref{sec:approach} describes in detail how \Modelizer works, including input generation, dataset preprocessing, model learning, and model deployment. \Cref{sec:implementation} covers the implementation aspects of \Modelizer.  In \Cref{sec:evaluation}, we evaluate \Modelizer on a variety of text-to-text translators, such as Markdown to HTML, query languages, and more.
\Cref{sec:limitations} discusses the limitations of \Modelizer;
\Cref{sec:related-work} details the related work. %, \Cref{sec:conclusion} closes with a conclusion and future work.
\Modelizer and all experimental data are available as open source; see \Cref{sec:conclusion} for details.

\section{Approach}
\label{sec:approach}

\begin{wrapfigure}[16]{r}{0.6\textwidth}
  \centering
  \vspace{-\baselineskip}
  \includegraphics[width=\linewidth]{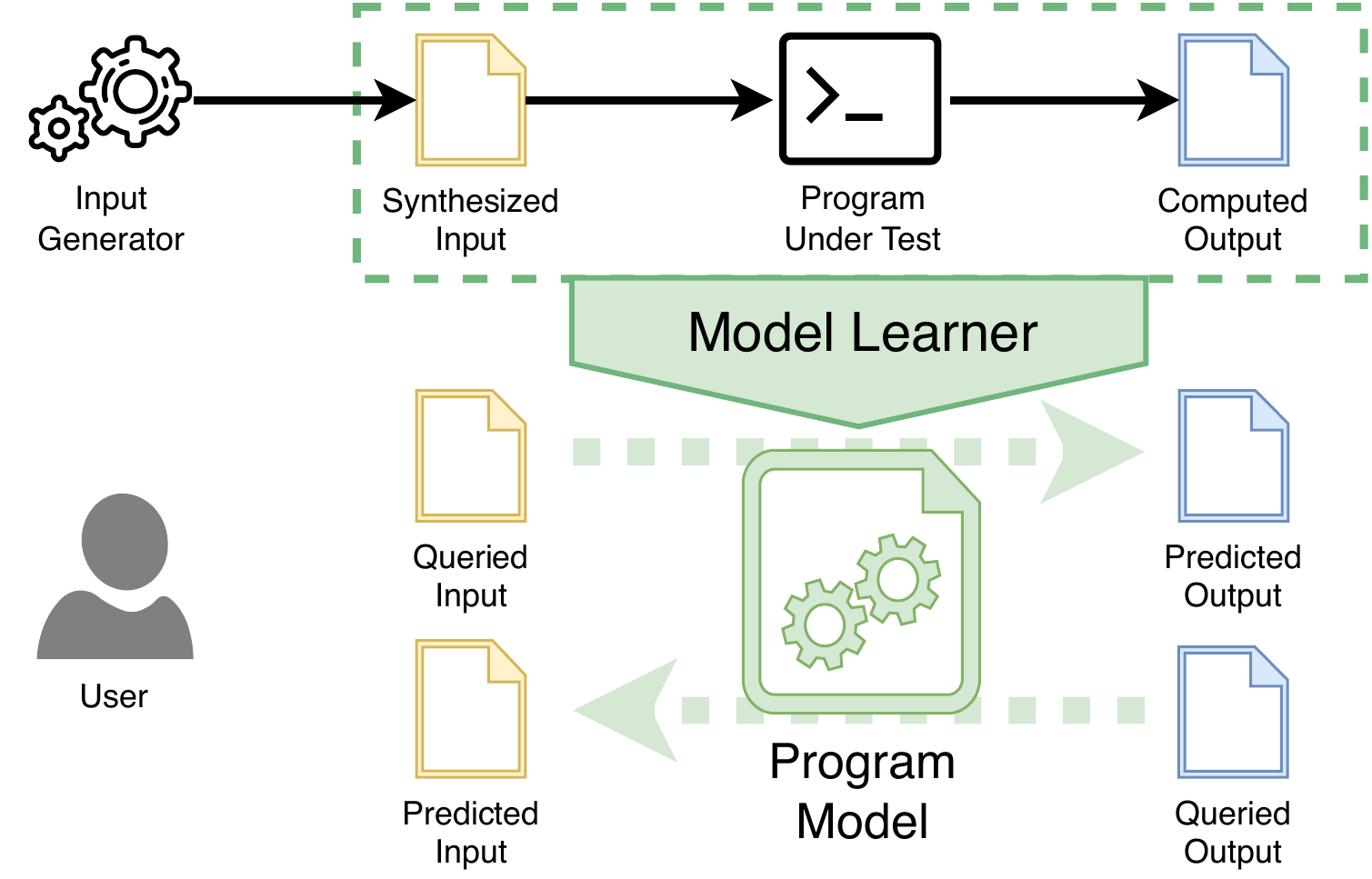}
  % \captionsetup{width=0.7\textwidth} 
  \caption[How \Modelizer works]{How \Modelizer works. \Modelizer tests the program with \emph{synthesized inputs} and automatically learns a \emph{reversible program behavior model}
  from extracted input-output pairs.}
  \Description{How \Modelizer works. \Modelizer tests the program with synthesized inputs and automatically learns a \emph{reversible program behavior model}
  from extracted input-output pairs.}
  \label{fig:overview}
\end{wrapfigure}

\Modelizer is a generic framework for \emph{learning program behavior.}
It uses \emph{synthesized data} to automatically learn \emph{reversible generative models} that can replicate the behavior of real-world programs with high accuracy.
\Cref{fig:overview} shows a general overview of the framework. Given the input specification of the selected PUT encoded as grammar, \Modelizer automatically generates the required number of inputs and passes them to the PUT for processing. Once the PUT produces the corresponding output, \Modelizer extracts these input-output pairs and uses them to train a reversible neural machine translation model. The trained model can be further deployed on modern computer architectures. 
We have automated \emph{program behavior mocking} using the following steps: Input Generation, Dataset Preprocessing, Model Learning, and Model Deployment. The following sections will cover each step in detail.

\subsection{Input Generation}
While developing the \Modelizer framework, we aimed to make it interoperable with different types of programs. 
Even though all programs are written in different programming languages, differ in implementation logic, and target different hardware architectures and operating systems, most programs are made to process input coming from users or other programs. Thus, we have decided to learn the program behavior by training our behavior-mocking models from the \emph{program inputs and processed outputs.}

To learn an accurate model, we need sufficiently many diverse input-output pairs.
Since public test datasets are typically biased towards commonly used features, and we do not know in advance how many inputs are required for learning the behavior of an unseen program, \Modelizer assumes that inputs are \emph{generated automatically}.

\subsubsection{Input Generation with Grammars}
\label{sec:grammars}

\begin{wrapfigure}[15]{r}{0.6\textwidth}
  \vspace{-\baselineskip}
  \centering
  \includegraphics[width=\linewidth]{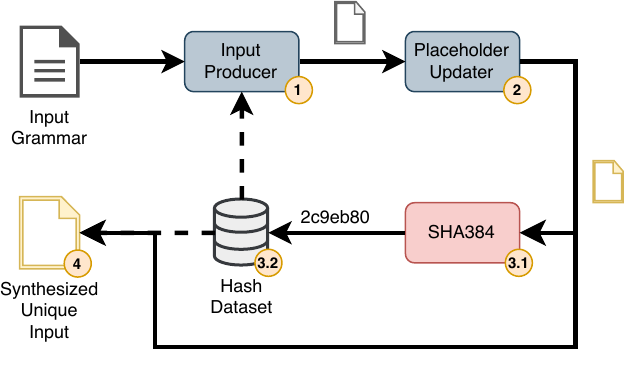}
  % \captionsetup{width=0.6\textwidth} 
  \caption[Input generation pipeline]{Input generation pipeline. \Modelizer automatically synthesizes and validates unique inputs from given specification. Input generation steps: (1)~\emph{Synthesis}, (2)~\emph{Post-Processing}, (3)~\emph{Hashing}, and (4)~\emph{Validation}.}
  \Description{Input generation pipeline. \Modelizer automatically synthesizes and validates unique inputs from given specification.}
  \label{fig:input-fuzzing}
\end{wrapfigure}

To cover the entire input space (and hence increase our chances to also cover the behavior and output spaces), we use \emph{grammar-based} input producers~\cite{fuzzingbook2023}.
These encode input specifications as context-free grammars (CFG)~\cite{1056813}---a finite set of recursive string generation rules.
For some input formats, such CFGs already exist and are publicly available.
For example, the Grammar Zoo~\cite{ZAYTSEV201528} repository contains definitions of 1756~grammars of mainstream programming languages mined from standard documentation or community-created resources. Developers could handwrite such grammars using the knowledge provided as protocol specifications such as RFCs. We cover alternative ways of automatically extracting grammars from existing systems in \Cref{rel:mining}.
% However, third-party grammar definitions may mismatch with the parsing rules implemented in the PUT.
% In such cases, the PUT can reject inputs generated with the help of crawled grammar.
% Then, the grammar definition has to be refined until it satisfies the PUT expectations.

The Input Generator Module is the first component of the \Modelizer framework. It acts as a data processing pipeline, as shown in \Cref{fig:input-fuzzing}, consisting of four steps: (1)~\emph{Synthesis}, (2)~\emph{Post-Processing}, (3)~\emph{Hashing}, and (4)~\emph{Validation}. 
At the synthesis step, the input producer synthesizes new input samples according to rules encoded in grammar. Then, the synthesized samples are refined to better align with real-world data. Such refinement helps map synthesized input features to real-world data features without increasing grammar size, thereby simplifying the model learning process. The refined samples are then checked for uniqueness by computing a hash of the synthesized sample. If the hash is absent in the Hashes Dataset, the sample is passed to the PUT for further processing. Otherwise, all steps are repeated until a unique sample is generated. The PUT produces the corresponding output, which is saved in persistent storage with the input. The collected input-output pairs are later used to train a program behavior model.

\begin{figure}[t]
\begin{center}
\begin{lstlisting}[caption=Markdown grammar used for input generation (excerpt), label={lst:markdown-grammar}, style=BNF]
<start> ::= <blocks>
<blocks> ::= <element> | <element>\n\n<blocks>
<element> ::= <heading> | <paragraph_block> | <blockquote> | 
                <horizontal_rules> | <ordered_list> | <unordered_list>
<heading> ::= # <paragraph_text> | ## <paragraph_text> | ### <paragraph_text> | 
                #### <paragraph_text> | ##### <paragraph_text> | ###### <paragraph_text> | 
                <paragraph_text>\n<double_equal_sign> | <paragraph_text>\n<double_dash_sign>
<paragraph_block> ::= <paragraph_text>\n | <paragraph_text>\n<paragraph_block>
<paragraph_text> ::= <formatted_text> | [<formatted_text>](<url>) |
                     <code_block> | <paragraph_text> <paragraph_text>
<formatted_text> ::= <text> | <bold_text> | <italic_text> | <bold_italic_text> | <quoted_text>
<bold_text> ::= **<text>** | __<text>__
<italic_text> ::= *<text>* | _<text>_
<bold_italic_text> ::= ***<text>*** | ___<text>___ | **_<text>_** | __*<text>*__
<code_block> ::= `<text>` | ```<text>``` |  <empty_space><text>
<quoted_text> ::= '<text>' | "<text>"
<empty_space> ::= "    "
<text> ::= TEXT
<url> ::= URL
\end{lstlisting}  
\end{center}
\end{figure}

Let us reuse our handwritten Markdown grammar (\Cref{lst:markdown-grammar}) to demonstrate the synthetic input generation process.
We will follow the steps that \Modelizer takes to produce inputs.
\Cref{lst:generate} shows the high-level implementation of the input generation pipeline written in Python. Once the input producer is configured, it is ready to synthesize as many inputs as needed (Step~1). Here is an example of synthesized input that was generated by the producer:
\begin{center}
  \verb|TEXT [TEXT](URL) TEXT **TEXT** TEXT `TEXT` TEXT\n|
\end{center}

\begin{figure}[t]
  \centering
  \begin{minipage}{0.85\textwidth}
\begin{lstlisting}[caption=Input generation in \Modelizer (pseudocode), label={lst:generate}]
def generate_data(grammar, min_non_term, max_non_term, total_elements, program_handler):
    dataset = list()
    seen_inputs = set()
    processor = PlaceholderProcessor(grammar)
    fuzzer = GrammarFuzzer(grammar, min_non_term, max_non_term)
    while(len(dataset) < total_elements):
        input_data = fuzzer.fuzz()                                   # Step 1
        input_data = processor.deduplicate(input_data)               # Step 2
        input_hash = sha384(input_data.encode("utf-8")).hexdigest()  # Step 3.1
        if input_hash not in seen_inputs:                            
            seen_inputs.add(input_hash)                              # Step 3.2
            output_data = exec(program_handler, input_data)          # Step 4
            dataset.append((input_data, output_data))
    return dataset
\end{lstlisting}
  \end{minipage}
\end{figure}

The producer has synthesized one top-level Markdown element, a paragraph block. The paragraph block was expanded to a single paragraph text element, which contains a sequence of formatted text, hyperlink, and code block elements. One of the formatted text elements was further expanded to a bold text element.
The synthesized input contains TEXT and URL placeholders instead of real character sequences. Instead of encoding the whole language vocabulary into our grammar and synthesizing natural language sentences, we have decided to define placeholder tokens to replace the real-world data. (We will elaborate more about placeholders in the follow-up section.)

\subsubsection{Placeholders}
\label{sec:placeholders}

Synthesizing data that is equivalent to real-world counterparts is a challenging task. As different programs can process the same input in various ways, the input grammar must be specified or refined according to program expectations. 
Depending on the complexity of the input specification, the number of required input samples for efficient behavior learning can also become extremely large. 
In their work on abstracting failure-inducing inputs, Gopinath et al.~\cite{failure-inducing-inputs} found that some input fragments are not affected or used by underlying implementation logic and can be abstracted for certain data types and programs. For example, many text format converters only process the document structure, leaving the document content untouched.
To benefit from similar effects for behavior extraction, we integrated the following two strategies in our input generation pipeline:
\begin{itemize}
    \item We suggest \emph{generalizing} certain low-level non-terminal expansion rules by replacing a fraction of non-terminal to terminal expansions with generic placeholders.
    \item We have added an optional \emph{post-processing step} that augments placeholders with unique identifiers.
\end{itemize}

Both strategies will help capture large portions of possible input space with few samples and correctly map real-world inputs to generic representations and vice versa during the inference phase.
Instead of learning the behavior from real-world examples, we learn how the PUT processes placeholders.
(In the current \Modelizer implementation, developers have to specify how to abstract grammar rules; future revisions will automate this process.)

Let us refer back to the Markdown synthesis example and follow the input generation phases. The Placeholder Processor at Step~2 will replace these placeholders with equivalents that additionally carry unique identifiers. The output of the Placeholder Processor is shown below:
\begin{center}
  \verb|TEXT_1 [TEXT_2](URL_1) TEXT_3 **TEXT_4** TEXT_5 `TEXT_6` TEXT_7\n|
\end{center}

We assume that a TEXT placeholder can represent any character sequence in a real Markdown document. 
Similarly, a URL placeholder can substitute an arbitrary definition. 
Learning from the small number of synthesized structures, our models will still be capable of processing principally equivalent real-world inputs.
While the input producer's output is well-formed and passes parsing guards, aligning real-world Markdown documents with the synthesized structure is challenging because it is hard to distinguish the generated \emph{placeholders}. 
We have implemented a post-processing phase in the input generation pipeline as a generic Placeholder Processor module. It initializes with a list of encoded placeholders like \verb|[TEXT, URL]| and augments each placeholder with a unique identifier. 
% Of course, it is possible to inject identifiers during the input generation. However, grammar-based input generators do not synthesize integers incrementally. They cannot guarantee that the generated identifier will be used in a synthesized sequence of characters only once. It is much faster and easier to generate a simple candidate sequence and then refine it using custom string-replacement or constraint-solving rules. 
Such identifiers enhance the accuracy of bidirectional translations, especially when elements need reordering. 
% During tokenization, which we'll discuss later, data portions receive a matching placeholder type with a unique identifier. Optionally, this unique identifier assignment can be disabled.

%Since grammar-based producers do not maintain a list of generated samples, we have implemented a sample validation procedure which first computes an HMAC-SHA-384~\cite{rfc6234} value (Step~3.1) for the synthesized sample and then checks whether the computed hash value is stored in the Hashes Dataset (Step~3.2). If a computed hash value is absent in the dataset, the dataset gets updated with a new record (Step 4), and the synthesized input is passed to the PUT to compute output. Otherwise, the synthesized sample is dropped, and all input generation steps are repeated till the unique sample is synthesized. Corresponding input-output pairs are saved in a persistent storage for later processing. The collected input-output pairs are saved in a persistent storage for later processing. 

Since grammar-based producers do not maintain a list of generated samples, we implemented a sample validation procedure. This involves computing an HMAC-SHA-384~\cite{rfc6234} value (Step~3.1) for the synthesized sample and checking if it's stored in the Hashes Dataset (Step~3.2). If absent, the dataset gets updated with a new record (Step 4), and the synthesized input is passed to the PUT for output computation. Otherwise, the sample is dropped, and input generation repeats until a unique sample is created. Input-output pairs are saved in persistent storage for later processing.

\subsection{Dataset Pre-processing}
\label{sec:pre-processing}
\label{sec:tokenizers}
Neural machine translation models are artificial intelligence models that use deep neural networks to learn dependencies between source and target language sentences. 
Their success in automated machine translation tasks inspired us to model program behavior as a sequence-to-sequence translation task. 
In principle, current neural machine translation architectures can process arbitrarily complex inputs, but the problem's computational complexity scales quadratically with the input length. 
So, practical implementations partition data into token sequences and process them in fragments. 
Then, models learn how to generate sequences of tokens in a target language by mapping the cross-token dependencies from a source language to a target language. 
Similarly, the program input and output must be converted into sequences of tokens for processing. 
While text partitioning is well-studied in natural language processing ~\cite{manning2008introduction, honnibal2017spacy}, tokenizing the program input and output could be a challenging task.

\subsubsection{Masked Tokenization}
% While our approach works with generic tokenizers, we obtain better results with \emph{custom tokenizers} adapted to the languages at hand.
% Our \verb|HTMLTokenizer| adapted to HTML, for instance, takes an HTML string such as
While our approach also works with unsupervised tokenizers, such as Google SentencePiece~\cite{kudo-richardson-2018-sentencepiece}, we obtain better results with \emph{custom tokenizers} adapted for the languages at hand.
For instance, our \verb|HTMLTokenizer| handles an HTML string such as:
\begin{center}
  \verb|<h1>Foobar</h1><p>Foobar is a <em>Python</em> library</p>|
\end{center}

The result of the tokenization looks like the following:
\begin{center}
  \verb|[ <h1>, TEXT_1, </h1>, <p>, TEXT_2, <em>, TEXT_3, </em>, TEXT_4, </p> ]|
  \verb|{ TEXT_1: Foobar, TEXT_2: Foobar is a, TEXT_3: Python, TEXT_4: library }|
\end{center}

If we include the placeholder mapping as input to the reconstruction routine for the selected example, it will output the original HTML string. Otherwise, we will get the following output:
\begin{center}
  \verb|<h1>TEXT_1</h1><p>TEXT_2<em>TEXT_3</em>TEXT_4</p>|
\end{center}

\Modelizer supports both early and late token instantiations, meaning that abstract tokens will be replaced with real-world values before or after the token concatenation.
Furthermore, as well as different \emph{mapping policies;}. See \Cref{sec:tokenize-placeholder} for details of our implementation, including the pseudocode for HTML tokenization.

Full control over the tokenization process allows us to control the tokenization granularity, impacting prediction accuracy and model learning time. The vocabulary size grows with to the number of unique tokens in the training set, which impacts the overall number of model parameters. It is worth combining tokens that sequentially appear in the same context. With this measure, we reduce
\begin{enumerate}
    \item the chances of breaking the syntactical correctness of the generated sequences;
    \item the time required for learning the model; and
    \item the time required for sequence generation.
\end{enumerate}

\Modelizer users can implement a domain-specific tokenizer by extending the functionality of parsers that are already available or could be derived from the grammar, for example, with the help of the Earley~\cite{earley-parser} parser algorithm.
In this case, one can reuse the grammar that was used for synthesizing inputs.
Once we can automatically and correctly extract output grammars from PUTs, we will be able to infer parsers and domain-specific tokenizers for them.
See \Cref{sec:tokenize-placeholder} for more information on implementing domain-specific tokenizers.

\subsection{Model Learning}
\label{sec:model-learning}
Our Learner is based on the \emph{Transformer} sequence-to-sequence neural machine translation model~\cite{NIPS2017_3f5ee243}. It learns a program-specific behavior mocking model from selected input-output pairs. We present the high-level representation of the learning pipeline in \Cref{lst:learn} and cover the steps taken in detail in the following paragraphs. 
The Model Learner accepts the input-output data formats as input and loads the corresponding synthesized training sample pairs from the dataset. The loaded samples get automatically split into training, validation, and test sets. 

\begin{figure}[h]
\begin{center}
  \begin{minipage}{0.85\textwidth}
\begin{lstlisting}[caption=The model learning pipeline in \Modelizer (pseudocode), label={lst:learn}]
def learn_model(dataset, input_type, output_type, train_fraction, num_epochs):
    dataloader = DataLoader(dataset, input_type, output_type)
    train_split, validation_split, test_split = dataloader.random_split(train_fraction)
    src_vocab, trg_vocab = dataloader.get_vocabularies()
    param_searcher = HyperparameterSearcher(src_vocab, trg_vocab,
                                                model_trials=250, lr_trials=100)
    model_params = param_searcher.search(test_split)
    model = Learner(model_params, src_vocab, trg_vocab)
    for _ in range(num_epochs):
        model.train(train_split)
        model.validate(validation_split)
    return model
\end{lstlisting}
  \end{minipage}
\end{center}
\end{figure}

\Modelizer additionally \emph{automates the hyperparameter tuning process.}  We have implemented an Optimizer that tries to automatically find the best combination of configurable parameter values using a small disjoint set of test samples, which, in our case, typically consists of 10,000 synthesized samples. 
The optimizer is built upon the \emph{Optuna}~\cite{optuna_2019} framework, which automates hyperparameter search.
Parameter search is conducted in two phases: \emph{Model Parameter Search} and \emph{Learning Rate Search}.
During Model Parameter Search, the best neural network configuration is sought for a fixed number of iterations~\cite{NIPS2017_3f5ee243}.
While using a fixed learning rate, the Optimizer computes the cross-entropy~loss~\cite{6773024} metric for various combinations of Linear Layer size, Embedding size, number of Attention Heads, and Encoder-Decoder layers.
The combination with the lowest cross-entropy value proceeds to the Learning Rate Search phase, where the initial learning rate (LR), weight decay, and learning rate adjustment policy are determined.
In our experiments, the Optimizer executes the Model Parameter Search for 250 trials and, afterward, the Learning Rate Search for 100 trials.
Automated hyperparameter search facilitates a faster and more effective exploration of the model configuration landscape, which is essential for optimal model performance and training efficiency.
It saves the time and effort needed for manual tuning, allowing potential \Modelizer users to focus on strategic aspects of program behavior extraction.

After hyperparameter search, the \verb|Learner| is initialized with the discovered model configuration, and learning runs for the specified epochs. 
We use the cross-entropy loss function to evaluate training, measuring the difference between expected and predicted labels. 
The \emph{AdamW} optimization algorithm~\cite{loshchilov2019decoupled} adjusts model parameters to minimize training loss.
The validation split is used to evaluate the model's performance on unseen data to prevent overfitting.
Trained models are ``lightweight'', which, for example, require less than 6.3~million parameters to learn the Markdown to HTML conversion behavior of the Pandoc markup converter~\cite{pandoc}.
Meanwhile, we are able to quickly and accurately mock program-specific behavior on commodity hardware.
The training time varies from several minutes to a couple of hours, depending on the subject.
For the subjects from the evaluation set, the forward models contained 3~to~7.7~million parameters, and inverse models required 2.8 to 10.3~million parameters to learn the hypothetical inverse behavior. 
Since leveraging the knowledge embedded in pre-trained LLMs could be beneficial for mocking program behavior, we additionally provide an adaptor for interacting with LLMs as an alternative to training task-specific models from scratch. 
We relied on the Unsloth~\cite{unsloth} framework to support resource-efficient local LLM querying and fine-tuning.

\subsection{Model Deployment}
\label{sec:deployment}

\begin{figure}[t]
  \centering
  \includegraphics[width=\linewidth]{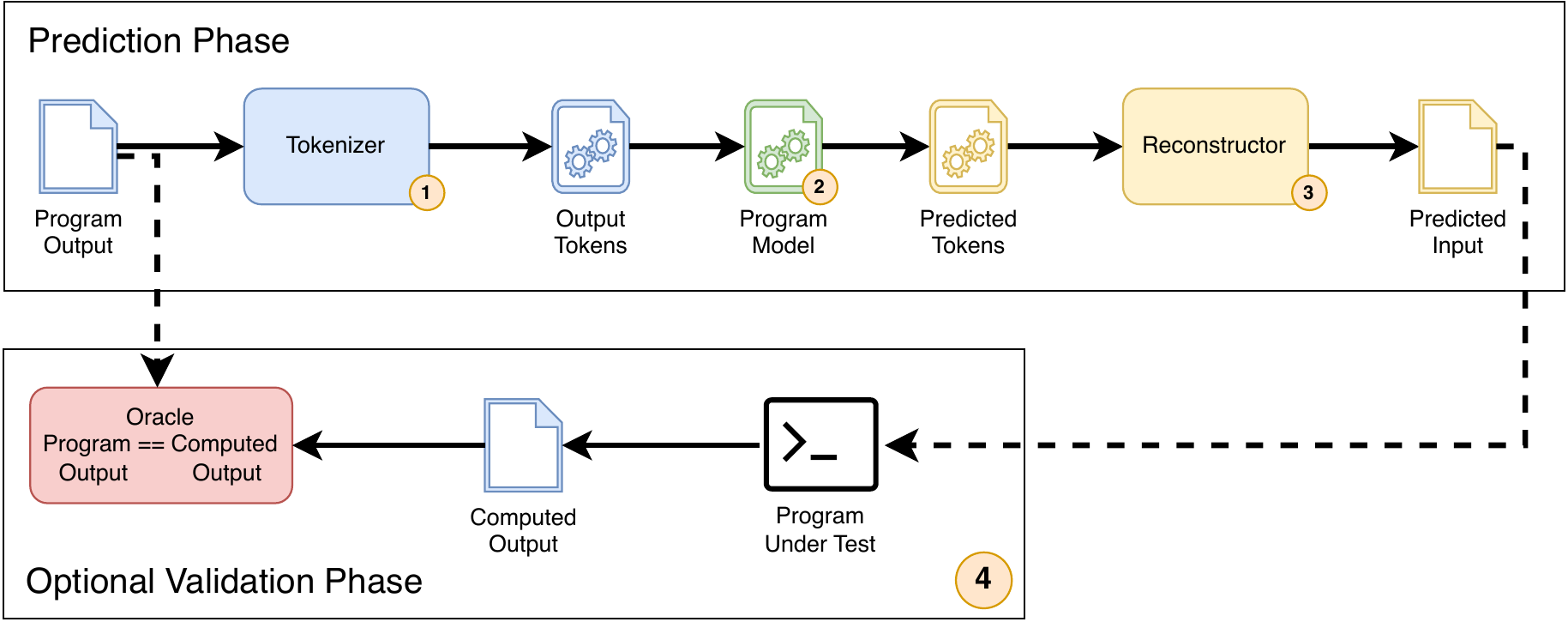}
  \caption[Model Deployment]{Model Deployment. An example scenario of the Behavior model predicting program input given the program output.\\Phases: (1) \emph{Input tokenization}, (2) \emph{Prediction generation}, (3) \emph{Output reconstruction}, (4) \emph{Prediction validation}.
  %The model prediction gets validated for correctness in the optional validation phase.
  }
  \Description{Model Deployment. An example scenario of the Learned Behavior Model predicting Program Input given the Program Output. The model prediction gets validated for correctness in the optional validation phase.}
  \label{fig:deployment}
\end{figure}

The Model Deployment phase, summarized in \Cref{fig:deployment}, includes the Prediction and optional Validation phases. Depending on the source data type, which can be either program input or output, the respective model, source tokenizer, and target tokenizer get initialized. From this point, model inference operates similarly to the model training phase. The source data gets tokenized~(1) and passed to the Program Model~(2), which predicts a target token sequence. Then, the target reconstructor component~(3) lifts the predicted sequence from token representation to real-world format.
The translation routine supports an optional max output sequence length parameter. If the framework user does not specify this argument, our default implementation dynamically estimates the hypothetical maximum output sequence length by applying a 1.25 multiplier to the length of the currently processed input sequence.

Lastly, the predicted data can be passed to an optional Validation phase (4), which requires access to the respective PUT. 
It aligns with the model's operation mode. 
The predicted and computed outputs are tokenized in behavior-mocking mode (predicting output from input), with the latter treated as a reference sequence. 
Then, we calculate the sequence similarity metrics such as Levenshtein distance~\cite{Levenshtein1965BinaryCC} and the BLEU~\cite{metric-bleu} score. 
The Levenshtein distance measures the difference between sequences by counting the number of minimal edits needed to transform the predicted sequence into the reference. 
The next section will discuss the BLEU score and sequence similarity metrics in more detail.

\begin{figure}[t]
\begin{center}
  \begin{minipage}{0.85\textwidth}
\begin{lstlisting}[caption=Model deployment with the help of \Modelizer (pseudocode), label={lst:deploy}]
def deploy(model, grammar, test_case, program_handler=None):    
    # Test the Model
    in_tokenizer  = Tokenizer(grammar)
    out_tokenizer = CustomTokenizer()
    test_tokens, test_mapping = in_tokenizer.mapped_feed(test_case)             # Step 1
    prediction = model.translate(test_tokens)                                   # Step 2
    reconstructed = out_tokenizer.mapped_reconstruct(prediction, test_mapping)  # Step 3

    # Evaluate the Translation
    if program_handler is not None:                                             # Step 4
        reference = exec(program_handler, test_case)
        if reconstructed == reference:
            print("Model correctly predicted", reconstructed)
        else:
            reference_tokens, _ = out_tokenizer.mapped_feed(reference)
            bleu_score = compute_bleu(reference_tokens, prediction)
            edit_distance = compute_distance(reference, reconstructed)
            print("Model failed to correctly predict the output")
            print("Expected:", reference, "\nPrediction:", reconstructed)
            print("BLEU Score:", bleu_score, "Edit Distance:", edit_distance)
    else:
        print("Model predicted", reconstructed)
\end{lstlisting}
  \end{minipage}
\end{center}
\end{figure}

In the Inverse-mocking mode (finding the program input used to compute the given program output), the computed output collected by passing the predicted input to the PUT gets compared with the initial program output. Since the actual program input (ground truth) is unknown, we can compute sequence similarity score metrics in cases when the program accepts the predicted input and does not fail while processing it. A failure, of course, will indicate the incorrectness of the predicted input. It is also challenging to validate the correctness of the prediction in the inverse-mocking mode if the PUT computes multiple valid inputs to the same output. Such programs have to be modeled by introducing an additional abstraction layer that defines the set of semantically equivalent inputs using formal rules.

We demonstrate the results of model deployment shown in \Cref{lst:deploy} using the example from \Cref{tab:translation-example}. 
In this example, we used the HTML to Markdown conversion model learned from the Pandoc program, which was trained on 1,000,000 synthesized document pairs.
First, the input HTML string is tokenized, and the content gets replaced with placeholders (Step~1).
The model uses the produced input tokens to predict a sequence of Markdown tokens (Step~2), and placeholders are replaced with real data during output reconstruction (Step~3). 
Since the Pandoc program was locally installed on the same machine that ran the HTML to Markdown conversion model, we can validate the correctness of the predicted output by comparing it with the computed output. We measured the BLEU score at 1.0 and the Levenshtein distance at 0.0, which indicates that the predicted output is identical to the computed output.

\begin{table}[t]
  \small
  \centering
  \caption{\label{tab:translation-example}A conversion example from HTML to Markdown.}
  \begin{tabular}{ll}
    \rowcolor{gray!50} 
    1. Input HTML & 4. Reconstructed Markdown \\
    \begin{minipage}{0.45\linewidth}
      \leavevmode
      \ttfamily
      \begin{verbatim}

<p>Note that <a href="/docs/
transformers/v4.34.0/en/
model_doc/tapas#transformers.
TapasTokenizer">TapasTokenizer</a> 
expects the data of the table 
to be <strong>text-only</strong>. 
You can use <code>.astype(str)
</code> on a dataframe to turn it 
into text-only data. Of course, 
this only shows how to encode a 
single training example. It is 
advised to create a dataloader 
to iterate over batches.</p>       
      \end{verbatim}
      \end{minipage}
      & \begin{minipage}{0.45\linewidth}
        \ttfamily
        \leavevmode
        \begin{verbatim}

Note that [TapasTokenizer]
(/docs/transformers/v4.34.0/en/
model_doc/tapas#transformers.
TapasTokenizer) expects the data 
of the table to be **text-only**. 
You can use `.astype(str)` on a 
dataframe to turn it into 
text-only data. Of course, this 
only shows how to encode a single 
training example. It is advised 
to create a dataloader to 
iterate over batches.         
        \end{verbatim}
        \end{minipage}                                                                                                           \\
    \rowcolor{gray!50}
    2. Input Tokens & 3. Predicted Tokens \\

\begin{minipage}{0.45\linewidth}
\leavevmode
\ttfamily
\begin{verbatim}
  
<p> TEXT_1 <a  href="URL_1" > 
TEXT_2 </a> TEXT_3 <strong> 
TEXT_4 </strong> TEXT_5 <code>
TEXT_6 </code> TEXT_7 </p>
 
\end{verbatim}
\end{minipage}                                                                                                &

\begin{minipage}{0.45\linewidth} 
\ttfamily     
\begin{verbatim}
TEXT_1 [ TEXT_2 ]( URL_1 ) TEXT_3 ** 
TEXT_4 ** TEXT_5 ` TEXT_6 ` TEXT_7 \n
\end{verbatim}
\end{minipage}                                                                                                     \\
\end{tabular}
\parbox{\linewidth}{\centering \small \textit{
%The conversion was performed by a model of the Pandoc program trained on 1,000,000 synthesized sequence pairs. BLEU score = 1.0, Levenshtein distance = 0.0
The conversion used a Pandoc model trained on 1,000,000 sequence pairs, achieving a BLEU score of 1.0 and a Levenshtein distance of 0.0.
}}
\end{table}

\section{Implementation}
\label{sec:implementation}

\Modelizer implements grammar-based input generators using grammar fuzzers from The Fuzzing Book~\cite{fuzzingbook2023}.
While the Fuzzing Book \emph{GrammarFuzzer} class randomly chooses the expansion rule from the set of available rules, its \emph{ProbabilisticGrammarFuzzer} variant allows the developer to control the input generation by assigning \emph{probabilities} to certain expansion rules.
In this case, the non-attributed expansion rules will be assigned by equally distributed remaining probabilities.
This allows us to
\begin{enumerate*}
    \item \emph{automatically generate} a dataset of syntactically \emph{valid} inputs;
    \item customize the feature selection procedure in the sample generation process; and
    \item ensure the representation of specific and \emph{corner samples} in the data corpus.
\end{enumerate*}

\subsection{System Requirements}

We implemented \Modelizer model learning in Python using the \emph{PyTorch} 2.0 Machine Learning Framework~\cite{NEURIPS2019_bdbca288}. Our framework currently supports the following hardware backends: 
\begin{itemize}
  \item NVIDIA GPUs with support for CUDA 11.7 or higher (Fastest)
  \item Apple Silicon processors with at least 16GB of system memory
  \item Systems with a x64 processor and 32~GB of system memory (Slowest)
\end{itemize}

Model learning is a highly parallelizable task, and it can benefit from the availability of dedicated hardware accelerators; however, the remaining tasks can also be scaled according to the number of available CPU cores.
For example, to reduce the total runtime, our generators support the execution of multiple grammar-based fuzzers with the same configuration in parallel.
The number of Fuzzers that were initialized depends on the number of available CPU cores.
\Modelizer is responsible for distributing the total workload among all fuzzers by assigning every fuzzer with its own budget.
Then, each fuzzer produces samples in batches until it runs out of budget.
\Modelizer systematically collects the generated samples and checks for their uniqueness.
If there was no previously unseen sample generated within a single batch or a configurable parameter of failing attempts is reached, the value of the minimum and maximum number of non-terminal expansions gets automatically incremented.
To simplify the usability and applicability of our framework, we include a base class generator that can be extended to support new subjects.

\subsection{Tokenization with Placeholders}
\label{sec:tokenize-placeholder}

Even though our models are trained on synthesized data, they must still be able to handle real-world data. Since synthesized data may include placeholders instead of actual values, % tokenizers' implementation needs to account for it. 
\Modelizer provides a base implementation and abstract interfaces for mapped tokenization and reconstruction routines.
Users can wrap unsupervised tokenizer functions with our mapped tokenization algorithm or implement a format-specific tokenizer in a subclass for new data types.

\subsubsection {Abstract Tokenizers}
Assuming the availability of suitable parsers for input and output formats, we implemented \verb|AbstractTokenizer|—a generic mapped tokenization-reconstruction algorithm leveraging format structure and content fragment replications in both input and output. While this is not a fully functional tokenizer, it offers a high-level, extendable implementation for mapped tokenization-reconstruction tasks, aiding the mapping of real-world content to abstract placeholders and vice-versa. \Cref{lst:tokenize-base} shows a part of the mapped tokenization logic.

To correctly map real content to abstract placeholders (references as mask tokens in the pseudocode), the parser-tokenizer process calls the \verb|mask_token| method once it handles the data fragment. The developer needs to implement a format-specific \verb|tokenize| logic in a subclass. Then, \Modelizer can replace content fragments with placeholders during tokenization and recover the original sequence during reconstruction. The tokenizer can learn placeholders from the grammar definition or refer to the user-defined ones, allowing it to handle both synthesized and real-world data.

\begin{figure}[t]
\begin{center}
  \begin{minipage}{0.85\textwidth}
\begin{lstlisting}[caption=Tokenization in \Modelizer (pseudocode), label={lst:tokenize-base}]
class AbstractTokenizer:
    def mapped_tokenize(self, data: str):
        self.clear_mappings()
        self.mask_mapping = True
        self.tokenize(data)
        self.mask_mapping = False
        return self.buffer, self.mappings
    
    def mask_token(self, token: str, placeholder: str) -> str:
        tag = placeholder + "_" + str(len(self.mappings[placeholder]) + 1)
        if self.mapping_policy == MappingPolicy.SIMPLIFIED:
            self.mappings[placeholder][tag] = token
            tag = placeholder
        elif self.mapping_policy == MappingPolicy.OPTIMIZING:
            if token in self.inverse_mapping:
                tag = self.inverse_mapping[token]
            else:
                self.mappings[placeholder][tag] = token
                self.inverse_mapping[token] = tag
        elif self.mapping_policy == MappingPolicy.EXHAUSTIVE:
            self.mappings[placeholder][tag] = token
            self.inverse_mapping[token] = tag
        return tag
    
    def set_mapping_policy(self, policy: MappingPolicy):
        self.mapping_policy = policy
    
    def tokenize(self, data: str):
        raise NotImplementedError
\end{lstlisting}
  \end{minipage}
\end{center}
\end{figure}

\subsubsection{A HTML Tokenizer}
\verb|HTMLTokenizer| class given below demonstrates the procedure of masking HTML content by extending the functionality of the \verb|AbstractTokenizer| class. Once the \verb|mapped_tokenize| method is called, the tokenizer clears all internal buffers and starts processing the given string using the selected parser. The \texttt{HTMLParser} module from the Python Standard Library, for instance, requires the user to implement handler methods to process HTML tags and document content (\verb|handle_data| method). Depending on the operation mode of the tokenizer, which is determined by the \verb|mask_mapping| flag, the tokenizer either replaces the document content with a matching placeholder or performs a sequence of format-specific string splitting operations, which are abstracted in this example.\footnotetext{
  The full implementation of HTMLTokenizer, which decomposes data using Python built-in HTMLParser, can be found in our GitHub repository (see \Cref{sec:conclusion} for details).  This repository also provides additional examples of tokenizer implementations that rely on PUT's functionality to parse inputs or implement a custom parser using regular expressions.
}

\begin{figure}[t]
\begin{center}
  \begin{minipage}{0.85\textwidth}
\begin{lstlisting}[caption=Tokenization in \Modelizer (pseudocode), label={lst:tokenize-html}]
from html.parser import HTMLParser

class HTMLTokenizer(AbstractTokenizer, HTMLParser):    
    def tokenize(self, data: str):
        self.buffer.clear()
        HTMLParser.feed(self, data)

    def handle_data(self, data: str):
        if self.mask_mapping == True:  # True only if called via mapped_tokenize(data)
            token = self.mask_token(data, self.placeholders[1])
            self.buffer.append(token)
        else:
            tokens = self.split_and_align(data)
            self.buffer.extend(tokens)


    
\end{lstlisting}
  \end{minipage}
\end{center}
\end{figure}

\subsubsection {Token Masking Strategies}

The token masking algorithm behaves differently according to the selected token mapping policy strategies:

\begin{itemize}
  \item The \verb|SIMPLIFIED| policy reduces model vocabulary size by removing identifiers from placeholder tokens.
  \item The \verb|OPTIMIZING| policy, our default, uses a minimal number of placeholder tokens to encode data.
  \item  The \verb|EXHAUSTIVE| policy assigns every content fragment with a unique placeholder token.
\end{itemize}

% We use the extended version of the HTML document from \Cref{fig:example} to illustrate the effect of using different token masking strategies in tokenization:
We use the extended HTML document from \Cref{fig:example} to show the impact of all token masking strategies in tokenization:
\begin{center}
  \verb|<h1>Foobar</h1><p>Foobar is a <em>Python</em> library. Find| 
  \verb|more on <a href="127.0.0.1"><i>Foobar</i>.com</a></p>|
\end{center}

When the tokenizer is running in \verb|SIMPLIFIED| token masking mode, it removes every unique identifier assigned to a placeholder. It significantly reduces the model's vocabulary size, but increases the chances of producing errors during the reconstruction phase. The tokenized sequence looks like the following:

\begin{center}
  \verb|[ <h1>, TEXT, </h1>, <p>, TEXT, <em>, TEXT, </em>, TEXT,|
  \\\verb|<a, href="URL">, <i>, TEXT, </i>, TEXT, </a>, </p> ]|
\end{center}

When configured with the \verb|OPTIMIZING| token masking strategy, the tokenizer uses a minimal number of placeholders to encode real-world data, reusing the same placeholder for repetitive content. This is the default mode, and it was used for the entire evaluation. The tokenized sequence looks like the following:
\begin{center}
  \verb|[ <h1>, TEXT_1, </h1>, <p>, TEXT_2, <em>, TEXT_3, </em>, TEXT_4,|
  \\\verb|<a, href="URL_1">, <i>, TEXT_1, </i>, TEXT_5, </a>, </p> ]|
\end{center}

The \verb|EXHAUSTIVE| token masking strategy increases the model's vocabulary size and negatively affects the model learning time. The tokenizer maps every content piece in the processed input to a unique identifier. However, this policy is useful for programs that reorder elements during processing. The tokenized sequence looks like the following:

\begin{center}
  \verb|[ <h1>, TEXT_1, </h1>, <p>, TEXT_2, <em>, TEXT_3, </em>, TEXT_4,|
  \\\verb|<a, href="URL_1">, <i>, TEXT_5, </i>, TEXT_6, </a>, </p> ]|
\end{center}

\section{Evaluation}
\label{sec:evaluation}

We evaluated \Modelizer by learning the execution behavior of real-world programs. We collected two datasets of input-output pairs for every PUT by executing each program with synthesized and real-world inputs, respectively. For every evaluation subject, \Modelizer additionally learned an inverse model ~$M^{-1}$ that corresponds to a hypothetical program \PUTinverse, which is expected to implement an inverse of the behavior of the original program \PUT. We also equipped our framework with the ability to fine-tune LLMs to capture the behavior of \PUT\hspace{0.5mm}and its inverse\hspace{0.5mm}\PUTinverse and compared this approach against custom-trained models.

\subsection{Research Questions}

In the evaluation, we address the following research questions:

\begin{enumerate}[label=\textbf{RQ\arabic*.},ref={RQ\arabic*}]

\item \textbf{How accurate are the \emph{\Modelizer models} in mocking the behavior of the \PUT\hspace{0.5mm}and its inverse \PUTinverse?}
\label{rq:mocking}
This question is at the heart of \Modelizer: How well do the learned models perform in predicting outputs from inputs and vice-versa?

\item \textbf{How well do the \emph{synthesized samples} represent the real-world data and how much \emph{synthetic data} is required for mocking program behavior?}
\label{rq:real-world}
With \ref{rq:real-world}, we investigate the quality of the synthesized training set, its correlation to real-world data, and investigate the impact of the size of the synthesized training set on the accuracy of the learned model.

\item \textbf{How efficiently can \Modelizer learn program models?}
\label{rq:learn}
This question evaluates \Modelizer's input generation and model learning efficiency.

\item \textbf{How does \Modelizer learn the \emph{hyperparameters} for each data pair combination?}
\label{rq:hyperparameters}
This question assesses the impact of \emph{hyperparameters} (\Cref{sec:model-learning}) on the performance of \Modelizer. We also check the \emph{reusability} of found hyperparameters with forward models for training inverse models.

\item \textbf{What is the effect of \emph{tokenizer} selection on learning performance?}
\label{rq:tokenizer}
With \ref{rq:tokenizer}, we investigate the impact of \emph{tokenizers} (\Cref{sec:tokenizers}) on the performance of \Modelizer.

\item \textbf{Could LLMs pre-trained on code mock programs?}
\label{rq:llm-comparison}
With \ref{rq:llm-comparison}, we investigate the ability of LLMs pre-trained on code replicating program behavior using zero-shot evaluation (without task-specific model adaptation) or after additional fine-tuning and compare their performance with custom-trained models.

\end{enumerate}

\subsection{Evaluation Subjects}
We conducted the evaluation using the four programs listed in \Cref{tab:eval-subjects}:

\begin{table}[h]
  \rowcolors{2}{gray!25}{white}
  \caption{\label{tab:eval-subjects} The list of evaluated programs and their respective input-output types.}
  \small
  \begin{tabular}{llll}
  \rowcolor{gray!50} 
  Program      & Input Type             & Output Type & Version \\
  Pandoc       & Markdown               & HTML        & 2.19.2  \\
  msticpy      & SQL                    & KQL         & 2.9.0   \\
  latexify-py  & Python Math Expression & \LaTeX{}       & 0.2.0   \\
  py-asciimath & MathML                 & \LaTeX{}       & 0.3.0   \\
  py-asciimath & \LaTeX{}               & AsciiMath   & 0.3.0   \\
  py-asciimath & AsciiMath              & MathML      & 0.3.0   \\
  \end{tabular}
\end{table}

\begin{description}
\item[Pandoc]\cite{pandoc} is an open-source library and command-line tool for converting markups from one format to another. While the library natively supports the translation in both directions(Markdown to HTML and HTML to Markdown), we have tested it only with Markdown to HTML translation functionality. We manually extracted a Markdown Grammar from the \textit{CommonMark Spec Version 0.29}~\cite{commonmark} specification. We used \texttt{Pandoc} to convert Markdown to HTML in our case study to validate neural behavior modeling of black-box systems and its applicability for inverse behavior modeling, where the model predicts input from output. The promising results of our first experiments encouraged us to test the approach with other subjects.

\item[Microsoft Threat Intelligence Python Security Tools] (\texttt{msticpy})~\cite{msticpy} is a library for InfoSec investigations, data analysis, and hunting in Jupyter Notebooks for Azure Sentinel, which is a cloud-native security information and event management solution offered by Microsoft for intelligent security analytics and threat intelligence. \Modelizer was evaluated on MSTIC's SQL to KQL Conversion feature, which converts \texttt{SELECT} SQL queries, using a subset of \textit{ANSI SQL-92}~\cite{iso9075}, into Kusto Query Language (KQL) - a proprietary query language developed by Microsoft for analyzing data in Azure Cloud. We encoded the supported \textit{SQL-92} subset as a CFG grammar.

\item[latexify-py]\cite{latexify-py} is a Python library that converts Python math expressions to \LaTeX{} code. The grammar for synthesizing Python math expressions encodes the math module~\cite{python3-math} functionality from the Python 3.10 standard library. 

\item[py-asciimath]\cite{pyasciimath} is a Python library for converting AsciiMath expressions to MathML and \LaTeX{} formats. The MathML grammar was manually extracted from MathML 2.0 Specification~\cite{mathml2}. Training pairs were gathered by converting synthesized MathML formulas to \LaTeX{} and then to AsciiMath. This process yielded three sets of input-output pairs, allowing us to test cross-protocol translations in directions unsupported by the original program.
\end{description}

\subsection{Training Setup}
\label{sec:eval-setup}

For every subject, we generated training sets of different sizes (5k, 10k, 50k, 100k, 250k, 500k, 1M). The input generation starts with the minimal number of non-terminal expansions (\emph{MinNonTerminals}) set to 10 and maximal (\emph{MaxNonTerminals}) set to 20. Once configured, the generator has 100,000 attempts to generate unique samples using a fixed configuration. If the generator runs out of attempts, the values of \emph{MinNonTerminals} and \emph{MaxNonTerminals} parameters are increased by 10, and the generator is restarted. The generation process runs till the generator produces the required number of unique samples. Also, we configured the generators to produce a separate set of input samples that had been synthesized with an increasing complexity of the generated samples. It was achieved by implementing the sliding window modification of \emph{MinNonTerminals} and \emph{MaxNonTerminals} values. In this case, the generator uses at least four different non-terminal expansion configurations to generate the required number of samples. Inputs were generated on an Ubuntu Linux 22.04 workstation equipped with a 24-core AMD Ryzen Threadripper 3960X CPU and 128 GB of memory.

Each model was evaluated on 10,000 unseen synthesized samples generated with a fixed number of non-terminal expansions between 10 and 20. Additionally, we evaluated models using the following datasets with real-world inputs:

\begin{description}
    \item[Pandoc.] We evaluated both \texttt{Pandoc} models with real-world data. We extracted 12,783 top-level elements (tree-like structures nesting other Markdown definitions) from 100 Markdown files that were collected from the open-access Markdown dataset~\cite{dataset-md} containing \texttt{README} files of projects hosted at the HuggingFace Hub~\cite{huggingface-hub}.
    \item[msticpy.] We tested \texttt{msticpy} models on 1,000 samples from the ``sql-create-context'' dataset~\cite{dataset-sql}, featuring natural language and SQL SELECT query pairs. We ensured that \PUT produced non-empty outputs for every selected SQL query.
    \item[latexify-py.] The evaluation of \texttt{latexify-py} was challenging due to the lack of datasets with mathematical expressions written in Python using the \emph{math} module from the Python Standard Library. Thus, we synthesized 1,000 additional samples using our Python Expression grammar and replaced placeholders with random values of the appropriate type.
    \item[py-asciimath.] The evaluation of \texttt{py-asciimath} involved 1,000 real-world MathML formulas from the NTCIR-12 dataset~\cite{dataset-mathml}.
\end{description}

While \Modelizer's translation routines support a configurable decoder beam size, we fixed the beam size for our experiments to 1. All models were trained for 10 epochs with a context window of 5000 tokens. In each configuration, 80\% of synthesized data was used for training, and the training quality was validated with the remaining 20\%. 
All models were trained on a single NVIDIA Geforce RTX 4090 GPU with 24 GB memory. During learning, \Modelizer allocated 650 MB to 22 GB of memory based on hyperparameters and vocabulary size values. All models trained from scratch can be deployed on modern computers with at least 8 GB memory that can execute a Python interpreter version 3.10 or newer. Unlike real-world programs, these models require no external dependencies or system-wide installation permissions. The necessary Python interpreter and modules can be pre-configured and shipped with the model. A single environment can run multiple model instances in parallel due to a lower memory footprint at the inference time.

Additionally, we experimented with querying and fine-tuning two commonly used LLMs pre-trained on code: \emph{CodeLlama}~\cite{roziere2024code} and \emph{CodeGemma}~\cite{codegemma}. The selected models were instruction tuned by their vendors, meaning that vendors have fine-tuned them on natural language instruction followed by input and expected output pairs. This process improves the model's general utility and performance in practical applications. Both models have 7 billion trainable parameters, and their weights were additionally quantized to 4bit to reduce memory requirements. Experiments involving the usage of LLMs were performed on the data-center grade NVIDIA A100 Tensor Core GPU with 40GB memory. At first, models were queried using the same real-world test samples from previous experiments. After that, we performed fine-tuning on synthesized data and repeated the evaluation with test samples. We fine-tuned an individual model for 10 epochs for each input-output pair to exclude potential interference between subjects and simplify comparison with custom-trained models. During fine-tuning, we reused the synthesized training sets with 10,000 samples generated using the \emph{MinNonTerminals}) set to 10 and \emph{MaxNonTerminals} set to 20.

\subsection{Evaluation Metrics}

To evaluate the quality of the learned models, we have used the following evaluation metrics:
\begin{description}
  \item[BLEU score] (bilingual evaluation understudy)~\cite{metric-bleu} is a commonly used metric for the evaluation of machine translation systems that calculates the geometric average precision of single, double, triple, and quadruple token overlap cases between reference and hypothesis token sequences multiplied by brevity penalty. The score output is always a number between 0 and 1, with 1 indicating the perfect translation case. The BLEU score was computed on the whole corpus of test cases using the implementation provided by the Python NLTK package~\cite{nltk-lib}.
  \item[BLEU Error] stands for the Standard Error computed using the BLEU score. 
  \item[NIST score]~\cite{metric-nist} is a metric developed by the US National Institute of Standards and Technology for evaluating the quality of machine-translated texts, which is also based on the BLEU score and additionally calculates the frequency of n-grams and gives a higher score to n-grams that are considered to less likely occur. We calculated the metric using the implementation provided by the Python NLTK package~\cite{nltk-lib}.
  \item[Word Error Rate] (WER)~\cite{morris04_interspeech} measures prediction accuracy as a ratio of token insertions, deletions, and substitutions needed to convert the predicted token sequence into the reference sequence and the total reference token count. A lower WER indicates better performance, with a score of 0 showing no translation mistakes.
  \item[Word Information Lost] (WIL)~\cite{morris04_interspeech} measures the percentage of total tokens incorrectly predicted between reference and hypothesis token sequences. The lower the value, the better the model performs, with a score of 0 indicating that the model has not lost any information during the translation process.  
  \item[Exact Match] measures the percentage of evaluation samples correctly translated by the model. A higher value indicates better performance, with a score of 100 showing no translation errors. This metric uses Levenshtein distance to assess the similarity between predicted and reference sequences.
  \item[Close Match] is a metric that reflects the percentage of evaluation samples that were translated by the model with a single error. We treat such cases as potentially recoverable at a post-processing stage. This metric relies on the Levenshtein distance to compare the similarity between the predicted and reference sequences.
\end{description}

\subsection{Results}
\label{sec:results}
Let us now turn to the evaluation results.

\begin{table}[]
  \small
  \centering
  \caption{\label{tab:eval-results} Evaluation results for models mocking Pandoc, msticpy, latexify-py, and py-asciimath behaviors.}
  \begin{tabular}{|l|llccrrrrrrr|}
      \hline
      \rowcolor{lightgrey} 
      Program                                                 & \multicolumn{1}{l|}{\cellcolor{lightgrey}Source}                         & \multicolumn{1}{l|}{\cellcolor{lightgrey}Target}                         & \multicolumn{1}{c|}{\cellcolor{lightgrey}\begin{tabular}[c]{@{}c@{}}Model\\ Type\end{tabular}}  & \multicolumn{1}{c|}{\cellcolor{lightgrey}Size} & \multicolumn{1}{c|}{\cellcolor{lightgrey}BLEU}   & \multicolumn{1}{c|}{\cellcolor{lightgrey}\begin{tabular}[c]{@{}c@{}}BLEU\\ Error\end{tabular}} & \multicolumn{1}{c|}{\cellcolor{lightgrey}NIST}    & \multicolumn{1}{c|}{\cellcolor{lightgrey}WER}   & \multicolumn{1}{c|}{\cellcolor{lightgrey}WIL}   & \multicolumn{1}{c|}{\cellcolor{lightgrey}\begin{tabular}[c]{@{}c@{}}Exact\\ Match\end{tabular}} & \multicolumn{1}{c|}{\cellcolor{lightgrey}\begin{tabular}[c]{@{}c@{}}Close\\ Match\end{tabular}} \\ \hline
      \cellcolor{htmlwhite}                                & \multicolumn{1}{l|}{\cellcolor{htmlwhite}}                               & \multicolumn{1}{l|}{\cellcolor{htmlwhite}}                               & \multicolumn{1}{c|}{}                                                                              & \multicolumn{1}{c|}{\cellcolor{lightyellow}1M}   & \multicolumn{1}{r|}{\cellcolor{lightyellow}0.9986} & \multicolumn{1}{r|}{\cellcolor{lightyellow}0.0033}                                               & \multicolumn{1}{r|}{\cellcolor{lightyellow}9.9102}  & \multicolumn{1}{r|}{\cellcolor{lightyellow}0.04}  & \multicolumn{1}{r|}{\cellcolor{lightyellow}0.05}  & \multicolumn{1}{r|}{\cellcolor{lightyellow}98.69}                                                 & \cellcolor{lightyellow}99.39                                                                      \\ \cline{5-12} 
      \cellcolor{htmlwhite}                                & \multicolumn{1}{l|}{\cellcolor{htmlwhite}}                               & \multicolumn{1}{l|}{\cellcolor{htmlwhite}}                               & \multicolumn{1}{c|}{}                                                                              & \multicolumn{1}{c|}{\cellcolor{lightblue}100k} & \multicolumn{1}{r|}{\cellcolor{lightblue}0.2923} & \multicolumn{1}{r|}{\cellcolor{lightblue}0.3025}                                               & \multicolumn{1}{r|}{\cellcolor{lightblue}4.2689}  & \multicolumn{1}{r|}{\cellcolor{lightblue}54.20}  & \multicolumn{1}{r|}{\cellcolor{lightblue}72.13} & \multicolumn{1}{r|}{\cellcolor{lightblue}51.72}                                                 & \cellcolor{lightblue}56.03                                                                      \\ \cline{5-12} 
      \cellcolor{htmlwhite}                                & \multicolumn{1}{l|}{\multirow{-3}{*}{\cellcolor{htmlwhite}Markdown}}     & \multicolumn{1}{l|}{\multirow{-3}{*}{\cellcolor{htmlwhite}HTML}}         & \multicolumn{1}{c|}{\multirow{-3}{*}{Forward}}                                                     & \multicolumn{1}{c|}{\cellcolor{lightpurple}500k} & \multicolumn{1}{r|}{\cellcolor{lightpurple}0.7128} & \multicolumn{1}{r|}{\cellcolor{lightpurple}0.2337}                                               & \multicolumn{1}{r|}{\cellcolor{lightpurple}7.5700}  & \multicolumn{1}{r|}{\cellcolor{lightpurple}23.19} & \multicolumn{1}{r|}{\cellcolor{lightpurple}35.30} & \multicolumn{1}{r|}{\cellcolor{lightpurple}90.57}                                                 & \cellcolor{lightpurple}94.79                                                                      \\ \cline{2-12} 
      \cellcolor{htmlwhite}                                & \multicolumn{1}{l|}{\cellcolor{htmlwhite}}                               & \multicolumn{1}{l|}{\cellcolor{htmlwhite}}                               & \multicolumn{1}{c|}{}                                                                              & \multicolumn{1}{c|}{\cellcolor{lightyellow}1M}   & \multicolumn{1}{r|}{\cellcolor{lightyellow}0.9988} & \multicolumn{1}{r|}{\cellcolor{lightyellow}0.0045}                                               & \multicolumn{1}{r|}{\cellcolor{lightyellow}10.9328} & \multicolumn{1}{r|}{\cellcolor{lightyellow}0.08}  & \multicolumn{1}{r|}{\cellcolor{lightyellow}0.11}  & \multicolumn{1}{r|}{\cellcolor{lightyellow}99.27}                                                 & \cellcolor{lightyellow}99.55                                                                      \\ \cline{5-12} 
      \cellcolor{htmlwhite}                                & \multicolumn{1}{l|}{\cellcolor{htmlwhite}}                               & \multicolumn{1}{l|}{\cellcolor{htmlwhite}}                               & \multicolumn{1}{c|}{}                                                                              & \multicolumn{1}{c|}{\cellcolor{lightblue}250k} & \multicolumn{1}{r|}{\cellcolor{lightblue}0.2352} & \multicolumn{1}{r|}{\cellcolor{lightblue}0.2319}                                               & \multicolumn{1}{r|}{\cellcolor{lightblue}4.0123}  & \multicolumn{1}{r|}{\cellcolor{lightblue}52.62} & \multicolumn{1}{r|}{\cellcolor{lightblue}66.00} & \multicolumn{1}{r|}{\cellcolor{lightblue}55.78}                                                 & \cellcolor{lightblue}57.51                                                                      \\ \cline{5-12} 
      \multirow{-6}{*}{\cellcolor{htmlwhite}Pandoc}        & \multicolumn{1}{l|}{\multirow{-3}{*}{\cellcolor{htmlwhite}HTML}}         & \multicolumn{1}{l|}{\multirow{-3}{*}{\cellcolor{htmlwhite}Markdown}}     & \multicolumn{1}{c|}{\multirow{-3}{*}{Backward}}                                                     & \multicolumn{1}{c|}{\cellcolor{lightpurple}1M}   & \multicolumn{1}{r|}{\cellcolor{lightpurple}0.7036} & \multicolumn{1}{r|}{\cellcolor{lightpurple}0.1554}                                               & \multicolumn{1}{r|}{\cellcolor{lightpurple}7.5202}  & \multicolumn{1}{r|}{\cellcolor{lightpurple}23.74} & \multicolumn{1}{r|}{\cellcolor{lightpurple}34.48} & \multicolumn{1}{r|}{\cellcolor{lightpurple}93.27}                                                 & \cellcolor{lightpurple}93.58                                                                      \\ \hline
      \cellcolor{htmlwhite}                                & \multicolumn{1}{l|}{\cellcolor{htmlwhite}}                               & \multicolumn{1}{l|}{\cellcolor{htmlwhite}}                               & \multicolumn{1}{c|}{}                                                                              & \multicolumn{1}{c|}{\cellcolor{lightyellow}500k} & \multicolumn{1}{r|}{\cellcolor{lightyellow}0.9997} & \multicolumn{1}{r|}{\cellcolor{lightyellow}0.0035}                                               & \multicolumn{1}{r|}{\cellcolor{lightyellow}11.0359} & \multicolumn{1}{r|}{\cellcolor{lightyellow}0.03}  & \multicolumn{1}{r|}{\cellcolor{lightyellow}0.03}  & \multicolumn{1}{r|}{\cellcolor{lightyellow}99.58}                                                 & \cellcolor{lightyellow}99.69                                                                      \\ \cline{5-12} 
      \cellcolor{htmlwhite}                                & \multicolumn{1}{l|}{\cellcolor{htmlwhite}}                               & \multicolumn{1}{l|}{\cellcolor{htmlwhite}}                               & \multicolumn{1}{c|}{}                                                                              & \multicolumn{1}{c|}{\cellcolor{lightblue}100k} & \multicolumn{1}{r|}{\cellcolor{lightblue}0.5040} & \multicolumn{1}{r|}{\cellcolor{lightblue}0.0497}                                               & \multicolumn{1}{r|}{\cellcolor{lightblue}5.0374}  & \multicolumn{1}{r|}{\cellcolor{lightblue}21.48} & \multicolumn{1}{r|}{\cellcolor{lightblue}38.07} & \multicolumn{1}{r|}{\cellcolor{lightblue}1.60}                                                  & \cellcolor{lightblue}1.90                                                                       \\ \cline{5-12} 
      \cellcolor{htmlwhite}                                & \multicolumn{1}{l|}{\multirow{-3}{*}{\cellcolor{htmlwhite}SQL}}          & \multicolumn{1}{l|}{\multirow{-3}{*}{\cellcolor{htmlwhite}KQL}}          & \multicolumn{1}{c|}{\multirow{-3}{*}{Forward}}                                                     & \multicolumn{1}{c|}{\cellcolor{lightpurple}100k} & \multicolumn{1}{r|}{\cellcolor{lightpurple}0.9791} & \multicolumn{1}{r|}{\cellcolor{lightpurple}0.0397}                                               & \multicolumn{1}{r|}{\cellcolor{lightpurple}6.4760}  & \multicolumn{1}{r|}{\cellcolor{lightpurple}1.51}  & \multicolumn{1}{r|}{\cellcolor{lightpurple}2.09}  & \multicolumn{1}{r|}{\cellcolor{lightpurple}95.40}                                                 & \cellcolor{lightpurple}96.90                                                                      \\ \cline{2-12} 
      \cellcolor{htmlwhite}                                & \multicolumn{1}{l|}{\cellcolor{htmlwhite}}                               & \multicolumn{1}{l|}{\cellcolor{htmlwhite}}                               & \multicolumn{1}{c|}{}                                                                              & \multicolumn{1}{c|}{\cellcolor{lightyellow}250k} & \multicolumn{1}{r|}{\cellcolor{lightyellow}0.9206} & \multicolumn{1}{r|}{\cellcolor{lightyellow}0.0374}                                               & \multicolumn{1}{r|}{\cellcolor{lightyellow}11.0730} & \multicolumn{1}{r|}{\cellcolor{lightyellow}6.41}  & \multicolumn{1}{r|}{\cellcolor{lightyellow}7.48}  & \multicolumn{1}{r|}{\cellcolor{lightyellow}27.51}                                                 & \cellcolor{lightyellow}32.64                                                                      \\ \cline{5-12} 
      \cellcolor{htmlwhite}                                & \multicolumn{1}{l|}{\cellcolor{htmlwhite}}                               & \multicolumn{1}{l|}{\cellcolor{htmlwhite}}                               & \multicolumn{1}{c|}{}                                                                              & \multicolumn{1}{c|}{\cellcolor{lightblue}50k}  & \multicolumn{1}{r|}{\cellcolor{lightblue}0.9566} & \multicolumn{1}{r|}{\cellcolor{lightblue}0.0388}                                               & \multicolumn{1}{r|}{\cellcolor{lightblue}5.0318}  & \multicolumn{1}{r|}{\cellcolor{lightblue}2.92}  & \multicolumn{1}{r|}{\cellcolor{lightblue}4.51}  & \multicolumn{1}{r|}{\cellcolor{lightblue}89.00}                                                 & \cellcolor{lightblue}93.90                                                                      \\ \cline{5-12} 
      \multirow{-6}{*}{\cellcolor{htmlwhite}msticpy}       & \multicolumn{1}{l|}{\multirow{-3}{*}{\cellcolor{htmlwhite}KQL}}          & \multicolumn{1}{l|}{\multirow{-3}{*}{\cellcolor{htmlwhite}SQL}}          & \multicolumn{1}{c|}{\multirow{-3}{*}{Backward}}                                                     & \multicolumn{1}{c|}{\cellcolor{lightpurple}100k} & \multicolumn{1}{r|}{\cellcolor{lightpurple}0.9725} & \multicolumn{1}{r|}{\cellcolor{lightpurple}0.0379}                                               & \multicolumn{1}{r|}{\cellcolor{lightpurple}5.1913}  & \multicolumn{1}{r|}{\cellcolor{lightpurple}1.94}  & \multicolumn{1}{r|}{\cellcolor{lightpurple}2.82}  & \multicolumn{1}{r|}{\cellcolor{lightpurple}94.60}                                                 & \cellcolor{lightpurple}95.80                                                                      \\ \hline
      \cellcolor{htmlwhite}                                & \multicolumn{1}{l|}{\cellcolor{htmlwhite}}                               & \multicolumn{1}{l|}{\cellcolor{htmlwhite}}                               & \multicolumn{1}{c|}{}                                                                              & \multicolumn{1}{c|}{\cellcolor{lightyellow}1M}   & \multicolumn{1}{r|}{\cellcolor{lightyellow}0.9969} & \multicolumn{1}{r|}{\cellcolor{lightyellow}0.0543}                                               & \multicolumn{1}{r|}{\cellcolor{lightyellow}12.9138} & \multicolumn{1}{r|}{\cellcolor{lightyellow}0.45}  & \multicolumn{1}{r|}{\cellcolor{lightyellow}0.48}  & \multicolumn{1}{r|}{\cellcolor{lightyellow}88.47}                                                 & \cellcolor{lightyellow}88.86                                                                      \\ \cline{5-12} 
      \cellcolor{htmlwhite}                                & \multicolumn{1}{l|}{\cellcolor{htmlwhite}}                               & \multicolumn{1}{l|}{\cellcolor{htmlwhite}}                               & \multicolumn{1}{c|}{}                                                                              & \multicolumn{1}{c|}{\cellcolor{lightblue}50k}  & \multicolumn{1}{r|}{\cellcolor{lightblue}0.5380} & \multicolumn{1}{r|}{\cellcolor{lightblue}0.1796}                                               & \multicolumn{1}{r|}{\cellcolor{lightblue}6.2148}  & \multicolumn{1}{r|}{\cellcolor{lightblue}23.30} & \multicolumn{1}{r|}{\cellcolor{lightblue}23.60} & \multicolumn{1}{r|}{\cellcolor{lightblue}0.00}                                                  & \cellcolor{lightblue}0.00                                                                       \\ \cline{5-12} 
      \cellcolor{htmlwhite}                                & \multicolumn{1}{l|}{\multirow{-3}{*}{\cellcolor{htmlwhite}PyExpres.}} & \multicolumn{1}{l|}{\multirow{-3}{*}{\cellcolor{htmlwhite}\LaTeX{}}}        & \multicolumn{1}{c|}{\multirow{-3}{*}{Forward}}                                                     & \multicolumn{1}{c|}{\cellcolor{lightpurple}100k} & \multicolumn{1}{r|}{\cellcolor{lightpurple}0.9219} & \multicolumn{1}{r|}{\cellcolor{lightpurple}0.2806}                                               & \multicolumn{1}{r|}{\cellcolor{lightpurple}9.8394}  & \multicolumn{1}{r|}{\cellcolor{lightpurple}8.48}  & \multicolumn{1}{r|}{\cellcolor{lightpurple}9.59}  & \multicolumn{1}{r|}{\cellcolor{lightpurple}48.50}                                                 & \cellcolor{lightpurple}53.50                                                                      \\ \cline{2-12} 
      \cellcolor{htmlwhite}                                & \multicolumn{1}{l|}{\cellcolor{htmlwhite}}                               & \multicolumn{1}{l|}{\cellcolor{htmlwhite}}                               & \multicolumn{1}{c|}{}                                                                              & \multicolumn{1}{c|}{\cellcolor{lightyellow}250k} & \multicolumn{1}{r|}{\cellcolor{lightyellow}0.9612} & \multicolumn{1}{r|}{\cellcolor{lightyellow}0.1039}                                               & \multicolumn{1}{r|}{\cellcolor{lightyellow}15.9279} & \multicolumn{1}{r|}{\cellcolor{lightyellow}3.39}  & \multicolumn{1}{r|}{\cellcolor{lightyellow}4.12}  & \multicolumn{1}{r|}{\cellcolor{lightyellow}40.69}                                                 & \cellcolor{lightyellow}43.98                                                                      \\ \cline{5-12} 
      \cellcolor{htmlwhite}                                & \multicolumn{1}{l|}{\cellcolor{htmlwhite}}                               & \multicolumn{1}{l|}{\cellcolor{htmlwhite}}                               & \multicolumn{1}{c|}{}                                                                              & \multicolumn{1}{c|}{\cellcolor{lightblue}10k}  & \multicolumn{1}{r|}{\cellcolor{lightblue}0.8595} & \multicolumn{1}{r|}{\cellcolor{lightblue}0.0908}                                               & \multicolumn{1}{r|}{\cellcolor{lightblue}10.7279} & \multicolumn{1}{r|}{\cellcolor{lightblue}9.43}  & \multicolumn{1}{r|}{\cellcolor{lightblue}12.48} & \multicolumn{1}{r|}{\cellcolor{lightblue}31.60}                                                 & \cellcolor{lightblue}52.40                                                                      \\ \cline{5-12} 
      \multirow{-6}{*}{\cellcolor{htmlwhite}latexify-py}   & \multicolumn{1}{l|}{\multirow{-3}{*}{\cellcolor{htmlwhite}\LaTeX{}}}        & \multicolumn{1}{l|}{\multirow{-3}{*}{\cellcolor{htmlwhite}PyExpres.}} & \multicolumn{1}{c|}{\multirow{-3}{*}{Backward}}                                                     & \multicolumn{1}{c|}{\cellcolor{lightpurple}100k} & \multicolumn{1}{r|}{\cellcolor{lightpurple}0.8747} & \multicolumn{1}{r|}{\cellcolor{lightpurple}0.1240}                                               & \multicolumn{1}{r|}{\cellcolor{lightpurple}11.0931} & \multicolumn{1}{r|}{\cellcolor{lightpurple}10.20} & \multicolumn{1}{r|}{\cellcolor{lightpurple}12.21} & \multicolumn{1}{r|}{\cellcolor{lightpurple}44.60}                                                 & \cellcolor{lightpurple}56.00                                                                      \\ \hline
      \cellcolor{htmlwhite}                                & \multicolumn{1}{l|}{\cellcolor{htmlwhite}}                               & \multicolumn{1}{l|}{\cellcolor{htmlwhite}}                               & \multicolumn{1}{c|}{}                                                                              & \multicolumn{1}{c|}{\cellcolor{lightyellow}1M}   & \multicolumn{1}{r|}{\cellcolor{lightyellow}0.9999} & \multicolumn{1}{r|}{\cellcolor{lightyellow}0.0008}                                               & \multicolumn{1}{r|}{\cellcolor{lightyellow}11.6252} & \multicolumn{1}{r|}{\cellcolor{lightyellow}0.01}  & \multicolumn{1}{r|}{\cellcolor{lightyellow}0.01}  & \multicolumn{1}{r|}{\cellcolor{lightyellow}99.82}                                                 & \cellcolor{lightyellow}98.92                                                                      \\ \cline{5-12} 
      \cellcolor{htmlwhite}                                & \multicolumn{1}{l|}{\cellcolor{htmlwhite}}                               & \multicolumn{1}{l|}{\cellcolor{htmlwhite}}                               & \multicolumn{1}{c|}{}                                                                              & \multicolumn{1}{c|}{\cellcolor{lightblue}250k} & \multicolumn{1}{r|}{\cellcolor{lightblue}0.3764} & \multicolumn{1}{r|}{\cellcolor{lightblue}0.3522}                                               & \multicolumn{1}{r|}{\cellcolor{lightblue}4.7470}  & \multicolumn{1}{r|}{\cellcolor{lightblue}48.28} & \multicolumn{1}{r|}{\cellcolor{lightblue}58.93} & \multicolumn{1}{r|}{\cellcolor{lightblue}0.00}                                                  & \cellcolor{lightblue}1.40                                                                       \\ \cline{5-12} 
      \cellcolor{htmlwhite}                                & \multicolumn{1}{l|}{\multirow{-3}{*}{\cellcolor{htmlwhite}\LaTeX{}}}        & \multicolumn{1}{l|}{\multirow{-3}{*}{\cellcolor{htmlwhite}AsciiMath}}    & \multicolumn{1}{c|}{\multirow{-3}{*}{Forward}}                                                     & \multicolumn{1}{c|}{\cellcolor{lightpurple}1M}   & \multicolumn{1}{r|}{\cellcolor{lightpurple}0.7553} & \multicolumn{1}{r|}{\cellcolor{lightpurple}0.2994}                                               & \multicolumn{1}{r|}{\cellcolor{lightpurple}8.4978}  & \multicolumn{1}{r|}{\cellcolor{lightpurple}19.24} & \multicolumn{1}{r|}{\cellcolor{lightpurple}24.15} & \multicolumn{1}{r|}{\cellcolor{lightpurple}35.70}                                                 & \cellcolor{lightpurple}35.00                                                                      \\ \cline{2-12} 
      \cellcolor{htmlwhite}                                & \multicolumn{1}{l|}{\cellcolor{htmlwhite}}                               & \multicolumn{1}{l|}{\cellcolor{htmlwhite}}                               & \multicolumn{1}{c|}{}                                                                              & \multicolumn{1}{c|}{\cellcolor{lightyellow}1M}   & \multicolumn{1}{r|}{\cellcolor{lightyellow}0.9983} & \multicolumn{1}{r|}{\cellcolor{lightyellow}0.0030}                                               & \multicolumn{1}{r|}{\cellcolor{lightyellow}12.1796} & \multicolumn{1}{r|}{\cellcolor{lightyellow}0.11}  & \multicolumn{1}{r|}{\cellcolor{lightyellow}0.12}  & \multicolumn{1}{r|}{\cellcolor{lightyellow}98.86}                                                 & \cellcolor{lightyellow}99.86                                                                      \\ \cline{5-12} 
      \cellcolor{htmlwhite}                                & \multicolumn{1}{l|}{\cellcolor{htmlwhite}}                               & \multicolumn{1}{l|}{\cellcolor{htmlwhite}}                               & \multicolumn{1}{c|}{}                                                                              & \multicolumn{1}{c|}{\cellcolor{lightblue}50k}  & \multicolumn{1}{r|}{\cellcolor{lightblue}0.3025} & \multicolumn{1}{r|}{\cellcolor{lightblue}0.4990}                                               & \multicolumn{1}{r|}{\cellcolor{lightblue}4.4004}  & \multicolumn{1}{r|}{\cellcolor{lightblue}55.25} & \multicolumn{1}{r|}{\cellcolor{lightblue}67.26} & \multicolumn{1}{r|}{\cellcolor{lightblue}0.10}                                                  & \cellcolor{lightblue}0.40                                                                       \\ \cline{5-12} 
      \cellcolor{htmlwhite}                                & \multicolumn{1}{l|}{\multirow{-3}{*}{\cellcolor{htmlwhite}AsciiMath}}    & \multicolumn{1}{l|}{\multirow{-3}{*}{\cellcolor{htmlwhite}\LaTeX{}}}        & \multicolumn{1}{c|}{\multirow{-3}{*}{Backward}}                                                     & \multicolumn{1}{c|}{\cellcolor{lightpurple}1M}   & \multicolumn{1}{r|}{\cellcolor{lightpurple}0.6899} & \multicolumn{1}{r|}{\cellcolor{lightpurple}0.3470}                                               & \multicolumn{1}{r|}{\cellcolor{lightpurple}8.1353}  & \multicolumn{1}{r|}{\cellcolor{lightpurple}24.57} & \multicolumn{1}{r|}{\cellcolor{lightpurple}32.19} & \multicolumn{1}{r|}{\cellcolor{lightpurple}30.40}                                                 & \cellcolor{lightpurple}42.50                                                                      \\ \cline{2-12} 
      \rowcolor[HTML]{C0C0C0} 
      \cellcolor{htmlwhite}                                & \multicolumn{11}{c|}{\cellcolor[HTML]{C0C0C0}Outliers}                                                                                                                                                                                                                                                                                                                                                                                                                                                                                                                                                                                                                                                                                                                                                                                                  \\ \cline{2-12} 
      \cellcolor{htmlwhite}                                & \multicolumn{1}{l|}{\cellcolor{htmlwhite}}                               & \multicolumn{1}{l|}{\cellcolor{htmlwhite}}                               & \multicolumn{1}{c|}{}                                                                              & \multicolumn{1}{c|}{\cellcolor{lightyellow}500k} & \multicolumn{1}{r|}{\cellcolor{lightyellow}0.9846} & \multicolumn{1}{r|}{\cellcolor{lightyellow}0.0086}                                               & \multicolumn{1}{r|}{\cellcolor{lightyellow}11.9758} & \multicolumn{1}{r|}{\cellcolor{lightyellow}0.93}  & \multicolumn{1}{r|}{\cellcolor{lightyellow}1.36}  & \multicolumn{1}{r|}{\cellcolor{lightyellow}93.33}                                                 & \cellcolor{lightyellow}93.81                                                                      \\ \cline{5-12} 
      \cellcolor{htmlwhite}                                & \multicolumn{1}{l|}{\cellcolor{htmlwhite}}                               & \multicolumn{1}{l|}{\cellcolor{htmlwhite}}                               & \multicolumn{1}{c|}{}                                                                              & \multicolumn{1}{c|}{\cellcolor{lightblue}100k} & \multicolumn{1}{r|}{\cellcolor{lightblue}0.0568} & \multicolumn{1}{r|}{\cellcolor{lightblue}3.6482}                                               & \multicolumn{1}{r|}{\cellcolor{lightblue}1.6700}  & \multicolumn{1}{r|}{\cellcolor{lightblue}84.26} & \multicolumn{1}{r|}{\cellcolor{lightblue}94.50} & \multicolumn{1}{r|}{\cellcolor{lightblue}0.00}                                                  & \cellcolor{lightblue}0.00                                                                       \\ \cline{5-12} 
      \cellcolor{htmlwhite}                                & \multicolumn{1}{l|}{\multirow{-3}{*}{\cellcolor{htmlwhite}MathML}}       & \multicolumn{1}{l|}{\multirow{-3}{*}{\cellcolor{htmlwhite}\LaTeX{}}}        & \multicolumn{1}{c|}{\multirow{-3}{*}{Forward}}                                                     & \multicolumn{1}{c|}{\cellcolor{lightpurple}5k}   & \multicolumn{1}{r|}{\cellcolor{lightpurple}0.2674} & \multicolumn{1}{r|}{\cellcolor{lightpurple}1.8474}                                               & \multicolumn{1}{r|}{\cellcolor{lightpurple}3.3514}  & \multicolumn{1}{r|}{\cellcolor{lightpurple}58.75} & \multicolumn{1}{r|}{\cellcolor{lightpurple}71.58} & \multicolumn{1}{r|}{\cellcolor{lightpurple}3.00}                                                  & \cellcolor{lightpurple}1.40                                                                       \\ \cline{2-12} 
      \cellcolor{htmlwhite}                                & \multicolumn{1}{l|}{\cellcolor{htmlwhite}}                               & \multicolumn{1}{l|}{\cellcolor{htmlwhite}}                               & \multicolumn{1}{c|}{}                                                                              & \multicolumn{1}{c|}{\cellcolor{lightyellow}500k} & \multicolumn{1}{r|}{\cellcolor{lightyellow}0.9782} & \multicolumn{1}{r|}{\cellcolor{lightyellow}0.0107}                                               & \multicolumn{1}{r|}{\cellcolor{lightyellow}9.2103}  & \multicolumn{1}{r|}{\cellcolor{lightyellow}0.88}  & \multicolumn{1}{r|}{\cellcolor{lightyellow}1.71}  & \multicolumn{1}{r|}{\cellcolor{lightyellow}61.09}                                                 & \cellcolor{lightyellow}93.44                                                                      \\ \cline{5-12} 
      \cellcolor{htmlwhite}                                & \multicolumn{1}{l|}{\cellcolor{htmlwhite}}                               & \multicolumn{1}{l|}{\cellcolor{htmlwhite}}                               & \multicolumn{1}{c|}{}                                                                              & \multicolumn{1}{c|}{\cellcolor{lightblue}500k} & \multicolumn{1}{r|}{\cellcolor{lightblue}0.0000} & \multicolumn{1}{r|}{\cellcolor{lightblue}0.5871}                                               & \multicolumn{1}{r|}{\cellcolor{lightblue}0.0171}  & \multicolumn{1}{r|}{\cellcolor{lightblue}97.16} & \multicolumn{1}{r|}{\cellcolor{lightblue}99.69} & \multicolumn{1}{r|}{\cellcolor{lightblue}0.00}                                                  & \cellcolor{lightblue}0.20                                                                       \\ \cline{5-12} 
      \cellcolor{htmlwhite}                                & \multicolumn{1}{l|}{\multirow{-3}{*}{\cellcolor{htmlwhite}\LaTeX{}}}        & \multicolumn{1}{l|}{\multirow{-3}{*}{\cellcolor{htmlwhite}MathML}}       & \multicolumn{1}{c|}{\multirow{-3}{*}{Backward}}                                                     & \multicolumn{1}{c|}{\cellcolor{lightpurple}10k}  & \multicolumn{1}{r|}{\cellcolor{lightpurple}0.5336} & \multicolumn{1}{r|}{\cellcolor{lightpurple}0.5286}                                               & \multicolumn{1}{r|}{\cellcolor{lightpurple}4.1001}  & \multicolumn{1}{r|}{\cellcolor{lightpurple}39.32} & \multicolumn{1}{r|}{\cellcolor{lightpurple}50.64} & \multicolumn{1}{r|}{\cellcolor{lightpurple}1.00}                                                  & \cellcolor{lightpurple}7.40                                                                       \\ \cline{2-12} 
      \cellcolor{htmlwhite}                                & \multicolumn{1}{l|}{\cellcolor{htmlwhite}}                               & \multicolumn{1}{l|}{\cellcolor{htmlwhite}}                               & \multicolumn{1}{c|}{}                                                                              & \multicolumn{1}{c|}{\cellcolor{lightyellow}1M}   & \multicolumn{1}{r|}{\cellcolor{lightyellow}0.9780} & \multicolumn{1}{r|}{\cellcolor{lightyellow}0.0101}                                               & \multicolumn{1}{r|}{\cellcolor{lightyellow}9.2108}  & \multicolumn{1}{r|}{\cellcolor{lightyellow}0.92}  & \multicolumn{1}{r|}{\cellcolor{lightyellow}1.74}  & \multicolumn{1}{r|}{\cellcolor{lightyellow}60.81}                                                 & \cellcolor{lightyellow}94.01                                                                      \\ \cline{5-12} 
      \cellcolor{htmlwhite}                                & \multicolumn{1}{l|}{\cellcolor{htmlwhite}}                               & \multicolumn{1}{l|}{\cellcolor{htmlwhite}}                               & \multicolumn{1}{c|}{}                                                                              & \multicolumn{1}{c|}{\cellcolor{lightblue}500k} & \multicolumn{1}{r|}{\cellcolor{lightblue}0.0000} & \multicolumn{1}{r|}{\cellcolor{lightblue}3.6580}                                               & \multicolumn{1}{r|}{\cellcolor{lightblue}0.0134}  & \multicolumn{1}{r|}{\cellcolor{lightblue}97.40} & \multicolumn{1}{r|}{\cellcolor{lightblue}99.72} & \multicolumn{1}{r|}{\cellcolor{lightblue}0.00}                                                  & \cellcolor{lightblue}0.00                                                                       \\ \cline{5-12} 
      \cellcolor{htmlwhite}                                & \multicolumn{1}{l|}{\multirow{-3}{*}{\cellcolor{htmlwhite}AsciiMath}}    & \multicolumn{1}{l|}{\multirow{-3}{*}{\cellcolor{htmlwhite}MathML}}       & \multicolumn{1}{c|}{\multirow{-3}{*}{\begin{tabular}[c]{@{}c@{}}Hypothet.\\ Forward\end{tabular}}} & \multicolumn{1}{c|}{\cellcolor{lightpurple}250k} & \multicolumn{1}{r|}{\cellcolor{lightpurple}0.3084} & \multicolumn{1}{r|}{\cellcolor{lightpurple}2.7773}                                               & \multicolumn{1}{r|}{\cellcolor{lightpurple}1.1503}  & \multicolumn{1}{r|}{\cellcolor{lightpurple}53.34} & \multicolumn{1}{r|}{\cellcolor{lightpurple}59.63} & \multicolumn{1}{r|}{\cellcolor{lightpurple}0.70}                                                  & \cellcolor{lightpurple}0.70                                                                       \\ \cline{2-12} 
      \cellcolor{htmlwhite}                                & \multicolumn{1}{l|}{\cellcolor{htmlwhite}}                               & \multicolumn{1}{l|}{\cellcolor{htmlwhite}}                               & \multicolumn{1}{c|}{}                                                                              & \multicolumn{1}{c|}{\cellcolor{lightyellow}500k} & \multicolumn{1}{r|}{\cellcolor{lightyellow}0.9844} & \multicolumn{1}{r|}{\cellcolor{lightyellow}0.0112}                                               & \multicolumn{1}{r|}{\cellcolor{lightyellow}11.4014} & \multicolumn{1}{r|}{\cellcolor{lightyellow}0.94}  & \multicolumn{1}{r|}{\cellcolor{lightyellow}1.37}  & \multicolumn{1}{r|}{\cellcolor{lightyellow}93.31}                                                 & \cellcolor{lightyellow}93.41                                                                      \\ \cline{5-12} 
      \cellcolor{htmlwhite}                                & \multicolumn{1}{l|}{\cellcolor{htmlwhite}}                               & \multicolumn{1}{l|}{\cellcolor{htmlwhite}}                               & \multicolumn{1}{c|}{}                                                                              & \multicolumn{1}{c|}{\cellcolor{lightblue}5k}   & \multicolumn{1}{r|}{\cellcolor{lightblue}0.1028} & \multicolumn{1}{r|}{\cellcolor{lightblue}0.6621}                                               & \multicolumn{1}{r|}{\cellcolor{lightblue}1.9223}  & \multicolumn{1}{r|}{\cellcolor{lightblue}74.02} & \multicolumn{1}{r|}{\cellcolor{lightblue}84.91} & \multicolumn{1}{r|}{\cellcolor{lightblue}0.00}                                                  & \cellcolor{lightblue}0.20                                                                       \\ \cline{5-12} 
      \multirow{-19}{*}{\cellcolor{htmlwhite}py-asciimath} & \multicolumn{1}{l|}{\multirow{-3}{*}{\cellcolor{htmlwhite}MathML}}       & \multicolumn{1}{l|}{\multirow{-3}{*}{\cellcolor{htmlwhite}AsciiMath}}    & \multicolumn{1}{c|}{\multirow{-3}{*}{\begin{tabular}[c]{@{}c@{}}Hypothet.\\ Backward\end{tabular}}} & \multicolumn{1}{c|}{\cellcolor{lightpurple}5k}   & \multicolumn{1}{r|}{\cellcolor{lightpurple}0.2963} & \multicolumn{1}{r|}{\cellcolor{lightpurple}0.5779}                                               & \multicolumn{1}{r|}{\cellcolor{lightpurple}3.6842}  & \multicolumn{1}{r|}{\cellcolor{lightpurple}54.06} & \multicolumn{1}{r|}{\cellcolor{lightpurple}67.69} & \multicolumn{1}{r|}{\cellcolor{lightpurple}3.40}                                                  & \cellcolor{lightpurple}8.80                                                                       \\ \hline
  \end{tabular}
  \parbox{\linewidth}{\centering \small \textit{Yellow cells \textcolor{lightyellow}{\fsquare} = evaluation with synthetic data; Blue cells \textcolor{lightblue}{\fsquare} = evaluation with real-world data;\\purple cells \textcolor{lightpurple}{\fsquare} = evaluation with real-world data after few-shot fine-tuning.}}
  % \parbox{\linewidth}{\centering \small \textit{Yellow cells \textcolor{lightyellow}{\fsquare} indicate an evaluation run with synthetic test data, blue cells \textcolor{lightblue}{\fsquare} indicate evaluation runs with the real-world data, and purple cells \textcolor{lightpurple}{\fsquare} indicate the evaluation runs on the real-world data performed after fine-tuning the baseline model on a few real-world data pairs.}}
\end{table}

\subsection*{\ref{rq:mocking}: How accurate are the \emph{\Modelizer models} in mocking the behavior of the \PUT\hspace{0.5mm}and its inverse \PUTinverse?}
  We start with addressing \ref{rq:mocking}: \emph{How well do \Modelizer models predict outputs from inputs?}
  To answer this question, we evaluated our models on sets of 10,000 synthesized test cases per subject that have no intersection with any of the training sets.
  The results of the evaluation are summarized in \Cref{tab:eval-results}.
  They show that the models trained on the synthetic data can achieve a high prediction accuracy for most subjects from the evaluation set. For most models and subjects, the BLEU score is above 0.98, the Word Information Lost score is below 0.02 and the Exact Match is above 93\%.

\begin{result}
  \Modelizer models mock the program behavior without making errors in more than 93\% of cases.
\end{result}

  Next, we evaluated the accuracy of \Modelizer when predicting \emph{inputs from outputs.}
  The prediction accuracy scores of HTML to Markdown, KQL to SQL, \LaTeX{} to PyExpression, and AsciiMath to \LaTeX{} models depicted in \Cref{tab:eval-results} show that the inverse model achieves a similar Exact Match accuracy as their counterpart forward conversion models. 
  The recorded BLEU Score values for both forward and inverse models are high and have insignificant margins between them. 
  While the models of the \texttt{Pandoc} and \texttt{msticpy} library achieve more than~90\% accuracy in both directions, the \texttt{latexify-py} and \texttt{py-asciimath} can produce fully correct results in half of the cases. 
  Between 20\%~to~35\% of predictions require up to 4 edit operations, including token insertions and deletions, to correct the prediction. Currently, \Modelizer can automatically detect these failures using the prediction validation mechanism discussed in \Cref{sec:deployment}. We plan to integrate the automatic prediction repair mechanism to tackle inaccurate prediction cases in our future work.

\begin{table}[]
  \small
  \centering
  \caption{\label{tab:pandoc_forward} Evaluation results of Markdown to HTML conversions done with the learned models of Pandoc converter.}
  \begin{tabular}{|lc|rrrr|rrrr|}
  \hline
  \rowcolor{lightgrey} 
  \multicolumn{2}{|c|}{\cellcolor{lightgrey}Dataset}                                              & \multicolumn{4}{c|}{\cellcolor{lightgrey}Trained on Synthetic Data}                                                                                                                                                                                                                                             & \multicolumn{4}{c|}{\cellcolor{lightgrey}Fine-Tuned on Real Data}                                                                                                                                                                                                                                               \\ \hline
  \rowcolor{lightgrey} 
  \multicolumn{1}{|c|}{\cellcolor{lightgrey}Size} & Partition                                     & \multicolumn{1}{c|}{\cellcolor{lightgrey}BLEU}   & \multicolumn{1}{c|}{\cellcolor{lightgrey}WIL}   & \multicolumn{1}{c|}{\cellcolor{lightgrey}\begin{tabular}[c]{@{}c@{}}Exact\\ Match\end{tabular}} & \multicolumn{1}{c|}{\cellcolor{lightgrey}\begin{tabular}[c]{@{}c@{}}Close\\ Match\end{tabular}} & \multicolumn{1}{c|}{\cellcolor{lightgrey}BLEU}   & \multicolumn{1}{c|}{\cellcolor{lightgrey}WIL}   & \multicolumn{1}{c|}{\cellcolor{lightgrey}\begin{tabular}[c]{@{}c@{}}Exact\\ Match\end{tabular}} & \multicolumn{1}{c|}{\cellcolor{lightgrey}\begin{tabular}[c]{@{}c@{}}Close\\ Match\end{tabular}} \\ \hline
  \multicolumn{1}{|l|}{}                             & \cellcolor{lightblue}\xmark & \multicolumn{1}{r|}{\cellcolor{lightblue}0.2576} & \multicolumn{1}{r|}{\cellcolor{lightblue}76.16} & \multicolumn{1}{r|}{\cellcolor{lightblue}51.40}                                                 & \cellcolor{lightblue}55.93                                                                      & \multicolumn{1}{r|}{\cellcolor{lightblue}0.6763} & \multicolumn{1}{r|}{\cellcolor{lightblue}36.97} & \multicolumn{1}{r|}{\cellcolor{lightblue}89.80}                                                 & \cellcolor{lightblue}94.01                                                                      \\ \cline{2-10} 
  \multicolumn{1}{|l|}{\multirow{-2}{*}{5k}}         & \cellcolor{lightpurple}\cmark & \multicolumn{1}{r|}{\cellcolor{lightpurple}0.2710} & \multicolumn{1}{r|}{\cellcolor{lightpurple}72.94} & \multicolumn{1}{r|}{\cellcolor{lightpurple}51.64}                                                 & \cellcolor{lightpurple}55.96                                                                      & \multicolumn{1}{r|}{\cellcolor{lightpurple}0.6813} & \multicolumn{1}{r|}{\cellcolor{lightpurple}42.80} & \multicolumn{1}{r|}{\cellcolor{lightpurple}89.35}                                                 & \cellcolor{lightpurple}93.61                                                                      \\ \hline
  \multicolumn{1}{|l|}{}                             & \cellcolor{lightblue}\xmark & \multicolumn{1}{r|}{\cellcolor{lightblue}0.2746} & \multicolumn{1}{r|}{\cellcolor{lightblue}74.28} & \multicolumn{1}{r|}{\cellcolor{lightblue}51.40}                                                 & \cellcolor{lightblue}55.93                                                                      & \multicolumn{1}{r|}{\cellcolor{lightblue}0.6934} & \multicolumn{1}{r|}{\cellcolor{lightblue}40.00} & \multicolumn{1}{r|}{\cellcolor{lightblue}88.53}                                                 & \cellcolor{lightblue}92.73                                                                      \\ \cline{2-10} 
  \multicolumn{1}{|l|}{\multirow{-2}{*}{10k}}        & \cellcolor{lightpurple}\cmark & \multicolumn{1}{r|}{\cellcolor{lightpurple}0.2723} & \multicolumn{1}{r|}{\cellcolor{lightpurple}75.36} & \multicolumn{1}{r|}{\cellcolor{lightpurple}44.43}                                                 & \cellcolor{lightpurple}52.11                                                                      & \multicolumn{1}{r|}{\cellcolor{lightpurple}0.6747} & \multicolumn{1}{r|}{\cellcolor{lightpurple}42.95} & \multicolumn{1}{r|}{\cellcolor{lightpurple}89.07}                                                 & \cellcolor{lightpurple}93.24                                                                      \\ \hline
  \multicolumn{1}{|l|}{}                             & \cellcolor{lightblue}\xmark & \multicolumn{1}{r|}{\cellcolor{lightblue}0.2851} & \multicolumn{1}{r|}{\cellcolor{lightblue}75.21} & \multicolumn{1}{r|}{\cellcolor{lightblue}51.71}                                                 & \cellcolor{lightblue}56.01                                                                      & \multicolumn{1}{r|}{\cellcolor{lightblue}0.6735} & \multicolumn{1}{r|}{\cellcolor{lightblue}43.97} & \multicolumn{1}{r|}{\cellcolor{lightblue}89.42}                                                 & \cellcolor{lightblue}93.80                                                                      \\ \cline{2-10} 
  \multicolumn{1}{|l|}{\multirow{-2}{*}{50k}}        & \cellcolor{lightpurple}\cmark & \multicolumn{1}{r|}{\cellcolor{lightpurple}0.2849} & \multicolumn{1}{r|}{\cellcolor{lightpurple}73.25} & \multicolumn{1}{r|}{\cellcolor{lightpurple}51.69}                                                 & \cellcolor{lightpurple}56.03                                                                      & \multicolumn{1}{r|}{\cellcolor{lightpurple}0.6874} & \multicolumn{1}{r|}{\cellcolor{lightpurple}40.49} & \multicolumn{1}{r|}{\cellcolor{lightpurple}89.72}                                                 & \cellcolor{lightpurple}93.84                                                                      \\ \hline
  \multicolumn{1}{|l|}{}                             & \cellcolor{lightblue}\cmark & \multicolumn{1}{r|}{\cellcolor{lightblue}0.2923} & \multicolumn{1}{r|}{\cellcolor{lightblue}72.13} & \multicolumn{1}{r|}{\cellcolor{lightblue}51.72}                                                 & \cellcolor{lightblue}56.03                                                                      & \multicolumn{1}{r|}{\cellcolor{lightblue}0.6966} & \multicolumn{1}{r|}{\cellcolor{lightblue}36.79} & \multicolumn{1}{r|}{\cellcolor{lightblue}88.66}                                                 & \cellcolor{lightblue}92.88                                                                      \\ \cline{2-10} 
  \multicolumn{1}{|l|}{\multirow{-2}{*}{100k}}       & \cellcolor{lightpurple}\cmark & \multicolumn{1}{r|}{\cellcolor{lightpurple}0.2704} & \multicolumn{1}{r|}{\cellcolor{lightpurple}73.26} & \multicolumn{1}{r|}{\cellcolor{lightpurple}51.61}                                                 & \cellcolor{lightpurple}55.89                                                                      & \multicolumn{1}{r|}{\cellcolor{lightpurple}0.6592} & \multicolumn{1}{r|}{\cellcolor{lightpurple}42.85} & \multicolumn{1}{r|}{\cellcolor{lightpurple}88.99}                                                 & \cellcolor{lightpurple}93.27                                                                      \\ \hline
  \multicolumn{1}{|l|}{}                             & \cellcolor{lightblue}\xmark & \multicolumn{1}{r|}{\cellcolor{lightblue}0.2898} & \multicolumn{1}{r|}{\cellcolor{lightblue}75.39} & \multicolumn{1}{r|}{\cellcolor{lightblue}51.72}                                                 & \cellcolor{lightblue}56.03                                                                      & \multicolumn{1}{r|}{\cellcolor{lightblue}0.6725} & \multicolumn{1}{r|}{\cellcolor{lightblue}38.03} & \multicolumn{1}{r|}{\cellcolor{lightblue}89.30}                                                 & \cellcolor{lightblue}93.56                                                                      \\ \cline{2-10} 
  \multicolumn{1}{|l|}{\multirow{-2}{*}{250k}}       & \cellcolor{lightpurple}\cmark & \multicolumn{1}{r|}{\cellcolor{lightpurple}0.2824} & \multicolumn{1}{r|}{\cellcolor{lightpurple}74.49} & \multicolumn{1}{r|}{\cellcolor{lightpurple}49.67}                                                 & \cellcolor{lightpurple}52.17                                                                      & \multicolumn{1}{r|}{\cellcolor{lightpurple}0.6753} & \multicolumn{1}{r|}{\cellcolor{lightpurple}37.61} & \multicolumn{1}{r|}{\cellcolor{lightpurple}89.03}                                                 & \cellcolor{lightpurple}93.23                                                                      \\ \hline
  \multicolumn{1}{|l|}{}                             & \cellcolor{lightblue}\xmark & \multicolumn{1}{r|}{\cellcolor{lightblue}0.2900} & \multicolumn{1}{r|}{\cellcolor{lightblue}70.90} & \multicolumn{1}{r|}{\cellcolor{lightblue}40.49}                                                 & \cellcolor{lightblue}55.78                                                                      & \multicolumn{1}{r|}{\cellcolor{lightblue}0.7128} & \multicolumn{1}{r|}{\cellcolor{lightblue}35.30} & \multicolumn{1}{r|}{\cellcolor{lightblue}90.57}                                                 & \cellcolor{lightblue}94.79                                                                      \\ \cline{2-10} 
  \multicolumn{1}{|l|}{\multirow{-2}{*}{500k}}       & \cellcolor{lightpurple}\cmark & \multicolumn{1}{r|}{\cellcolor{lightpurple}0.2950} & \multicolumn{1}{r|}{\cellcolor{lightpurple}68.56} & \multicolumn{1}{r|}{\cellcolor{lightpurple}51.71}                                                 & \cellcolor{lightpurple}56.02                                                                      & \multicolumn{1}{r|}{\cellcolor{lightpurple}0.6867} & \multicolumn{1}{r|}{\cellcolor{lightpurple}35.64} & \multicolumn{1}{r|}{\cellcolor{lightpurple}89.31}                                                 & \cellcolor{lightpurple}93.76                                                                      \\ \hline
  \multicolumn{1}{|l|}{}                             & \cellcolor{lightblue}\xmark & \multicolumn{1}{r|}{\cellcolor{lightblue}0.2535} & \multicolumn{1}{r|}{\cellcolor{lightblue}72.90} & \multicolumn{1}{r|}{\cellcolor{lightblue}24.38}                                                 & \cellcolor{lightblue}28.68                                                                      & \multicolumn{1}{r|}{\cellcolor{lightblue}0.6886} & \multicolumn{1}{r|}{\cellcolor{lightblue}33.95} & \multicolumn{1}{r|}{\cellcolor{lightblue}90.25}                                                 & \cellcolor{lightblue}94.56                                                                      \\ \cline{2-10} 
  \multicolumn{1}{|l|}{\multirow{-2}{*}{1000k}}      & \cellcolor{lightpurple}\cmark & \multicolumn{1}{r|}{\cellcolor{lightpurple}0.2940} & \multicolumn{1}{r|}{\cellcolor{lightpurple}70.49} & \multicolumn{1}{r|}{\cellcolor{lightpurple}51.61}                                                 & \cellcolor{lightpurple}55.91                                                                      & \multicolumn{1}{r|}{\cellcolor{lightpurple}0.7208} & \multicolumn{1}{r|}{\cellcolor{lightpurple}36.64} & \multicolumn{1}{r|}{\cellcolor{lightpurple}90.35}                                                 & \cellcolor{lightpurple}94.69                                                                      \\ \hline
  \end{tabular}
  \parbox{0.72\linewidth}{\centering \small \textit{Blue cells \textcolor{lightblue}{\fsquare} = fixed number of expansions; purple cells \textcolor{lightpurple}{\fsquare} = increasing number of expansions.}}
  % \parbox{0.72\linewidth}{\small \textit{Models were trained on various dataset sizes with synthesized Markdown-to-HTML structures. Blue~\textcolor{lightblue}{\fsquare} cells mark models trained on data generated with a fixed number of non-terminal expansions, and purple~\textcolor{lightpurple}{\fsquare} cells mark training on the data with an increasing number of non-terminal expansions.}}
\end{table}

\begin{table}[]
  \small
  \centering
  \caption{\label{tab:pandoc_inverse} Evaluation results of HTML to Markdown conversions done with the learned inverse models of Pandoc.}
  \begin{tabular}{|lc|rrrr|rrrr|}
  \hline
  \rowcolor{lightgrey} 
  \multicolumn{2}{|c|}{\cellcolor{lightgrey}Dataset}                                              & \multicolumn{4}{c|}{\cellcolor{lightgrey}Trained on Synthetic Data}                                                                                                                                                                                                                                             & \multicolumn{4}{c|}{\cellcolor{lightgrey}Fine-Tuned on Real Data}                                                                                                                                                                                                                                               \\ \hline
  \rowcolor{lightgrey} 
  \multicolumn{1}{|c|}{\cellcolor{lightgrey}Size} & Partition                                     & \multicolumn{1}{c|}{\cellcolor{lightgrey}BLEU}   & \multicolumn{1}{c|}{\cellcolor{lightgrey}WIL}   & \multicolumn{1}{c|}{\cellcolor{lightgrey}\begin{tabular}[c]{@{}c@{}}Exact\\ Match\end{tabular}} & \multicolumn{1}{c|}{\cellcolor{lightgrey}\begin{tabular}[c]{@{}c@{}}Close\\ Match\end{tabular}} & \multicolumn{1}{c|}{\cellcolor{lightgrey}BLEU}   & \multicolumn{1}{c|}{\cellcolor{lightgrey}WIL}   & \multicolumn{1}{c|}{\cellcolor{lightgrey}\begin{tabular}[c]{@{}c@{}}Exact\\ Match\end{tabular}} & \multicolumn{1}{c|}{\cellcolor{lightgrey}\begin{tabular}[c]{@{}c@{}}Close\\ Match\end{tabular}} \\ \hline
  \multicolumn{1}{|l|}{}                             & \cellcolor{lightblue}\xmark & \multicolumn{1}{r|}{\cellcolor{lightblue}0.1988} & \multicolumn{1}{r|}{\cellcolor{lightblue}76.01} & \multicolumn{1}{r|}{\cellcolor{lightblue}55.08}                                                 & \cellcolor{lightblue}56.83                                                                      & \multicolumn{1}{r|}{\cellcolor{lightblue}0.6441} & \multicolumn{1}{r|}{\cellcolor{lightblue}43.73} & \multicolumn{1}{r|}{\cellcolor{lightblue}81.51}                                                 & \cellcolor{lightblue}90.85                                                                      \\ \cline{2-10} 
  \multicolumn{1}{|l|}{\multirow{-2}{*}{5k}}         & \cellcolor{lightpurple}\cmark & \multicolumn{1}{r|}{\cellcolor{lightpurple}0.1886} & \multicolumn{1}{r|}{\cellcolor{lightpurple}76.63} & \multicolumn{1}{r|}{\cellcolor{lightpurple}28.36}                                                 & \cellcolor{lightpurple}57.40                                                                      & \multicolumn{1}{r|}{\cellcolor{lightpurple}0.6550} & \multicolumn{1}{r|}{\cellcolor{lightpurple}42.45} & \multicolumn{1}{r|}{\cellcolor{lightpurple}91.14}                                                 & \cellcolor{lightpurple}91.97                                                                      \\ \hline
  \multicolumn{1}{|l|}{}                             & \cellcolor{lightblue}\xmark & \multicolumn{1}{r|}{\cellcolor{lightblue}0.2006} & \multicolumn{1}{r|}{\cellcolor{lightblue}74.47} & \multicolumn{1}{r|}{\cellcolor{lightblue}55.71}                                                 & \cellcolor{lightblue}57.46                                                                      & \multicolumn{1}{r|}{\cellcolor{lightblue}0.6441} & \multicolumn{1}{r|}{\cellcolor{lightblue}44.38} & \multicolumn{1}{r|}{\cellcolor{lightblue}90.39}                                                 & \cellcolor{lightblue}91.26                                                                      \\ \cline{2-10} 
  \multicolumn{1}{|l|}{\multirow{-2}{*}{10k}}        & \cellcolor{lightpurple}\cmark & \multicolumn{1}{r|}{\cellcolor{lightpurple}0.2316} & \multicolumn{1}{r|}{\cellcolor{lightpurple}72.49} & \multicolumn{1}{r|}{\cellcolor{lightpurple}28.39}                                                 & \cellcolor{lightpurple}57.47                                                                      & \multicolumn{1}{r|}{\cellcolor{lightpurple}0.6280} & \multicolumn{1}{r|}{\cellcolor{lightpurple}45.16} & \multicolumn{1}{r|}{\cellcolor{lightpurple}87.64}                                                 & \cellcolor{lightpurple}88.59                                                                      \\ \hline
  \multicolumn{1}{|l|}{}                             & \cellcolor{lightblue}\xmark & \multicolumn{1}{r|}{\cellcolor{lightblue}0.2115} & \multicolumn{1}{r|}{\cellcolor{lightblue}70.59} & \multicolumn{1}{r|}{\cellcolor{lightblue}53.72}                                                 & \cellcolor{lightblue}55.25                                                                      & \multicolumn{1}{r|}{\cellcolor{lightblue}0.6282} & \multicolumn{1}{r|}{\cellcolor{lightblue}43.94} & \multicolumn{1}{r|}{\cellcolor{lightblue}92.15}                                                 & \cellcolor{lightblue}92.93                                                                      \\ \cline{2-10} 
  \multicolumn{1}{|l|}{\multirow{-2}{*}{50k}}        & \cellcolor{lightpurple}\cmark & \multicolumn{1}{r|}{\cellcolor{lightpurple}0.2503} & \multicolumn{1}{r|}{\cellcolor{lightpurple}65.47} & \multicolumn{1}{r|}{\cellcolor{lightpurple}55.75}                                                 & \cellcolor{lightpurple}56.84                                                                      & \multicolumn{1}{r|}{\cellcolor{lightpurple}0.6170} & \multicolumn{1}{r|}{\cellcolor{lightpurple}46.18} & \multicolumn{1}{r|}{\cellcolor{lightpurple}88.61}                                                 & \cellcolor{lightpurple}88.97                                                                      \\ \hline
  \multicolumn{1}{|l|}{}                             & \cellcolor{lightblue}\cmark & \multicolumn{1}{r|}{\cellcolor{lightblue}0.2096} & \multicolumn{1}{r|}{\cellcolor{lightblue}69.33} & \multicolumn{1}{r|}{\cellcolor{lightblue}50.58}                                                 & \cellcolor{lightblue}57.31                                                                      & \multicolumn{1}{r|}{\cellcolor{lightblue}0.6119} & \multicolumn{1}{r|}{\cellcolor{lightblue}45.50} & \multicolumn{1}{r|}{\cellcolor{lightblue}88.60}                                                 & \cellcolor{lightblue}88.95                                                                      \\ \cline{2-10} 
  \multicolumn{1}{|l|}{\multirow{-2}{*}{100k}}       & \cellcolor{lightpurple}\cmark & \multicolumn{1}{r|}{\cellcolor{lightpurple}0.2112} & \multicolumn{1}{r|}{\cellcolor{lightpurple}70.16} & \multicolumn{1}{r|}{\cellcolor{lightpurple}28.62}                                                 & \cellcolor{lightpurple}56.03                                                                      & \multicolumn{1}{r|}{\cellcolor{lightpurple}0.6471} & \multicolumn{1}{r|}{\cellcolor{lightpurple}37.28} & \multicolumn{1}{r|}{\cellcolor{lightpurple}89.24}                                                 & \cellcolor{lightpurple}91.39                                                                      \\ \hline
  \multicolumn{1}{|l|}{}                             & \cellcolor{lightblue}\xmark & \multicolumn{1}{r|}{\cellcolor{lightblue}0.2218} & \multicolumn{1}{r|}{\cellcolor{lightblue}65.89} & \multicolumn{1}{r|}{\cellcolor{lightblue}55.78}                                                 & \cellcolor{lightblue}57.31                                                                      & \multicolumn{1}{r|}{\cellcolor{lightblue}0.6499} & \multicolumn{1}{r|}{\cellcolor{lightblue}40.72} & \multicolumn{1}{r|}{\cellcolor{lightblue}90.57}                                                 & \cellcolor{lightblue}90.97                                                                      \\ \cline{2-10} 
  \multicolumn{1}{|l|}{\multirow{-2}{*}{250k}}       & \cellcolor{lightpurple}\cmark & \multicolumn{1}{r|}{\cellcolor{lightpurple}0.2352} & \multicolumn{1}{r|}{\cellcolor{lightpurple}66.00} & \multicolumn{1}{r|}{\cellcolor{lightpurple}55.78}                                                 & \cellcolor{lightpurple}57.51                                                                      & \multicolumn{1}{r|}{\cellcolor{lightpurple}0.6778} & \multicolumn{1}{r|}{\cellcolor{lightpurple}39.30} & \multicolumn{1}{r|}{\cellcolor{lightpurple}92.20}                                                 & \cellcolor{lightpurple}92.52                                                                      \\ \hline
  \multicolumn{1}{|l|}{}                             & \cellcolor{lightblue}\xmark & \multicolumn{1}{r|}{\cellcolor{lightblue}0.2251} & \multicolumn{1}{r|}{\cellcolor{lightblue}66.88} & \multicolumn{1}{r|}{\cellcolor{lightblue}55.78}                                                 & \cellcolor{lightblue}55.89                                                                      & \multicolumn{1}{r|}{\cellcolor{lightblue}0.6793} & \multicolumn{1}{r|}{\cellcolor{lightblue}36.75} & \multicolumn{1}{r|}{\cellcolor{lightblue}91.51}                                                 & \cellcolor{lightblue}92.72                                                                      \\ \cline{2-10} 
  \multicolumn{1}{|l|}{\multirow{-2}{*}{500k}}       & \cellcolor{lightpurple}\cmark & \multicolumn{1}{r|}{\cellcolor{lightpurple}0.2145} & \multicolumn{1}{r|}{\cellcolor{lightpurple}68.21} & \multicolumn{1}{r|}{\cellcolor{lightpurple}19.78}                                                 & \cellcolor{lightpurple}47.65                                                                      & \multicolumn{1}{r|}{\cellcolor{lightpurple}0.6646} & \multicolumn{1}{r|}{\cellcolor{lightpurple}40.97} & \multicolumn{1}{r|}{\cellcolor{lightpurple}89.02}                                                 & \cellcolor{lightpurple}89.56                                                                      \\ \hline
  \multicolumn{1}{|l|}{}                             & \cellcolor{lightblue}\xmark & \multicolumn{1}{r|}{\cellcolor{lightblue}0.2326} & \multicolumn{1}{r|}{\cellcolor{lightblue}67.26} & \multicolumn{1}{r|}{\cellcolor{lightblue}54.97}                                                 & \cellcolor{lightblue}56.31                                                                      & \multicolumn{1}{r|}{\cellcolor{lightblue}0.6749} & \multicolumn{1}{r|}{\cellcolor{lightblue}39.36} & \multicolumn{1}{r|}{\cellcolor{lightblue}91.08}                                                 & \cellcolor{lightblue}91.36                                                                      \\ \cline{2-10} 
  \multicolumn{1}{|l|}{\multirow{-2}{*}{1000k}}      & \cellcolor{lightpurple}\cmark & \multicolumn{1}{r|}{\cellcolor{lightpurple}0.2408} & \multicolumn{1}{r|}{\cellcolor{lightpurple}64.17} & \multicolumn{1}{r|}{\cellcolor{lightpurple}55.75}                                                 & \cellcolor{lightpurple}57.31                                                                      & \multicolumn{1}{r|}{\cellcolor{lightpurple}0.7036} & \multicolumn{1}{r|}{\cellcolor{lightpurple}34.48} & \multicolumn{1}{r|}{\cellcolor{lightpurple}93.27}                                                 & \cellcolor{lightpurple}93.58                                                                      \\ \hline
  \end{tabular}
  \parbox{0.72\linewidth}{\centering \small \textit{Blue cells \textcolor{lightblue}{\fsquare} = fixed number of expansions; purple cells \textcolor{lightpurple}{\fsquare} = increasing number of expansions.}}
  %\parbox{0.72\linewidth}{\small \textit{Models were trained on various dataset sizes with synthesized Markdown-to-HTML structures. Blue~\textcolor{lightblue}{\fsquare} cells mark models trained on data generated with a fixed number of non-terminal expansions, and purple~\textcolor{lightpurple}{\fsquare} cells mark training on the data with an increasing number of non-terminal expansions.}}
\end{table}

\begin{result}
  \Modelizer models trained on synthetic data are also accurate in predicting~inputs~from~outputs.
\end{result}

According to the representative results for Pandoc models depicted in \Cref{tab:pandoc_forward} and \Cref{tab:pandoc_inverse}, models trained on the medium training set size can already achieve a high prediction accuracy.
Using the input generation configuration specified in \Cref{sec:eval-setup}, the data generator can produce the vast majority of input feature combinations within the 100,000--250,000 samples range.
By further increasing the number of generated samples while eliminating duplicates, the generator starts to produce more unnatural input feature combinations, which tends to negatively affect the model accuracy by making the learning more costly and hitting the prediction performance.
\begin{result}
  Medium datasets with 100,000 to 250,000 synthesized samples suffice to obtain good \Modelizer models.
\end{result}

\subsection*{\ref{rq:real-world}: How well do the \emph{synthesized samples} represent the real-world data and how much \emph{synthetic data} is required for mocking program behavior?}
  To answer this question, we evaluated our models on sets of 1,000 real-world test cases per subject, except Markdown and HTML, where, as already mentioned, we took 100 real-world \emph{readme} documents and extracted 12,783~top-level elements (basic blocks) from them. 
  These test sets do not intersect with any training sets. 
  According to \Cref{tab:eval-results}, while all models achieved high accuracy with synthetic test data, they performed noticeably worse when tested with real data.

  A probable cause for this effect is that synthesized and real-world test sets have different distributions of input-output features.
  In input generation, we intentionally avoided learning feature distribution from specific test cases to prevent overfitting and to assess \Modelizer's ability to learn program behavior without extra knowledge of the test subject or processed data.
  As a result, the feature distribution between real-world and synthesized samples did not align, causing a significant performance drop.
  This negative effect can be fixed with \emph{few-shot fine-tuning}~\cite{few-shot, few-shot2}, a widely used technique for the adaptation of pre-trained models to domain-specific tasks at a small cost.
  We applied it to models trained on synthetic data, which helped to mitigate feature distribution issues and enhanced prediction accuracy.
  Performance improved significantly after fine-tuning each model with 10 new randomly selected input-output pairs from source datasets. Purple cells in \Cref{tab:eval-results} reflect the performance numbers after few-shot fine-tuning.

\begin{figure}[]
  \centering
  \includegraphics[scale=0.5]{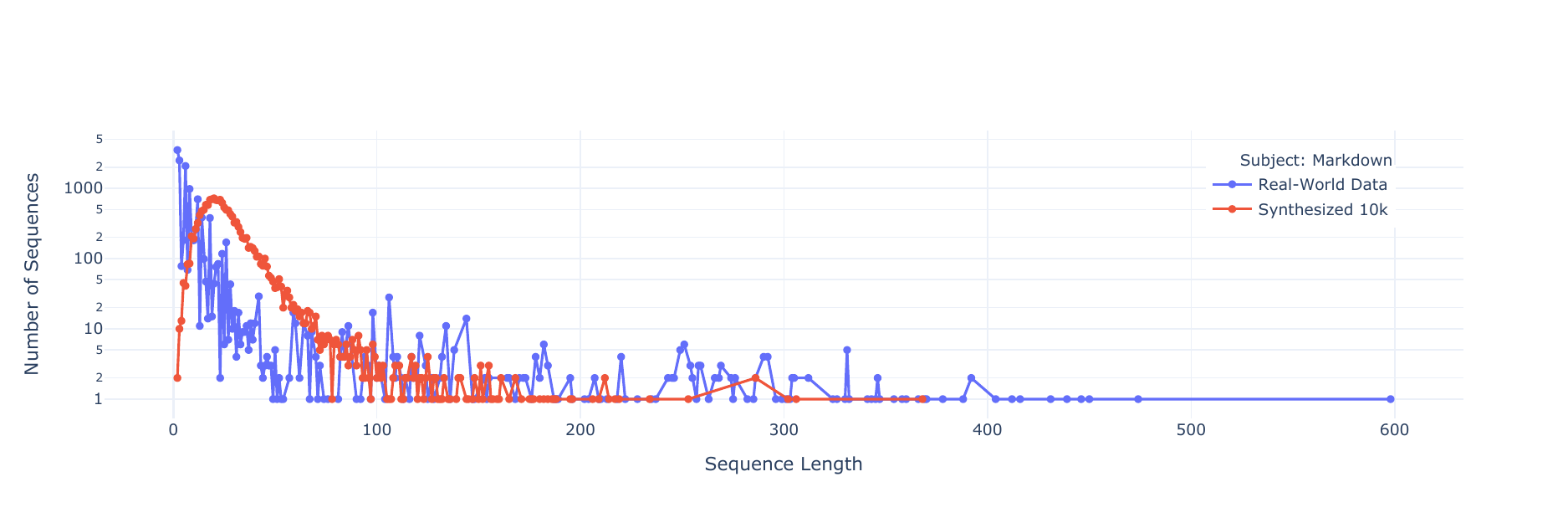}
  \caption[Markdown Sequence Length Frequency Distribution]{Markdown Sequence Length Frequency Distribution. The x-axis represents the length of different token sequences, and the y-axis denotes the frequency count of each token length. Every data split is represented by a different color.}
  \Description{Markdown Sequence Length Frequency Distribution. The x-axis represents the length of different token sequences, and the y-axis denotes the frequency count of each token length. Every data split is represented by a different color.}
  \label{fig:md-seq-len-plot}
\end{figure}

\begin{figure}[]
  \centering
  \includegraphics[scale=0.5]{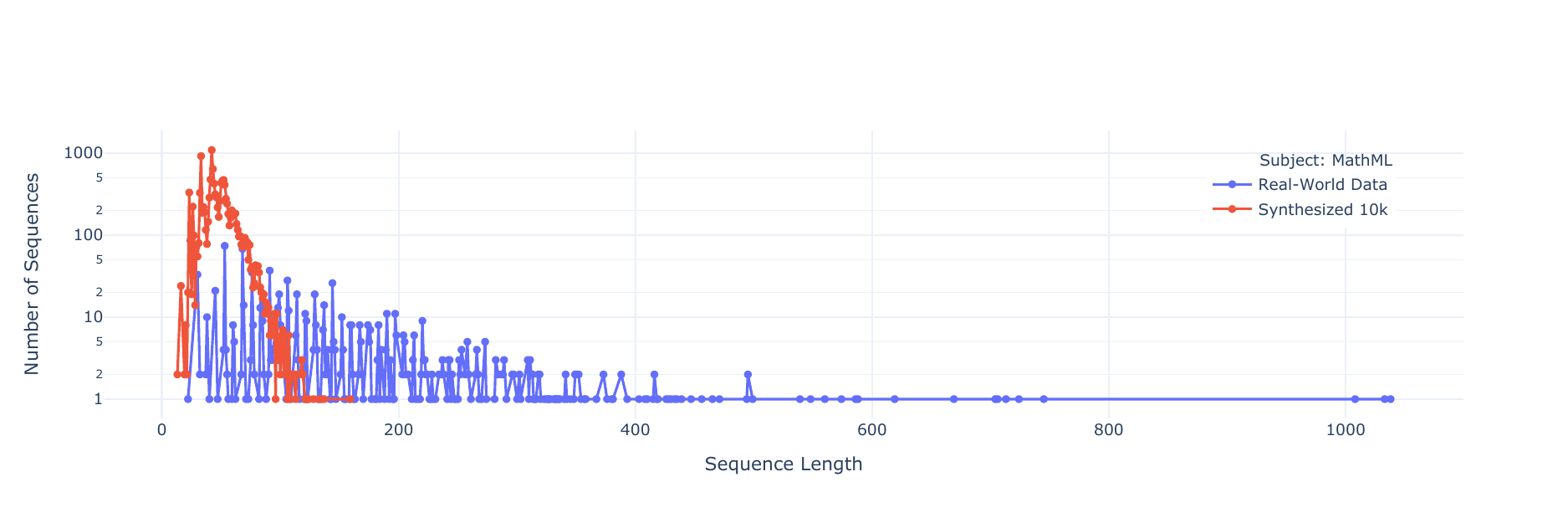}
  \caption[MathML Sequence Length Frequency Distribution]{MathML Sequence Length Frequency Distribution. The x-axis represents the length of different token sequences, and the y-axis denotes the frequency count of each token length. Every data split is represented by a different color.}
  \Description{MathML Sequence Length Frequency Distribution. The x-axis represents the length of different token sequences, and the y-axis denotes the frequency count of each token length. Every data split is represented by a different color.}
  \label{fig:mathml-seq-len-plot}
\end{figure}

  To explain such a performance boost we examined the token sequence length frequency for data structures used in our experiments. 
  As expected, the synthesized data differs from real-world data in terms of the number of tokens that these data structures contain. 
  For example, the token sequence length frequency distribution for Markdown that is depicted in \Cref{fig:md-seq-len-plot} shows that the longest recorded synthetic Markdown structure has 368~tokens, with a mean length of 26~tokens.
  When it comes to real-world data, the longest processed sequence has 598~tokens, with a mean length of 12~tokens. 
  Also, for MathML, whose token sequence length distribution is depicted in \Cref{fig:mathml-seq-len-plot}, we observed a similar pattern. 
  The synthetic data generator produced MathML structures that were tokenized into sequences of up to 159 tokens, while the mean sequence contained 48 tokens. 
  However, real-world MathML structures from the test set contained up to 1038 tokens with a mean MathML formula consisting of 151 tokens. 
  We conclude that synthetic inputs produced by \Modelizer generators are shorter than test samples, which affects the model's ability to generalize to longer real-world inputs.
  One way to solve this issue is to tweak the input generator's configuration to produce longer sequences by recursively applying more node expansions from the input grammar.
  However, a larger expansion budget can be wasted on irrelevant expansions due to random grammar expansion picks at the input generation time.
  While this process can be guided, for example, by assigning expansion probabilities to rules of interest, it still requires prior knowledge of input feature distributions inferred from at least a few program inputs and incorporation of this knowledge into grammar\cite{inputs-from-hell}.
  The generated dataset will then match the properties of seen inputs, and models trained on such a dataset will still struggle to generalize to dissimilar inputs.
  Alternatively, fine-tuning pre-trained models on real-world samples could be a more generic solution to the problem.
  It does not require the regeneration of training data followed by model re-training.
  Instead, it helps to adapt the capabilities of trained models that contain foundational knowledge about input-output relations
  to new conditions at a low cost. 

  \begin{result}
    Sequence length discrepancy between synthetic and real-world data can be smoothed using few-shot fine-tuning. 
  \end{result}

  For a small number of test samples, fine-tuned models still struggled to accurately convert input to output and vice-versa.
  The inspection of these cases has shown that they contain unseen features that are not part of the grammar used in input synthesis, for example, features present in the input syntax extensions such as \emph{Markdown Extra}~\cite{markdown-extra} or \emph{GitHub Flavored Markdown}~\cite{markdown-gfm}.
  Similarly, all models that had been trained on MathML Formulas used as model input or output (MathML to \LaTeX{}, \LaTeX{} to MathML, MathML to AsciiMath, AsciiMath to MathML) have shown the lowest prediction accuracy.
  When tested on real-world data, these models behaved as outliers because the MathML language allows multiple encodings for the semantically equivalent operators and operands, and the grammar we used for the input generation covers only a fraction of these definitions. 
  %The reason for this is the complexity of the MathML language and the absence of the corresponding features in the grammar we used for input generation. 
  %While the MathML grammar was used for synthetic input generation and the py-asciimath library was utilized for sequential conversion of first MathML to \LaTeX{} and then \LaTeX{} to AsciiMath during the training dataset formation, the accuracy of AsciiMath to \LaTeX{} and \LaTeX{} to AsciiMath models were less impacted by the inconsistency of the vocabularies. 
  %The accuracy drop was only caused by the lack of the specific unseen features, while the encoding variations of the same operators got mapped to the same command by the py-asciimath library in the \LaTeX{} and AsciiMath representations, thus the model's vocabulary had already contained them.
  This means that selected grammar does not cover all the capabilities of the \PUT, and thus, samples with such features are not present in the training data. 
  If a tokenizer relies on the subject's syntactical structure learned from grammar, it will also fail to parse these cases.

  We also evaluated the quality of the synthesized samples by checking for the inclusion of tokens collected from real-world samples into the vocabularies of learned models. Here are the statistics per data type used in our evaluation:
    \begin{description}
      \item[Markdown.] One top-level element out of 12,783 contained a high number of URL placeholders. Our generators did not create sequences with over 51 URL declarations per fuzzing round using the chosen non-terminal expansion configuration. Consequently, these placeholders were missing from the model's vocabulary, leading to their encoding as \verb|UNKNOWN| tokens and hampering accurate output prediction. 
      \item[HTML.] In one top-level HTML element, we observed the absence of matching URL placeholders in the model's vocabulary. Additionally, in 9 other top-level elements, we noticed the presence of extra HTML tags, like \verb|<head>, </head>, <title>, </title>|, and meta attributes like the \verb|data-cites| class, which were not seen during the synthetic data generation because input generator was removing HTML attributes that do not semantically affect data representation. In these cases, the model's predictions were semantically correct but slightly differed syntactically from the reference sequence.
      \item[SQL.] The test data contained three test cases that had been accessing more than three tables in a single query, whereas the synthesized queries lacked such complexity, limiting the model's ability to generate matching predictions.
      \item[KQL.] Two~out of~1,000 queries have accessed more than three~tables in a single request and used one~extra feature not seen during the data generation. The model was unable to accurately predict the output for these queries.
      \item[MathML Formulas.] In 115~cases out of 1,000~formulas, the token sequence contained the unseen operands and operators. 
      Our grammar for synthesizing MathML Formulas covers only a fraction of operands and specifies a single encoding per operand type.
      Thus, the MathML tokenizer also can not parse some operands found in test data due to unseen encoding. 
      An abstraction mechanism for operands, similar to numeric values and string literals, was not implemented.
      Another issue is the utilization of more than 18~identifiers in a single formula, which was uncovered by the selected input generation configuration.
      \item [LaTeX and AsciiMath Formulas.] We found inconsistencies in placeholder quantities between the train and test sets, similar to MathML, with unseen operands and operators in 70 and 73 test data token sequences, respectively.
      \item[Python and \LaTeX{} Expressions.] No unseen tokens were detected.
    \end{description}
  
  This investigation demonstrates that the quality of learned models depends on the completeness of synthesizing grammar and the similarity of input feature distributions between training and testing sets. 
  The more input features are covered in the grammar, the more program behavior is tested and leaked, and the better the model is at accurately mocking program behavior. 
  A fully trained model that frequently fails to predict the output acts as a marker of the missing features in the training data and, thus, in the input grammar. 
  This means that the grammar that is used in the data generation process is an estimate of the actual protocol specification, and it needs to be refined according to newly discovered features, which could also be mined automatically from the \PUT{} (see \Cref{rel:mining}),
  which also involves the generation of additional samples with new features and the population of the tokenizer vocabulary with new tokens. 
  Since grammar refinement is a significant change to the learning process, it also enforces either the model training restart or modifications to the embedding layer of the trained model to support new tokens and post-training on newly synthesized samples.
  Otherwise, if the grammar exactly follows the protocol specification it means that the program implementation does not. 
  For example, our case study of the  \texttt{Pandoc} markup converter with \Modelizer detected deviations from the \textit{CommonMark Spec Version 0.29}~\cite{commonmark}. 
  The model accuracy significantly improved once we extended our grammar declaration with extended syntax support. 
  Thus, we think that our models can assist in detecting behavior deviations from standard and protocol specifications refinement tasks since the learned model acts as a test oracle.
  Both types of failures can be automatically detected using a mechanism described in \Cref{sec:deployment}. 
  
\begin{figure}[]
  \centering
  \includegraphics[scale=0.24]{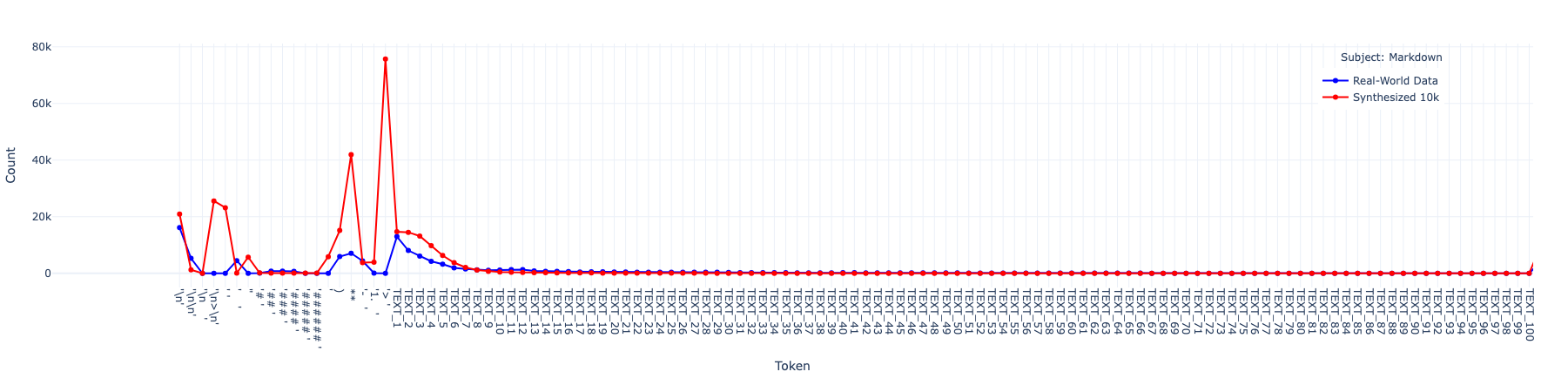}
  \caption[Markdown Token Frequency Distribution]{Markdown Token Frequency Distribution. The x-axis represents individual tokens, and the y-axis denotes the frequency of each token in the data split. Each split is represented by a different color.}
  \Description{Markdown Token Frequency Distribution. The x-axis represents individual tokens, and the y-axis denotes the frequency of each token in the data split. Each split is represented by a different color.}
  \label{fig:markdown-token-distribution}
\end{figure}

\begin{figure}[]
  \centering
  \includegraphics[scale=0.24]{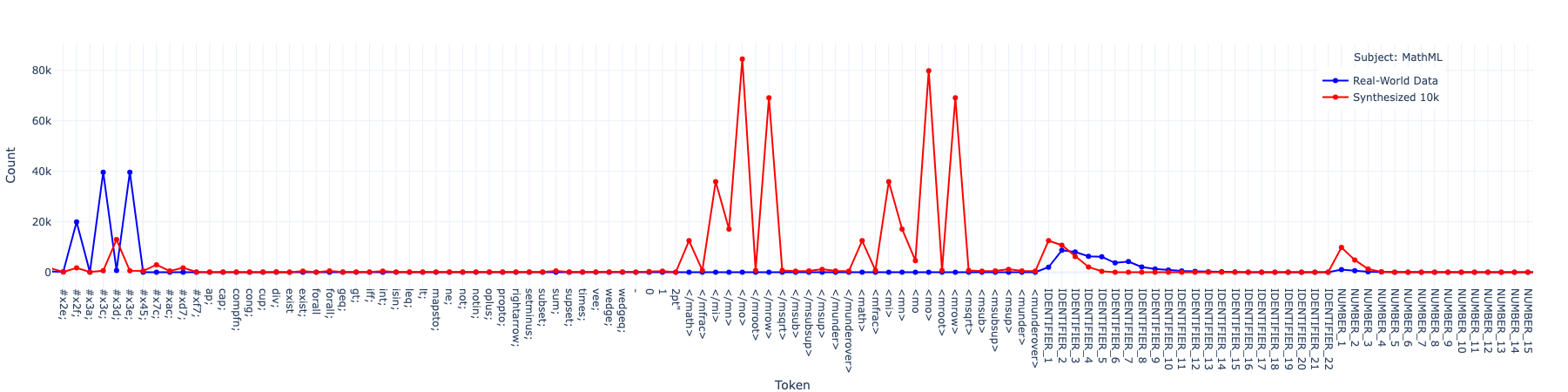}
  \caption[MathML Token Frequency Distribution]{MathML Token Frequency Distribution. The x-axis represents individual tokens, and the y-axis denotes the frequency of each token in the data split. Each split is represented by a different color.}
  \Description{MathML Token Frequency Distribution. The x-axis represents individual tokens, and the y-axis denotes the frequency of each token in the data split. Each split is represented by a different color.}
  \label{fig:mathml-token-distribution}
\end{figure}

  To further cover the input feature inconsistency problem, we plotted token frequency distribution for evaluated subjects.
  In contrast to the majority of synthetic data, a few real-world test inputs contained an unusual number of certain placeholder tokens. 
  For example, \Cref{fig:markdown-token-distribution} plot shows us that synthetic input generators do not generate more than 65 \verb|TEXT| placeholders in a single Markdown structure, with a significant fraction of synthesized structures containing only up to 10 \verb|TEXT| placeholders. 
  At the same time, our analysis has shown that there are 2 Markdown structures in the real-world test set with 100 \verb|TEXT| placeholders and almost 400 structures with 25 \verb|TEXT| placeholders. 
  Also, real-world structures contain from~4.2~to~54.8~times more alignment tokens (depending on the type) than synthetic equivalents, which are instead saturated more with blockquote elements. 
  Even worse disproportion is monitored with MathML (\Cref{fig:mathml-token-distribution}) where first synthetic structures contain fewer identifiers and numbers per synthesized formula than in real-world formulas, completely omit some operand types found in real-world formulas, and contain an enormous amount of MathML tag declarations mostly shaped as recursive structures consisting of opening and closing \verb|<mn>,| \verb|<mrow>,| \verb|<mi>| tags. While tweaks to the number of recursive expansions performed by a grammar fuzzer suffice to improve the quality of synthesized Markdown structures and minimize the gap between synthetic and real-world documents, this will have minimal to zero effect in the case of synthesizing MathML formulas. MathML grammar declaration used in our experiments is a significantly more complex structure with multiple mutually-recursive rules. As a result, simply increasing the number of expansion cycles will generate abnormal and unrealistic structures. Moreover, the given grammar does not define some operators and operands frequently seen in the test samples, which means that both grammar and input generator need adjustments to be able to generate higher-quality inputs. These results once more conclude that the quality of learned models depends on the quality and quantity of generated inputs; thus, it is important to have both correct grammar definitions and targeted configurations of input generators.
  
  \begin{result}
  %\Modelizer requires that the training data (and hence, the synthesizing grammar) covers the features found in the testing data.
  % The performance of \Modelizer depends on the accuracy and completeness of the producing grammar.
  \Modelizer requires the training data and synthesizing grammar to cover the features in the testing data.
  \end{result}

\subsection*{\ref{rq:learn}: How efficiently can \Modelizer learn program models?}

\begin{figure}[ht]
  \centering
  \includegraphics[width=1\linewidth]{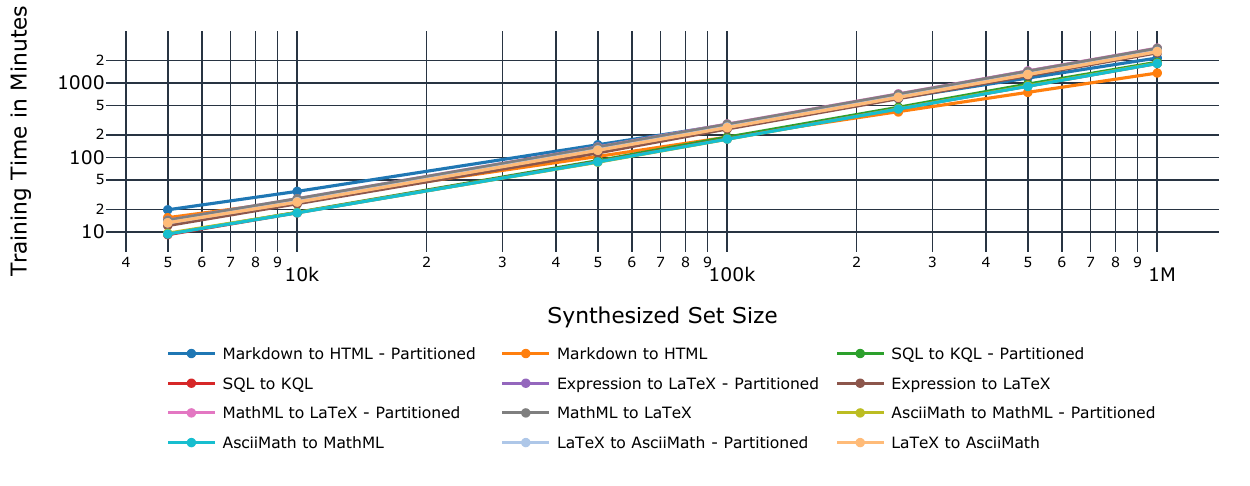}
  \caption[Speed of \Modelizer learning models from samples with a fixed and increasing complexity]{Speed of \Modelizer learning models from samples with a fixed and increasing complexity (Partitioned).\\All model learning experiments were done using a single NVIDIA Geforce RTX 4090 GPU.} 
  \Description{Speed of \Modelizer learning models from samples with a fixed and increasing complexity (Partitioned). All model learning was done using a single NVIDIA Geforce RTX 4090 GPU.} 
  \label{fig:learning_time}
\end{figure}
To answer this question, we need to measure \Modelizer's learning speed on commodity hardware by repeating our learning experiments multiple times on each subject with different training set sizes and fixed subject-specific model configurations\Cref{tab:hyperparameters}. According to timing records from the model learning log, \Modelizer needs less than half an hour to learn a model from 10,000 synthesized training pairs for 10 epochs using NVIDIA RTX 4090 GPU. The similar models will be trained within 7 to 12 hours using a mid-size dataset with 250,000 training samples. Lastly, \Modelizer needs 22 to 49 hours to train a model using a dataset with 1,000,000 records. \Cref{fig:learning_time} shows the model learning time distribution across subjects from the evaluation set, training set sizes, and input generation strategies. While an early program mock can be produced already within 30 minutes using a GPU, a significantly more accurate model will be trained within 10 hours. We also measured a slowdown from 1\% to 59\%, with an average of 11\%, while training models containing inputs generated with fixed or increasing complexity. The training time grows almost linearly with the increase of the training set size, given the automatically discovered hyperparameter configurations per each subject from the evaluation set.

\begin{result}
  Training a \Modelizer model takes a few hours on a PC equipped with a commodity GPU.
\end{result}

While looking for an optimal learning strategy, we studied the impact of the non-terminal expansion strategy on the accuracy of the learned models. 
As we can see in the representative results of Markdown to HTML~(\Cref{tab:pandoc_forward}) conversion and HTML to Markdown~(\Cref{tab:pandoc_inverse}) conversion, continuously increasing the complexity of synthesized tokens does not significantly improve the conversion accuracy neither when the model is trained purely on synthetic data nor after fine-tuning on real-world samples. 
Without providing the feature distribution as grammar rule expansion probabilities, the larger expansion budget still gets wasted on randomly chosen expansions that do not align with patterns observed in the real-world data.  
We conclude that the efficient learning strategy is generating the mid-size synthetic dataset with a fixed number of non-terminal expansions, which, according to our results, is sufficient for learning feature mapping between input and output languages. 
Then, if available, we perform a few-shot fine-tuning on the real-world data. 
It achieves near-optimal conversion accuracy with low resource use--a pattern we saw in half of the subjects we tested. 
More statistics for the remaining subjects can be found in \Cref{tab:eval-results}.

\begin{result}
Optimal \Modelizer performance is achieved with a medium-sized synthetic dataset and real-world fine-tuning.
\end{result}

\subsection*{\ref{rq:hyperparameters}: How does \Modelizer learn the \emph{hyperparameters} for each data pair combination?}
\begin{table}[]
  \rowcolors{2}{gray!25}{white}
  \small
  \centering
  \caption{\label{tab:hyperparameters}Automatically configured hyperparameter values using the Hyperparameter Search.}
  \begin{tabular}{lrrrrrrrcrrrcr}
  \rowcolor{gray!50}
  Model               & \rot{Source Vocab Size } & \rot{Target Vocab Size} & \rot{Encoder Layers} & \rot{Decoder Layers} & \rot{Embedding Size} & \rot{Feedforward Size} & \rot{Attention Heads} & \rot{Learning Policy} & \rot{Learning Rate} & \rot{Weight Decay} & \rot{Dropout} & \rot{Clip Gradients} & \rot{Parameter Count} \\
  Markdown HTML       & 292               & 309               & 1              & 3              & 256            & 2048             & 32              & cosine          & 0.0005        & 0.0100       & 0.0     & False          & 6285621    \\
  HTML to Markdown    & 309               & 292               & 1              & 3              & 256            & 2048             & 32              & step            & 0.0001        & 0.0100       & 0.0     & False          & 6281252    \\
  SQL to KQL          & 674               & 675               & 1              & 2              & 256            & 2048             & 64              & cosine          & 0.0001        & 0.0005       & 0.1     & True           & 4992419    \\
  KQL to SQL          & 675               & 674               & 1              & 1              & 256            & 2048             & 64              & cosine          & 0.0001        & 0.0001       & 0.0     & False          & 3413410    \\
  Expression to \LaTeX{} & 1501              & 656               & 1              & 3              & 256            & 2048             & 16              & cosine          & 0.0005        & 0.0005       & 0.1     & True           & 6773136    \\
  \LaTeX{} to Expression & 656               & 1501              & 1              & 1              & 256            & 1024             & 16              & multipl.        & 0.0005        & 0.0005       & 0.0     & True           & 2782173    \\
  AsciiMath \LaTeX{}     & 122               & 131               & 1              & 1              & 256            & 2048             & 32              & step            & 0.0001        & 0.0100       & 0.1     & False          & 2993283    \\
  \LaTeX{} to AsciiMath  & 131               & 122               & 1              & 4              & 256            & 2048             & 64              & step            & 0.0001        & 0.0100       & 0.1     & False          & 7727226    \\
  MathML to \LaTeX{}     & 174               & 131               & 1              & 4              & 256            & 1024             & 64              & cosine          & 0.0001        & 0.0005       & 0.0     & False          & 5116291    \\
  \LaTeX{} to MathML     & 131               & 174               & 1              & 3              & 256            & 4096             & 64              & cosine          & 0.0001        & 0.0100       & 0.1     & False          & 10377646   \\
  MathML to AsciiMath & 174               & 122               & 1              & 3              & 256            & 2048             & 64              & step            & 0.0001        & 0.0100       & 0.0     & False          & 6159482    \\
  AsciiMath to MathML & 122               & 174               & 1              & 2              & 256            & 1024             & 64              & step            & 0.0001        & 0.0100       & 0.0     & False          & 3018158    \\
  \end{tabular}
\end{table}

Our next research question concerns \emph{hyperparameters:} \emph{How does \Modelizer learn the hyperparameters for each data pair combination?} While \Modelizer makes no assumptions on the availability of data at the training time, it still can perform \emph{hyperparameter optimization.} For that, it needs a fixed set of test cases that can be generated using grammar, which is used to evaluate different model configurations over multiple trials. The optimizer tries to minimize the value of the loss criterion and automatically prunes non-promising configurations to further speed up the search process. In our experiments, we reused the same set of test cases to discover optimal configurations for both forward and inverse models, which allowed us to study the reusability of discovered hyperparameters for the inverse models as well. While the optimal embedding size value and the number of encoder layers were common for all input-output pairs, the rest of the hyperparameters differed. The best models were mainly trained with the smallest Learning Rate value; Linear Layer Size has to be at least four times larger than the Embedding Size; and, for the same data type pairs, inverse models require mostly two times more decoder layers than forward models. Thus, we can conclude that the hyperparameters for the forward models can serve as a lower bound for the hyperparameter search for the inverse models.
Given the behavior mocking accuracy we have observed during the evaluation of real-world data, we think that running a hyperparameter search on synthesized data is a suitable approach for cases when sufficient quality and quantity of real-world inputs are not available at the training time. According to the results of the hyperparameter search in \Cref{tab:hyperparameters}, the optimizer managed to find an individual configuration for every translation pair type, and the resulting number of model parameters is small, ranging between 2.7 and 10 million parameters for the given subjects. 

\begin{result}
  Hyperparameter optimization performed on synthetic data improves the quality of learned models.
\end{result}

\subsection*{\ref{rq:tokenizer}: What is the effect of \emph{tokenizer} selection on learning performance?}
Let us turn to \emph{tokenizers.} \emph{How should the model input/output be tokenized to achieve the best learning performance?}
Considering the higher development costs, should the pre-trained tokenizers be used instead of custom tokenizers? The model's performance heavily relies on token quality.
Longer token sequences necessitate the use of more complex models, which leads to longer output generation times and a rise in prediction errors.
Such models also need more processing power, while models trained from small vocabulary have limited learning flexibility. 
We propose using custom semantically aware tokenizers developed by domain experts to achieve optimal token sequence length—vocabulary size ratios by leveraging data structure knowledge.

\begin{table}[t]
  \small
  \caption{\label{tab:tokenizer-comparison} Comparison of the tokenization quality between custom-implemented tokenizers and selection of pre-trained tokenizers.}
  \begin{tabular}{|l|l|r|r|r|r|}
    \hline
    \rowcolor{lightgrey} 
    \multicolumn{1}{|c|}{Subject} & Tokenizer                      & \multicolumn{1}{c|}{Average Length} & \multicolumn{1}{c|}{Vocab Size} & \multicolumn{1}{c|}{Length Ratio} & \multicolumn{1}{c|}{Vocab Ratio} \\ \hline
                                                          & \cellcolor{lightyellow}Custom & \cellcolor{lightyellow}12                                & \cellcolor{lightyellow}176                           & \cellcolor{lightyellow}-                                 & \cellcolor{lightyellow}-                                \\ \cline{2-6} 
                                                          & \cellcolor{lightblue}BERT   & \cellcolor{lightblue}22                                & \cellcolor{lightblue}115                           & \cellcolor{lightblue}1.75                              & \cellcolor{lightblue}0.65                             \\ \cline{2-6} 
    \multirow{-3}{*}{Markdown}                            & \cellcolor{lightpurple}Llama3 & \cellcolor{lightpurple}22                                & \cellcolor{lightpurple}136                           & \cellcolor{lightpurple}1.84                              & \cellcolor{lightpurple}0.77                             \\ \hline
                                                          & \cellcolor{lightyellow}Custom & \cellcolor{lightyellow}16                                & \cellcolor{lightyellow}245                           & \cellcolor{lightyellow}-                                 & \cellcolor{lightyellow}-                                \\ \cline{2-6} 
                                                          & \cellcolor{lightblue}BERT   & \cellcolor{lightblue}49                                & \cellcolor{lightblue}178                           & \cellcolor{lightblue}3.10                               & \cellcolor{lightblue}0.73                             \\ \cline{2-6} 
    \multirow{-3}{*}{HTML}                                & \cellcolor{lightpurple}Llama3 & \cellcolor{lightpurple}39                                & \cellcolor{lightpurple}191                           & \cellcolor{lightpurple}2.43                              & \cellcolor{lightpurple}0.78                             \\ \hline
                                                          & \cellcolor{lightyellow}Custom & \cellcolor{lightyellow}11                                & \cellcolor{lightyellow}42                            & \cellcolor{lightyellow}-                                 & \cellcolor{lightyellow}-                                \\ \cline{2-6} 
                                                          & \cellcolor{lightblue}BERT   & \cellcolor{lightblue}20                                & \cellcolor{lightblue}38                            & \cellcolor{lightblue}1.93                              & \cellcolor{lightblue}0.90                              \\ \cline{2-6} 
    \multirow{-3}{*}{SQL}                                 & \cellcolor{lightpurple}Llama3 & \cellcolor{lightpurple}20                                & \cellcolor{lightpurple}42                            & \cellcolor{lightpurple}1.93                              & \cellcolor{lightpurple}1                              \\ \hline
                                                          & \cellcolor{lightyellow}Custom & \cellcolor{lightyellow}14                                & \cellcolor{lightyellow}42                            & \cellcolor{lightyellow}-                                 & \cellcolor{lightyellow}-                                \\ \cline{2-6} 
                                                          & \cellcolor{lightblue}BERT   & \cellcolor{lightblue}26                                & \cellcolor{lightblue}41                            & \cellcolor{lightblue}1.89                              & \cellcolor{lightblue}0.98                             \\ \cline{2-6} 
    \multirow{-3}{*}{KQL}                                 & \cellcolor{lightpurple}Llama3 & \cellcolor{lightpurple}27                                & \cellcolor{lightpurple}50                            & \cellcolor{lightpurple}1.97                              & \cellcolor{lightpurple}1.19                             \\ \hline
                                                          & \cellcolor{lightyellow}Custom & \cellcolor{lightyellow}25                                & \cellcolor{lightyellow}81                            & \cellcolor{lightyellow}-                                 & \cellcolor{lightyellow}-                                \\ \cline{2-6} 
                                                          & \cellcolor{lightblue}BERT   & \cellcolor{lightblue}47                                & \cellcolor{lightblue}102                           & \cellcolor{lightblue}1.87                              & \cellcolor{lightblue}1.26                             \\ \cline{2-6} 
    \multirow{-3}{*}{Expression}                          & \cellcolor{lightpurple}Llama3 & \cellcolor{lightpurple}42                                & \cellcolor{lightpurple}103                           & \cellcolor{lightpurple}1.69                              & \cellcolor{lightpurple}1.27                             \\ \hline
                                                          & \cellcolor{lightyellow}Custom & \cellcolor{lightyellow}57                                & \cellcolor{lightyellow}87                            & \cellcolor{lightyellow}-                                 & \cellcolor{lightyellow}-                                \\ \cline{2-6} 
                                                          & \cellcolor{lightblue}BERT   & \cellcolor{lightblue}90                                & \cellcolor{lightblue}108                           & \cellcolor{lightblue}1.58                              & \cellcolor{lightblue}1.24                             \\ \cline{2-6} 
    \multirow{-3}{*}{\LaTeX{} Expression}                    & \cellcolor{lightpurple}Llama3 & \cellcolor{lightpurple}66                                & \cellcolor{lightpurple}148                           & \cellcolor{lightpurple}1.16                              & \cellcolor{lightpurple}1.7                              \\ \hline
                                                          & \cellcolor{lightyellow}Custom & \cellcolor{lightyellow}151                               & \cellcolor{lightyellow}101                           & \cellcolor{lightyellow}-                                 & \cellcolor{lightyellow}-                                \\ \cline{2-6} 
                                                          & \cellcolor{lightblue}BERT   & \cellcolor{lightblue}342                               & \cellcolor{lightblue}77                            & \cellcolor{lightblue}2.30                               & \cellcolor{lightblue}0.76                             \\ \cline{2-6} 
    \multirow{-3}{*}{MathML}                              & \cellcolor{lightpurple}Llama3 & \cellcolor{lightpurple}296                               & \cellcolor{lightpurple}101                           & \cellcolor{lightpurple}1.99                              & \cellcolor{lightpurple}1                              \\ \hline
                                                          & \cellcolor{lightyellow}Custom & \cellcolor{lightyellow}27                                & \cellcolor{lightyellow}104                           & \cellcolor{lightyellow}-                                 & \cellcolor{lightyellow}-                                \\ \cline{2-6} 
                                                          & \cellcolor{lightblue}BERT   & \cellcolor{lightblue}56                                & \cellcolor{lightblue}145                           & \cellcolor{lightblue}2.20                               & \cellcolor{lightblue}1.39                             \\ \cline{2-6} 
    \multirow{-3}{*}{\LaTeX{} Formula}                       & \cellcolor{lightpurple}Llama3 & \cellcolor{lightpurple}42                                & \cellcolor{lightpurple}219                           & \cellcolor{lightpurple}1.69                              & \cellcolor{lightpurple}2.11                             \\ \hline
                                                          & \cellcolor{lightyellow}Custom & \cellcolor{lightyellow}26                                & \cellcolor{lightyellow}86                            & \cellcolor{lightyellow}-                                 & \cellcolor{lightyellow}-                                \\ \cline{2-6} 
                                                          & \cellcolor{lightblue}BERT   & \cellcolor{lightblue}51                                & \cellcolor{lightblue}111                           & \cellcolor{lightblue}2.14                              & \cellcolor{lightblue}1.29                             \\ \cline{2-6} 
    \multirow{-3}{*}{AsciiMath}                           & \cellcolor{lightpurple}Llama3 & \cellcolor{lightpurple}39                                & \cellcolor{lightpurple}229                           & \cellcolor{lightpurple}1.66                              & \cellcolor{lightpurple}2.66                             \\ \hline
  \end{tabular} 
  \parbox{0.75\linewidth}{\centering \small \textit{Yellow cells \textcolor{lightyellow}{\fsquare} = Custom Tokenizer; Blue cells \textcolor{lightblue}{\fsquare} = BERT Tokenizer; purple cells \textcolor{lightpurple}{\fsquare} = Llama3 Tokenizer.}}
\end{table}

We compared the tokenization quality of the pre-trained tokenizers that are shipped with the state-of-the-art NLP models BERT~\cite{Devlin2019BERTPO} and Llama3~\cite{llama3modelcard} with the custom structure-aware tokenizers that were implemented for each subject. 
We have computed the average length of the tokenized sequences and the vocabulary size per subject. 
To simplify the comparison, we derived additional metrics, such as the length ratio and vocabulary ratio, which are computed as the ratio of the average length and vocabulary size of the pre-trained tokenizer to the structure-aware tokenizer. 
The results of our experiments are presented in \Cref{tab:tokenizer-comparison} and show that the custom structure-aware tokenizers outperform the pre-trained tokenizers in terms of the average length of the tokenized sequences. 
On the other hand, the pre-trained and unsupervised tokenizers do not require additional development effort, can tokenize multiple subjects, and are easily interchangeable.
The evaluation was conducted on 1000 real-world samples for each subject, except for Markdown and HTML, where the tokenization was performed on all top-level elements extracted from 100 real-world documents. 
All inputs that were processed by each tokenizer were augmented by the \emph{PlaceholderProcessor} module, which replaced the numeric and string literals with the matching placeholders. 
We think that custom structure-aware tokenizers can achieve better tokenization granularity and are thus better suited for program behavior mocking tasks. 
% We suggest using the pre-trained or unsupervised tokenizers as guidance and a lower bound for the structure-aware tokenization or for immediate application in preliminary model learning.

While the pre-trained tokenizers managed to formulate a smaller vocabulary for half of the evaluated subjects, the average length of the tokenized sequences was significantly higher than the one produced by the structure-aware tokenizers. 
We demonstrate an example of the tokenization of the HTML element by the structure-aware and pre-trained tokenizers in \Cref{tab:tokenizer-example}. In this case, the structure-aware tokenizer produces a sequence consisting of 18~tokens, BERT produces 58~tokens, and Llama3 produces 48~tokens. 
Both BERT and Llama3 split the HTML tags and placeholders into multiple sub-tokens, increasing the tokenized sequence length. 
On the other hand, the structure-aware tokenizer produces a shorter sequence of tokens that are more meaningful and easier to interpret.

It is possible to dynamically infer a structure-aware tokenizer for many program input-output pairs by randomly mutating program inputs and analyzing changes in program execution and program output. 
We are working on automating this process to reduce the usage effort of such tokenizers while maintaining high tokenization quality.
\begin{table}[t]
  \small
  \caption{\label{tab:tokenizer-example} Example of the tokenization of the HTML element by the domain-specific and pre-trained tokenizers.}
  \rowcolors{2}{gray!25}{white}
  \begin{tabular}{cc}
  \cellcolor{lightgrey}Input & \begin{minipage}{0.75\linewidth}
    \centering
    \ttfamily
    \begin{verbatim}   <p>TEXT_1<a href="URL_1">TEXT_2</a>TEXT_3<strong>
   TEXT_4</strong>TEXT_5<code>TEXT_6</code>TEXT_7</p>\end{verbatim}
    \end{minipage}      \\
    \\
  \rowcolor{lightgrey}
  Tokenizer & Tokens \\
  Custom    & \begin{minipage}{0.75\linewidth}
    \centering
    \ttfamily
    \smallskip
    \begin{verbatim}
  <p>   TEXT_1   <a    href="URL_1"   >   TEXT_2   </a>   
  TEXT_3   <strong>   TEXT_4   </strong>   TEXT_5   
  <code>   TEXT_6   </code>   TEXT_7   </p>   \n\end{verbatim}
  \smallskip
  \end{minipage}      \\
  BERT      & \begin{minipage}{0.75\linewidth}
    \centering
    \ttfamily
    \smallskip
    \begin{verbatim}
  <   p   >   text   _   1   <   a   hr   ##ef   =   "   
  ur   ##l   _   1   "   >   text   _   2   <   /   a   >   
  text   _   3   <   strong   >   text   _   4   <   /   
  strong      >   text   _   5   <   code   >   text   _   
  6   <   /   code      >   text   _   7   <   /   p   >\end{verbatim}
  \smallskip
  \end{minipage}      \\
  Llama3    & \begin{minipage}{0.75\linewidth}
    \centering
    \ttfamily
    \smallskip
    \begin{verbatim}
  <p   >   TEXT   _   1   <a   Ġhref   ="   URL   _   1   
  ">   TEXT   _   2   </   a   >   TEXT   _   3   <strong   >     
  TEXT   _   4   </   strong   >   TEXT   _   5   <   code   >      
  TEXT   _   6   </   code   >   TEXT   _   7   </   p   >Ċ\end{verbatim}
  \smallskip
  \end{minipage}      \\ 
  \end{tabular}
\end{table}

\begin{result}
  Structure-aware tokenizers are more suitable for behavior learning than pre-trained or unsupervised tokenizers.
\end{result}

\subsection*{\ref{rq:llm-comparison}: Could LLMs pre-trained on code mock programs?}
In this research question, we examined the ability of LLMs to reproduce program behavior and compared their performance with custom domain-specific models that we have previously trained.   
At first, we decided to perform zero-shot prompting of LLMs pre-trained on code. However, selected LLMs, \emph{CodeLlama}~\cite{roziere2024code} and \emph{CodeGemma}~\cite{codegemma}, were instruction fine-tuned by their vendors, so we prepended every prompt with the following instruction template:
\begin{center}
  \verb|You are a converter program that translates {SOURCE} to {TARGET}.|
  \verb|Translate the following data: {DATA}| 
\end{center}
Due to the zero-shot prompting strategy, the instruction does not contain any task-specific examples but, instead, instruments the model with contextual knowledge. At the time of querying, \verb|{SOURCE}|, \verb|{TARGET}|, and \verb|{DATA}| placeholders were replaced with actual values. 
We summarized the zero-shot prompting results in \Cref{tab:llm-raw-inference}.
We recorded zero or close to zero Exact Match and Close Match scores for all evaluated subjects. 
It means that selected LLMs were unable to replicate neither actual program behavior nor the behavior of the hypothetical inverse program. 
From BLEU score values, we can infer that the \emph{CodeGemma} model has a limited understanding of Markdown to HTML and HTML to Markdown conversion tasks, while \emph{CodeLlama} models struggle to provide any valuable results for any data pair and conversion direction combinations without additional fine-tuning. 
We provide an example of interaction with selected LLMs, including the model outputs, in \Cref{tab:llm-raw-inference-example}.
Despite the ability to better understand code-specific tasks due to additional pre-training, these models can not act as program stubs without additional fine-tuning.

\begin{table}[h]
\small
\centering
\caption{\label{tab:llm-raw-inference-example} An example of zero-shot prompting of LLMs pre-trained on code converting Markdown to HTML}
\begin{tabular}{llll}
\rowcolor[HTML]{C0C0C0} 
\multicolumn{1}{c}{\cellcolor[HTML]{C0C0C0}Model Input}                                                                                                        & \multicolumn{1}{c}{\cellcolor[HTML]{C0C0C0}Expected Output}          & \multicolumn{1}{c}{\cellcolor[HTML]{C0C0C0}CodeLlama Output}                                                                             & \multicolumn{1}{c}{\cellcolor[HTML]{C0C0C0}CodeGemma Output}                                                                                                                                                                           \\
\begin{tabular}[c]{@{}l@{}}You are a converter program \\ that translates Markdown to HTML.\\ Translate the following data: \\ \#\# BarkProcessor\end{tabular} & \textless{}h2\textgreater{}BarkProcessor\textless{}/h2\textgreater{} & \begin{tabular}[c]{@{}l@{}}\#\#\# BarkProcessor.cs\\ \\ ```csharp\\ using System;\\ using System.Collections.Generic;\\ ...\end{tabular} & \begin{tabular}[c]{@{}l@{}}\textless{}h2\textgreater{}\textless{}strong\textgreater{}\textless{}em\textgreater\\ BarkProcessor\\ \textless{}/em\textgreater{}\textless{}/strong\textgreater{}\textless{}/h2\textgreater{}\end{tabular}
\end{tabular}
\end{table}

\begin{table}[t]
  \small
  \centering
  \caption{\label{tab:llm-raw-inference} Evaluation results for models mimicking Pandoc, msticpy, latexify-py, and py-asciimath via LLMs pre-trained on code.}

  \begin{tabular}{|l|l|l|c|r|r|r|r|r|r|r|r|}
  \hline
  \rowcolor{lightgrey} 
  \multicolumn{1}{|c|}{\cellcolor{lightgrey}Program} & \multicolumn{1}{c|}{\cellcolor{lightgrey}Source} & \multicolumn{1}{c|}{\cellcolor{lightgrey}Target} & \begin{tabular}[c]{@{}c@{}}Model\\ Type\end{tabular}                          & \multicolumn{1}{c|}{\cellcolor{lightgrey}Model} & \multicolumn{1}{c|}{\cellcolor{lightgrey}BLEU} & \multicolumn{1}{c|}{\cellcolor{lightgrey}\begin{tabular}[c]{@{}c@{}}BLEU\\ Error\end{tabular}} & \multicolumn{1}{c|}{\cellcolor{lightgrey}NIST} & \multicolumn{1}{c|}{\cellcolor{lightgrey}WER} & \multicolumn{1}{c|}{\cellcolor{lightgrey}WIL} & \multicolumn{1}{c|}{\cellcolor{lightgrey}\begin{tabular}[c]{@{}c@{}}Exact\\ Match\end{tabular}} & \multicolumn{1}{c|}{\cellcolor{lightgrey}\begin{tabular}[c]{@{}c@{}}Close\\ Match\end{tabular}} \\ \hline
                                                        &                                                     &                                                     &                                                                               & \cellcolor{lightblue}CodeLlama               & \cellcolor{lightblue}0.1070                    & \cellcolor{lightblue}0.0010                                                                    & \cellcolor{lightblue}1.8602                    & \cellcolor{lightblue}174.23                   & \cellcolor{lightblue}97.23                    & \cellcolor{lightblue}0                                                                          & \cellcolor{lightblue}0                                                                          \\ \cline{5-12} 
                                                        & \multirow{-2}{*}{Markdown}                          & \multirow{-2}{*}{HTML}                              & \multirow{-2}{*}{Forward}                                                     & \cellcolor{lightpurple}CodeGemma               & \cellcolor{lightpurple}0.4366                    & \cellcolor{lightpurple}0.0029                                                                    & \cellcolor{lightpurple}7.4377                    & \cellcolor{lightpurple}68.24                    & \cellcolor{lightpurple}68.82                    & \cellcolor{lightpurple}0.61                                                                       & \cellcolor{lightpurple}2.17                                                                       \\ \cline{2-12} 
                                                        &                                                     &                                                     &                                                                               & \cellcolor{lightblue}CodeLlama               & \cellcolor{lightblue}0.0875                    & \cellcolor{lightblue}0.0009                                                                    & \cellcolor{lightblue}1.4325                    & \cellcolor{lightblue}266.98                   & \cellcolor{lightblue}97.02                    & \cellcolor{lightblue}0                                                                          & \cellcolor{lightblue}0                                                                          \\ \cline{5-12} 
  \multirow{-4}{*}{Pandoc}                              & \multirow{-2}{*}{HTML}                              & \multirow{-2}{*}{Markdown}                          & \multirow{-2}{*}{Backward}                                                     & \cellcolor{lightpurple}CodeGemma               & \cellcolor{lightpurple}0.5348                    & \cellcolor{lightpurple}0.0028                                                                    & \cellcolor{lightpurple}9.1093                    & \cellcolor{lightpurple}49.84                    & \cellcolor{lightpurple}49.67                    & \cellcolor{lightpurple}0                                                                          & \cellcolor{lightpurple}5.70                                                                       \\ \hline
                                                        &                                                     &                                                     &                                                                               & \cellcolor{lightblue}CodeLlama               & \cellcolor{lightblue}0.0542                    & \cellcolor{lightblue}0.0009                                                                    & \cellcolor{lightblue}1.0152                    & \cellcolor{lightblue}662.57                   & \cellcolor{lightblue}93.47                    & \cellcolor{lightblue}0                                                                          & \cellcolor{lightblue}0                                                                          \\ \cline{5-12} 
                                                        & \multirow{-2}{*}{SQL}                               & \multirow{-2}{*}{KQL}                               & \multirow{-2}{*}{Forward}                                                     & \cellcolor{lightpurple}CodeGemma               & \cellcolor{lightpurple}0.1546                    & \cellcolor{lightpurple}0.0039                                                                    & \cellcolor{lightpurple}2.5782                    & \cellcolor{lightpurple}250.21                   & \cellcolor{lightpurple}84.32                    & \cellcolor{lightpurple}0                                                                          & \cellcolor{lightpurple}0                                                                          \\ \cline{2-12} 
                                                        &                                                     &                                                     &                                                                               & \cellcolor{lightblue}CodeLlama               & \cellcolor{lightblue}0.0338                    & \cellcolor{lightblue}0.0010                                                                    & \cellcolor{lightblue}0.6302                    & \cellcolor{lightblue}644.66                   & \cellcolor{lightblue}96.77                    & \cellcolor{lightblue}0                                                                          & \cellcolor{lightblue}0                                                                          \\ \cline{5-12} 
  \multirow{-4}{*}{msticpy}                             & \multirow{-2}{*}{KQL}                               & \multirow{-2}{*}{SQL}                               & \multirow{-2}{*}{Backward}                                                     & \cellcolor{lightpurple}CodeGemma               & \cellcolor{lightpurple}0.2960                    & \cellcolor{lightpurple}0.0043                                                                    & \cellcolor{lightpurple}5.0848                    & \cellcolor{lightpurple}67.94                    & \cellcolor{lightpurple}52.17                    & \cellcolor{lightpurple}0                                                                          & \cellcolor{lightpurple}0                                                                          \\ \hline
                                                        &                                                     &                                                     &                                                                               & \cellcolor{lightblue}CodeLlama               & \cellcolor{lightblue}0.0767                    & \cellcolor{lightblue}0.0011                                                                    & \cellcolor{lightblue}1.4521                    & \cellcolor{lightblue}266.39                   & \cellcolor{lightblue}94.65                    & \cellcolor{lightblue}0                                                                          & \cellcolor{lightblue}0                                                                          \\ \cline{5-12} 
                                                        & \multirow{-2}{*}{PyExpres.}                         & \multirow{-2}{*}{\LaTeX{}}                             & \multirow{-2}{*}{Forward}                                                     & \cellcolor{lightpurple}CodeGemma               & \cellcolor{lightpurple}0.1855                    & \cellcolor{lightpurple}0.0032                                                                    & \cellcolor{lightpurple}3.4188                    & \cellcolor{lightpurple}108.77                   & \cellcolor{lightpurple}83.83                    & \cellcolor{lightpurple}0                                                                          & \cellcolor{lightpurple}0                                                                          \\ \cline{2-12} 
                                                        &                                                     &                                                     &                                                                               & \cellcolor{lightblue}CodeLlama               & \cellcolor{lightblue}0.0545                    & \cellcolor{lightblue}0.0012                                                                    & \cellcolor{lightblue}0.9835                    & \cellcolor{lightblue}387.52                   & \cellcolor{lightblue}94.74                    & \cellcolor{lightblue}0                                                                          & \cellcolor{lightblue}0                                                                          \\ \cline{5-12} 
  \multirow{-4}{*}{latexify-py}                         & \multirow{-2}{*}{\LaTeX{}}                             & \multirow{-2}{*}{PyExpres.}                         & \multirow{-2}{*}{Backward}                                                     & \cellcolor{lightpurple}CodeGemma               & \cellcolor{lightpurple}0.2428                    & \cellcolor{lightpurple}0.0048                                                                    & \cellcolor{lightpurple}4.1169                    & \cellcolor{lightpurple}100.26                   & \cellcolor{lightpurple}76.04                    & \cellcolor{lightpurple}0                                                                          & \cellcolor{lightpurple}0                                                                          \\ \hline
                                                        &                                                     &                                                     &                                                                               & \cellcolor{lightblue}CodeLlama               & \cellcolor{lightblue}0.0076                    & \cellcolor{lightblue}0.0003                                                                    & \cellcolor{lightblue}0.3041                    & \cellcolor{lightblue}988.32                   & \cellcolor{lightblue}98.10                    & \cellcolor{lightblue}0                                                                          & \cellcolor{lightblue}0                                                                          \\ \cline{5-12} 
                                                        & \multirow{-2}{*}{\LaTeX{}}                             & \multirow{-2}{*}{AsciiMath}                         & \multirow{-2}{*}{Forward}                                                     & \cellcolor{lightpurple}CodeGemma               & \cellcolor{lightpurple}0.0737                    & \cellcolor{lightpurple}0.0047                                                                    & \cellcolor{lightpurple}2.3622                    & \cellcolor{lightpurple}121.12                   & \cellcolor{lightpurple}85.26                    & \cellcolor{lightpurple}0.70                                                                       & \cellcolor{lightpurple}1.20                                                                       \\ \cline{2-12} 
                                                        &                                                     &                                                     &                                                                               & \cellcolor{lightblue}CodeLlama               & \cellcolor{lightblue}0.0077                    & \cellcolor{lightblue}0.0004                                                                    & \cellcolor{lightblue}0.3127                    & \cellcolor{lightblue}894.36                   & \cellcolor{lightblue}98.37                    & \cellcolor{lightblue}0                                                                          & \cellcolor{lightblue}0                                                                          \\ \cline{5-12} 
                                                        & \multirow{-2}{*}{AsciiMath}                         & \multirow{-2}{*}{\LaTeX{}}                             & \multirow{-2}{*}{Backward}                                                     & \cellcolor{lightpurple}CodeGemma               & \cellcolor{lightpurple}0.1231                    & \cellcolor{lightpurple}0.0045                                                                    & \cellcolor{lightpurple}2.9856                    & \cellcolor{lightpurple}117.92                   & \cellcolor{lightpurple}82.96                    & \cellcolor{lightpurple}0.10                                                                       & \cellcolor{lightpurple}0.10                                                                       \\ \cline{2-12} 
                                                        &                                                     &                                                     &                                                                               & \cellcolor{lightblue}CodeLlama               & \cellcolor{lightblue}0.0023                    & \cellcolor{lightblue}0.0002                                                                    & \cellcolor{lightblue}0.2507                    & \cellcolor{lightblue}1026.46                  & \cellcolor{lightblue}98.62                    & \cellcolor{lightblue}0                                                                          & \cellcolor{lightblue}0                                                                          \\ \cline{5-12} 
                                                        & \multirow{-2}{*}{MathML}                            & \multirow{-2}{*}{\LaTeX{}}                             & \multirow{-2}{*}{Forward}                                                     & \cellcolor{lightpurple}CodeGemma               & \cellcolor{lightpurple}0.0853                    & \cellcolor{lightpurple}0.0030                                                                    & \cellcolor{lightpurple}2.5428                    & \cellcolor{lightpurple}109.35                   & \cellcolor{lightpurple}81.96                    & \cellcolor{lightpurple}0                                                                          & \cellcolor{lightpurple}0                                                                          \\ \cline{2-12} 
                                                        &                                                     &                                                     &                                                                               & \cellcolor{lightblue}CodeLlama               & \cellcolor{lightblue}0.0037                    & \cellcolor{lightblue}0.0002                                                                    & \cellcolor{lightblue}0.5105                    & \cellcolor{lightblue}155.67                   & \cellcolor{lightblue}99.34                    & \cellcolor{lightblue}0                                                                          & \cellcolor{lightblue}0                                                                          \\ \cline{5-12} 
                                                        & \multirow{-2}{*}{\LaTeX{}}                             & \multirow{-2}{*}{MathML}                            & \multirow{-2}{*}{Backward}                                                     & \cellcolor{lightpurple}CodeGemma               & \cellcolor{lightpurple}0.0172                    & \cellcolor{lightpurple}0.0006                                                                    & \cellcolor{lightpurple}0.9931                    & \cellcolor{lightpurple}80.27                    & \cellcolor{lightpurple}85.70                    & \cellcolor{lightpurple}0                                                                          & \cellcolor{lightpurple}0                                                                          \\ \cline{2-12} 
                                                        &                                                     &                                                     &                                                                               & \cellcolor{lightblue}CodeLlama               & \cellcolor{lightblue}0.0078                    & \cellcolor{lightblue}0.0005                                                                    & \cellcolor{lightblue}0.5531                    & \cellcolor{lightblue}154.20                   & \cellcolor{lightblue}99.26                    & \cellcolor{lightblue}0                                                                          & \cellcolor{lightblue}0                                                                          \\ \cline{5-12} 
                                                        & \multirow{-2}{*}{AsciiMath}                         & \multirow{-2}{*}{MathML}                            & \multirow{-2}{*}{\begin{tabular}[c]{@{}c@{}}Hypothet.\\ Forward\end{tabular}} & \cellcolor{lightpurple}CodeGemma               & \cellcolor{lightpurple}0.0190                     & \cellcolor{lightpurple}0.0006                                                                    & \cellcolor{lightpurple}1.0907                    & \cellcolor{lightpurple}78.96                    & \cellcolor{lightpurple}86.87                    & \cellcolor{lightpurple}0                                                                          & \cellcolor{lightpurple}0                                                                          \\ \cline{2-12} 
                                                        &                                                     &                                                     &                                                                               & \cellcolor{lightblue}CodeLlama               & \cellcolor{lightblue}0.0023                    & \cellcolor{lightblue}0.0002                                                                    & \cellcolor{lightblue}0.2408                    & \cellcolor{lightblue}1094.34                  & \cellcolor{lightblue}98.55                    & \cellcolor{lightblue}0                                                                          & \cellcolor{lightblue}0                                                                          \\ \cline{5-12} 
  \multirow{-12}{*}{py-asciimath}                       & \multirow{-2}{*}{MathML}                            & \multirow{-2}{*}{AsciiMath}                         & \multirow{-2}{*}{\begin{tabular}[c]{@{}c@{}}Hypothet.\\ Backward\end{tabular}} & \cellcolor{lightpurple}CodeGemma               & \cellcolor{lightpurple}0.0412                    & \cellcolor{lightpurple}0.0030                                                                    & \cellcolor{lightpurple}1.5875                    & \cellcolor{lightpurple}140.33                   & \cellcolor{lightpurple}88.26                    & \cellcolor{lightpurple}0                                                                        & \cellcolor{lightpurple}0.50                                                                       \\ \hline
  \end{tabular}

  \parbox{0.9\linewidth}{\centering \small \textit{Blue cells \textcolor{lightblue}{\fsquare} = CodeLlama 7B; purple cells \textcolor{lightpurple}{\fsquare} = CodeGemma 7B.}}
\end{table}

\begin{table}[h]
  \small
  \centering
  \caption{\label{tab:llm-raw-tuned} 
  Evaluation results for models mocking Pandoc, msticpy, latexify-py, and py-asciimath behaviors queried after fine-tuning.
  %Evaluation results for models mocking Pandoc, msticpy, latexify-py, and py-asciimath behaviors by querying LLMs pre-trained on code and further fine-tuned on synthetic data.
  }

  \begin{tabular}{|lllcrrrrrrrr|}
  \hline
  \rowcolor{lightgrey} 
  \multicolumn{1}{|c|}{\cellcolor{lightgrey}Program} & \multicolumn{1}{c|}{\cellcolor{lightgrey}Source} & \multicolumn{1}{c|}{\cellcolor{lightgrey}Target} & \multicolumn{1}{c|}{\cellcolor{lightgrey}\begin{tabular}[c]{@{}c@{}}Model\\ Type\end{tabular}}  & \multicolumn{1}{c|}{\cellcolor{lightgrey}Model}        & \multicolumn{1}{c|}{\cellcolor{lightgrey}BLEU}   & \multicolumn{1}{c|}{\cellcolor{lightgrey}\begin{tabular}[c]{@{}c@{}}BLEU\\ Error\end{tabular}} & \multicolumn{1}{c|}{\cellcolor{lightgrey}NIST}    & \multicolumn{1}{c|}{\cellcolor{lightgrey}WER}    & \multicolumn{1}{c|}{\cellcolor{lightgrey}WIL}   & \multicolumn{1}{c|}{\cellcolor{lightgrey}\begin{tabular}[c]{@{}c@{}}Exact\\ Match\end{tabular}} & \multicolumn{1}{c|}{\cellcolor{lightgrey}\begin{tabular}[c]{@{}c@{}}Close\\ Match\end{tabular}} \\ \hline
  \multicolumn{1}{|l|}{}                                & \multicolumn{1}{l|}{}                               & \multicolumn{1}{l|}{}                               & \multicolumn{1}{c|}{}                                                                              & \multicolumn{1}{r|}{\cellcolor{lightblue}CodeLlama} & \multicolumn{1}{r|}{\cellcolor{lightblue}0.4248} & \multicolumn{1}{r|}{\cellcolor{lightblue}0.0033}                                               & \multicolumn{1}{r|}{\cellcolor{lightblue}6.4448}  & \multicolumn{1}{r|}{\cellcolor{lightblue}87.7}   & \multicolumn{1}{r|}{\cellcolor{lightblue}71.89} & \multicolumn{1}{r|}{\cellcolor{lightblue}6.78}                                                  & \cellcolor{lightblue}12.16                                                                      \\ \cline{5-12} 
  \multicolumn{1}{|l|}{}                                & \multicolumn{1}{l|}{\multirow{-2}{*}{Markdown}}     & \multicolumn{1}{l|}{\multirow{-2}{*}{HTML}}         & \multicolumn{1}{c|}{\multirow{-2}{*}{Forward}}                                                     & \multicolumn{1}{r|}{\cellcolor{lightpurple}CodeGemma} & \multicolumn{1}{r|}{\cellcolor{lightpurple}0.3463} & \multicolumn{1}{r|}{\cellcolor{lightpurple}0.0027}                                               & \multicolumn{1}{r|}{\cellcolor{lightpurple}5.6852}  & \multicolumn{1}{r|}{\cellcolor{lightpurple}82.44}  & \multicolumn{1}{r|}{\cellcolor{lightpurple}76.61} & \multicolumn{1}{r|}{\cellcolor{lightpurple}0}                                                     & \cellcolor{lightpurple}0                                                                          \\ \cline{2-12} 
  \multicolumn{1}{|l|}{}                                & \multicolumn{1}{l|}{}                               & \multicolumn{1}{l|}{}                               & \multicolumn{1}{c|}{}                                                                              & \multicolumn{1}{r|}{\cellcolor{lightblue}CodeLlama} & \multicolumn{1}{r|}{\cellcolor{lightblue}0.4148} & \multicolumn{1}{r|}{\cellcolor{lightblue}0.0031}                                               & \multicolumn{1}{r|}{\cellcolor{lightblue}6.4789}  & \multicolumn{1}{r|}{\cellcolor{lightblue}95.00}  & \multicolumn{1}{r|}{\cellcolor{lightblue}64.99} & \multicolumn{1}{r|}{\cellcolor{lightblue}0}                                                     & \cellcolor{lightblue}13.05                                                                      \\ \cline{5-12} 
  \multicolumn{1}{|l|}{\multirow{-4}{*}{Pandoc}}        & \multicolumn{1}{l|}{\multirow{-2}{*}{HTML}}         & \multicolumn{1}{l|}{\multirow{-2}{*}{Markdown}}     & \multicolumn{1}{c|}{\multirow{-2}{*}{Backward}}                                                     & \multicolumn{1}{r|}{\cellcolor{lightpurple}CodeGemma} & \multicolumn{1}{r|}{\cellcolor{lightpurple}0.4701} & \multicolumn{1}{r|}{\cellcolor{lightpurple}0.0022}                                               & \multicolumn{1}{r|}{\cellcolor{lightpurple}7.9473}  & \multicolumn{1}{r|}{\cellcolor{lightpurple}51.40}  & \multicolumn{1}{r|}{\cellcolor{lightpurple}54.69} & \multicolumn{1}{r|}{\cellcolor{lightpurple}0}                                                     & \cellcolor{lightpurple}0                                                                          \\ \hline
  \multicolumn{1}{|l|}{}                                & \multicolumn{1}{l|}{}                               & \multicolumn{1}{l|}{}                               & \multicolumn{1}{c|}{}                                                                              & \multicolumn{1}{r|}{\cellcolor{lightblue}CodeLlama} & \multicolumn{1}{r|}{\cellcolor{lightblue}0.4240} & \multicolumn{1}{r|}{\cellcolor{lightblue}0.0071}                                               & \multicolumn{1}{r|}{\cellcolor{lightblue}6.6428}  & \multicolumn{1}{r|}{\cellcolor{lightblue}82.81}  & \multicolumn{1}{r|}{\cellcolor{lightblue}56.16} & \multicolumn{1}{r|}{\cellcolor{lightblue}0}                                                     & \cellcolor{lightblue}3.00                                                                       \\ \cline{5-12} 
  \multicolumn{1}{|l|}{}                                & \multicolumn{1}{l|}{\multirow{-2}{*}{SQL}}          & \multicolumn{1}{l|}{\multirow{-2}{*}{KQL}}          & \multicolumn{1}{c|}{\multirow{-2}{*}{Forward}}                                                     & \multicolumn{1}{r|}{\cellcolor{lightpurple}CodeGemma} & \multicolumn{1}{r|}{\cellcolor{lightpurple}0.7035} & \multicolumn{1}{r|}{\cellcolor{lightpurple}0.0045}                                               & \multicolumn{1}{r|}{\cellcolor{lightpurple}10.3251} & \multicolumn{1}{r|}{\cellcolor{lightpurple}22.59}  & \multicolumn{1}{r|}{\cellcolor{lightpurple}25.89} & \multicolumn{1}{r|}{\cellcolor{lightpurple}0}                                                     & \cellcolor{lightpurple}0                                                                          \\ \cline{2-12} 
  \multicolumn{1}{|l|}{}                                & \multicolumn{1}{l|}{}                               & \multicolumn{1}{l|}{}                               & \multicolumn{1}{c|}{}                                                                              & \multicolumn{1}{r|}{\cellcolor{lightblue}CodeLlama} & \multicolumn{1}{r|}{\cellcolor{lightblue}0.7787} & \multicolumn{1}{r|}{\cellcolor{lightblue}0.0055}                                               & \multicolumn{1}{r|}{\cellcolor{lightblue}11.0942} & \multicolumn{1}{r|}{\cellcolor{lightblue}14.42}  & \multicolumn{1}{r|}{\cellcolor{lightblue}21.95} & \multicolumn{1}{r|}{\cellcolor{lightblue}0}                                                   & \cellcolor{lightblue}15.20                                                                      \\ \cline{5-12} 
  \multicolumn{1}{|l|}{\multirow{-4}{*}{msticpy}}       & \multicolumn{1}{l|}{\multirow{-2}{*}{KQL}}          & \multicolumn{1}{l|}{\multirow{-2}{*}{SQL}}          & \multicolumn{1}{c|}{\multirow{-2}{*}{Backward}}                                                     & \multicolumn{1}{r|}{\cellcolor{lightpurple}CodeGemma} & \multicolumn{1}{r|}{\cellcolor{lightpurple}0.0896} & \multicolumn{1}{r|}{\cellcolor{lightpurple}0.0096}                                               & \multicolumn{1}{r|}{\cellcolor{lightpurple}1.2981}  & \multicolumn{1}{r|}{\cellcolor{lightpurple}427.72} & \multicolumn{1}{r|}{\cellcolor{lightpurple}91.98} & \multicolumn{1}{r|}{\cellcolor{lightpurple}0}                                                     & \cellcolor{lightpurple}0                                                                          \\ \hline
  \multicolumn{1}{|l|}{}                                & \multicolumn{1}{l|}{}                               & \multicolumn{1}{l|}{}                               & \multicolumn{1}{c|}{}                                                                              & \multicolumn{1}{r|}{\cellcolor{lightblue}CodeLlama} & \multicolumn{1}{r|}{\cellcolor{lightblue}0.4270} & \multicolumn{1}{r|}{\cellcolor{lightblue}0.0054}                                               & \multicolumn{1}{r|}{\cellcolor{lightblue}6.1379}  & \multicolumn{1}{r|}{\cellcolor{lightblue}59.58}  & \multicolumn{1}{r|}{\cellcolor{lightblue}52.22} & \multicolumn{1}{r|}{\cellcolor{lightblue}0}                                                     & \cellcolor{lightblue}0                                                                          \\ \cline{5-12} 
  \multicolumn{1}{|l|}{}                                & \multicolumn{1}{l|}{\multirow{-2}{*}{PyExpres.}}    & \multicolumn{1}{l|}{\multirow{-2}{*}{\LaTeX{}}}        & \multicolumn{1}{c|}{\multirow{-2}{*}{Forward}}                                                     & \multicolumn{1}{r|}{\cellcolor{lightpurple}CodeGemma} & \multicolumn{1}{r|}{\cellcolor{lightpurple}0.5250} & \multicolumn{1}{r|}{\cellcolor{lightpurple}0.0029}                                               & \multicolumn{1}{r|}{\cellcolor{lightpurple}7.7616}  & \multicolumn{1}{r|}{\cellcolor{lightpurple}27.17}  & \multicolumn{1}{r|}{\cellcolor{lightpurple}39.37} & \multicolumn{1}{r|}{\cellcolor{lightpurple}0}                                                     & \cellcolor{lightpurple}0                                                                          \\ \cline{2-12} 
  \multicolumn{1}{|l|}{}                                & \multicolumn{1}{l|}{}                               & \multicolumn{1}{l|}{}                               & \multicolumn{1}{c|}{}                                                                              & \multicolumn{1}{r|}{\cellcolor{lightblue}CodeLlama} & \multicolumn{1}{r|}{\cellcolor{lightblue}0.4777} & \multicolumn{1}{r|}{\cellcolor{lightblue}0.0103}                                               & \multicolumn{1}{r|}{\cellcolor{lightblue}6.7309}  & \multicolumn{1}{r|}{\cellcolor{lightblue}82.56}  & \multicolumn{1}{r|}{\cellcolor{lightblue}52.97} & \multicolumn{1}{r|}{\cellcolor{lightblue}10.10}                                                 & \cellcolor{lightblue}14.90                                                                      \\ \cline{5-12} 
  \multicolumn{1}{|l|}{\multirow{-4}{*}{latexify-py}}   & \multicolumn{1}{l|}{\multirow{-2}{*}{\LaTeX{}}}        & \multicolumn{1}{l|}{\multirow{-2}{*}{PyExpres.}}    & \multicolumn{1}{c|}{\multirow{-2}{*}{Backward}}                                                     & \multicolumn{1}{r|}{\cellcolor{lightpurple}CodeGemma} & \multicolumn{1}{r|}{\cellcolor{lightpurple}0.8204} & \multicolumn{1}{r|}{\cellcolor{lightpurple}0.0048}                                               & \multicolumn{1}{r|}{\cellcolor{lightpurple}11.3188} & \multicolumn{1}{r|}{\cellcolor{lightpurple}16.55}  & \multicolumn{1}{r|}{\cellcolor{lightpurple}18.91} & \multicolumn{1}{r|}{\cellcolor{lightpurple}0}                                                     & \cellcolor{lightpurple}0                                                                          \\ \hline
  \multicolumn{1}{|l|}{}                                & \multicolumn{1}{l|}{}                               & \multicolumn{1}{l|}{}                               & \multicolumn{1}{c|}{}                                                                              & \multicolumn{1}{r|}{\cellcolor{lightblue}CodeLlama} & \multicolumn{1}{r|}{\cellcolor{lightblue}0.5603} & \multicolumn{1}{r|}{\cellcolor{lightblue}0.0085}                                               & \multicolumn{1}{r|}{\cellcolor{lightblue}8.3336}  & \multicolumn{1}{r|}{\cellcolor{lightblue}43.06}  & \multicolumn{1}{r|}{\cellcolor{lightblue}41.39} & \multicolumn{1}{r|}{\cellcolor{lightblue}1.40}                                                  & \cellcolor{lightblue}21.20                                                                      \\ \cline{5-12} 
  \multicolumn{1}{|l|}{}                                & \multicolumn{1}{l|}{\multirow{-2}{*}{\LaTeX{}}}        & \multicolumn{1}{l|}{\multirow{-2}{*}{AsciiMath}}    & \multicolumn{1}{c|}{\multirow{-2}{*}{Forward}}                                                     & \multicolumn{1}{r|}{\cellcolor{lightpurple}CodeGemma} & \multicolumn{1}{r|}{\cellcolor{lightpurple}0.5500} & \multicolumn{1}{r|}{\cellcolor{lightpurple}0.0080}                                               & \multicolumn{1}{r|}{\cellcolor{lightpurple}8.2532}  & \multicolumn{1}{r|}{\cellcolor{lightpurple}37.55}  & \multicolumn{1}{r|}{\cellcolor{lightpurple}39.34} & \multicolumn{1}{r|}{\cellcolor{lightpurple}0}                                                     & \cellcolor{lightpurple}0                                                                          \\ \cline{2-12} 
  \multicolumn{1}{|l|}{}                                & \multicolumn{1}{l|}{}                               & \multicolumn{1}{l|}{}                               & \multicolumn{1}{c|}{}                                                                              & \multicolumn{1}{r|}{\cellcolor{lightblue}CodeLlama} & \multicolumn{1}{r|}{\cellcolor{lightblue}0.2036} & \multicolumn{1}{r|}{\cellcolor{lightblue}0.0084}                                               & \multicolumn{1}{r|}{\cellcolor{lightblue}3.6731}  & \multicolumn{1}{r|}{\cellcolor{lightblue}133.84} & \multicolumn{1}{r|}{\cellcolor{lightblue}76.99} & \multicolumn{1}{r|}{\cellcolor{lightblue}2.00}                                                  & \cellcolor{lightblue}2.60                                                                       \\ \cline{5-12} 
  \multicolumn{1}{|l|}{}                                & \multicolumn{1}{l|}{\multirow{-2}{*}{AsciiMath}}    & \multicolumn{1}{l|}{\multirow{-2}{*}{\LaTeX{}}}        & \multicolumn{1}{c|}{\multirow{-2}{*}{Backward}}                                                     & \multicolumn{1}{r|}{\cellcolor{lightpurple}CodeGemma} & \multicolumn{1}{r|}{\cellcolor{lightpurple}0.3616} & \multicolumn{1}{r|}{\cellcolor{lightpurple}0.0076}                                               & \multicolumn{1}{r|}{\cellcolor{lightpurple}6.0105}  & \multicolumn{1}{r|}{\cellcolor{lightpurple}64.78}  & \multicolumn{1}{r|}{\cellcolor{lightpurple}61.27} & \multicolumn{1}{r|}{\cellcolor{lightpurple}0}                                                     & \cellcolor{lightpurple}0                                                                          \\ \cline{2-12} 
  \multicolumn{1}{|l|}{}                                & \multicolumn{1}{l|}{}                               & \multicolumn{1}{l|}{}                               & \multicolumn{1}{c|}{}                                                                              & \multicolumn{1}{r|}{\cellcolor{lightblue}CodeLlama} & \multicolumn{1}{r|}{\cellcolor{lightblue}0.2441} & \multicolumn{1}{r|}{\cellcolor{lightblue}0.0066}                                               & \multicolumn{1}{r|}{\cellcolor{lightblue}4.9843}  & \multicolumn{1}{r|}{\cellcolor{lightblue}76.21}  & \multicolumn{1}{r|}{\cellcolor{lightblue}69.62} & \multicolumn{1}{r|}{\cellcolor{lightblue}0.50}                                                  & \cellcolor{lightblue}0.70                                                                       \\ \cline{5-12} 
  \multicolumn{1}{|l|}{}                                & \multicolumn{1}{l|}{\multirow{-2}{*}{MathML}}       & \multicolumn{1}{l|}{\multirow{-2}{*}{\LaTeX{}}}        & \multicolumn{1}{c|}{\multirow{-2}{*}{Forward}}                                                     & \multicolumn{1}{r|}{\cellcolor{lightpurple}CodeGemma} & \multicolumn{1}{r|}{\cellcolor{lightpurple}0.2991} & \multicolumn{1}{r|}{\cellcolor{lightpurple}0.0072}                                               & \multicolumn{1}{r|}{\cellcolor{lightpurple}5.4391}  & \multicolumn{1}{r|}{\cellcolor{lightpurple}65.28}  & \multicolumn{1}{r|}{\cellcolor{lightpurple}64.26} & \multicolumn{1}{r|}{\cellcolor{lightpurple}0}                                                     & \cellcolor{lightpurple}0                                                                          \\ \cline{2-12} 
  \multicolumn{1}{|l|}{}                                & \multicolumn{1}{l|}{}                               & \multicolumn{1}{l|}{}                               & \multicolumn{1}{c|}{}                                                                              & \multicolumn{1}{r|}{\cellcolor{lightblue}CodeLlama} & \multicolumn{1}{r|}{\cellcolor{lightblue}0.5168} & \multicolumn{1}{r|}{\cellcolor{lightblue}0.0050}                                               & \multicolumn{1}{r|}{\cellcolor{lightblue}4.0192}  & \multicolumn{1}{r|}{\cellcolor{lightblue}75.85}  & \multicolumn{1}{r|}{\cellcolor{lightblue}69.88} & \multicolumn{1}{r|}{\cellcolor{lightblue}0}                                                     & \cellcolor{lightblue}0                                                                          \\ \cline{5-12} 
  \multicolumn{1}{|l|}{}                                & \multicolumn{1}{l|}{\multirow{-2}{*}{\LaTeX{}}}        & \multicolumn{1}{l|}{\multirow{-2}{*}{MathML}}       & \multicolumn{1}{c|}{\multirow{-2}{*}{Backward}}                                                     & \multicolumn{1}{r|}{\cellcolor{lightpurple}CodeGemma} & \multicolumn{1}{r|}{\cellcolor{lightpurple}0.5542} & \multicolumn{1}{r|}{\cellcolor{lightpurple}0.0043}                                               & \multicolumn{1}{r|}{\cellcolor{lightpurple}4.4655}  & \multicolumn{1}{r|}{\cellcolor{lightpurple}63.05}  & \multicolumn{1}{r|}{\cellcolor{lightpurple}62.49} & \multicolumn{1}{r|}{\cellcolor{lightpurple}0}                                                     & \cellcolor{lightpurple}0                                                                          \\ \cline{2-12} 
  \multicolumn{1}{|l|}{}                                & \multicolumn{1}{l|}{}                               & \multicolumn{1}{l|}{}                               & \multicolumn{1}{c|}{}                                                                              & \multicolumn{1}{r|}{\cellcolor{lightblue}CodeLlama} & \multicolumn{1}{r|}{\cellcolor{lightblue}0.5176} & \multicolumn{1}{r|}{\cellcolor{lightblue}0.0049}                                               & \multicolumn{1}{r|}{\cellcolor{lightblue}3.9829}  & \multicolumn{1}{r|}{\cellcolor{lightblue}77.77}  & \multicolumn{1}{r|}{\cellcolor{lightblue}69.13} & \multicolumn{1}{r|}{\cellcolor{lightblue}0}                                                     & \cellcolor{lightblue}0                                                                          \\ \cline{5-12} 
  \multicolumn{1}{|l|}{}                                & \multicolumn{1}{l|}{\multirow{-2}{*}{AsciiMath}}    & \multicolumn{1}{l|}{\multirow{-2}{*}{MathML}}       & \multicolumn{1}{c|}{\multirow{-2}{*}{\begin{tabular}[c]{@{}c@{}}Hypothet.\\ Forward\end{tabular}}} & \multicolumn{1}{r|}{\cellcolor{lightpurple}CodeGemma} & \multicolumn{1}{r|}{\cellcolor{lightpurple}0.5210} & \multicolumn{1}{r|}{\cellcolor{lightpurple}0.0043}                                               & \multicolumn{1}{r|}{\cellcolor{lightpurple}4.0966}  & \multicolumn{1}{r|}{\cellcolor{lightpurple}68.45}  & \multicolumn{1}{r|}{\cellcolor{lightpurple}64.73} & \multicolumn{1}{r|}{\cellcolor{lightpurple}0}                                                     & \cellcolor{lightpurple}0                                                                          \\ \cline{2-12} 
  \multicolumn{1}{|l|}{}                                & \multicolumn{1}{l|}{}                               & \multicolumn{1}{l|}{}                               & \multicolumn{1}{c|}{}                                                                              & \multicolumn{1}{r|}{\cellcolor{lightblue}CodeLlama} & \multicolumn{1}{r|}{\cellcolor{lightblue}0.2892} & \multicolumn{1}{r|}{\cellcolor{lightblue}0.0071}                                               & \multicolumn{1}{r|}{\cellcolor{lightblue}5.3566}  & \multicolumn{1}{r|}{\cellcolor{lightblue}68.68}  & \multicolumn{1}{r|}{\cellcolor{lightblue}62.83} & \multicolumn{1}{r|}{\cellcolor{lightblue}0.30}                                                  & \cellcolor{lightblue}3.30                                                                       \\ \cline{5-12} 
  \multicolumn{1}{|l|}{\multirow{-12}{*}{py-asciimath}} & \multicolumn{1}{l|}{\multirow{-2}{*}{MathML}}       & \multicolumn{1}{l|}{\multirow{-2}{*}{AsciiMath}}    & \multicolumn{1}{c|}{\multirow{-2}{*}{\begin{tabular}[c]{@{}c@{}}Hypothet.\\ Backward\end{tabular}}} & \multicolumn{1}{r|}{\cellcolor{lightpurple}CodeGemma} & \multicolumn{1}{r|}{\cellcolor{lightpurple}0.3590} & \multicolumn{1}{r|}{\cellcolor{lightpurple}0.0073}                                               & \multicolumn{1}{r|}{\cellcolor{lightpurple}6.1543}  & \multicolumn{1}{r|}{\cellcolor{lightpurple}52.99}  & \multicolumn{1}{r|}{\cellcolor{lightpurple}57.63} & \multicolumn{1}{r|}{\cellcolor{lightpurple}0}                                                     & \cellcolor{lightpurple}0                                                                          \\ \hline
  \multicolumn{12}{|c|}{\cellcolor{lightgrey}Outliers}                                                                                                                                                                                                                                                                                                                                                                                                                                                                                                                                                                                                                                                                                                                                                                                                                                            \\ \hline
  \multicolumn{1}{|l|}{Pandoc}                          & \multicolumn{1}{l|}{Markdown}                       & \multicolumn{1}{l|}{HTML}                           & \multicolumn{1}{c|}{Forward}                                                                       & \multicolumn{1}{r|}{\cellcolor{lightyellow}CodeLlama} & \multicolumn{1}{r|}{\cellcolor{lightyellow}0.5645} & \multicolumn{1}{r|}{\cellcolor{lightyellow}0.0026}                                               & \multicolumn{1}{r|}{\cellcolor{lightyellow}4.9441}  & \multicolumn{1}{r|}{\cellcolor{lightyellow}73.09}  & \multicolumn{1}{r|}{\cellcolor{lightyellow}60.17} & \multicolumn{1}{r|}{\cellcolor{lightyellow}15.88}                                                 & \cellcolor{lightyellow}71.97                                                                      \\ \hline
  \multicolumn{1}{|l|}{Pandoc}                          & \multicolumn{1}{l|}{HTML}                           & \multicolumn{1}{l|}{Markdown}                       & \multicolumn{1}{c|}{Backward}                                                                       & \multicolumn{1}{r|}{\cellcolor{lightyellow}CodeLlama} & \multicolumn{1}{r|}{\cellcolor{lightyellow}0.2424} & \multicolumn{1}{r|}{\cellcolor{lightyellow}0.0028}                                               & \multicolumn{1}{r|}{\cellcolor{lightyellow}2.3148}  & \multicolumn{1}{r|}{\cellcolor{lightyellow}188.1}  & \multicolumn{1}{r|}{\cellcolor{lightyellow}78.01} & \multicolumn{1}{r|}{\cellcolor{lightyellow}0}                                                   & \cellcolor{lightyellow}40.18                                                                      \\ \hline
  \multicolumn{1}{|l|}{latexify-py}                     & \multicolumn{1}{l|}{\LaTeX{}}                          & \multicolumn{1}{l|}{PyExpres.}                      & \multicolumn{1}{c|}{Backward}                                                                       & \multicolumn{1}{r|}{\cellcolor{lightyellow}CodeLlama} & \multicolumn{1}{r|}{\cellcolor{lightyellow}0.5306} & \multicolumn{1}{r|}{\cellcolor{lightyellow}0.0100}                                               & \multicolumn{1}{r|}{\cellcolor{lightyellow}5.065}   & \multicolumn{1}{r|}{\cellcolor{lightyellow}73.24}  & \multicolumn{1}{r|}{\cellcolor{lightyellow}46.79} & \multicolumn{1}{r|}{\cellcolor{lightyellow}15.4}                                                  & \cellcolor{lightyellow}20.2                                                                       \\ \hline
  \multicolumn{1}{|l|}{py-asciimath}                    & \multicolumn{1}{l|}{\LaTeX{}}                          & \multicolumn{1}{l|}{AsciiMath}                      & \multicolumn{1}{c|}{Forward}                                                                       & \multicolumn{1}{r|}{\cellcolor{lightyellow}CodeLlama} & \multicolumn{1}{r|}{\cellcolor{lightyellow}0.8283} & \multicolumn{1}{r|}{\cellcolor{lightyellow}0.0058}                                               & \multicolumn{1}{r|}{\cellcolor{lightyellow}7.0132}  & \multicolumn{1}{r|}{\cellcolor{lightyellow}15.54}  & \multicolumn{1}{r|}{\cellcolor{lightyellow}18.09} & \multicolumn{1}{r|}{\cellcolor{lightyellow}1.9}                                                   & \cellcolor{lightyellow}30.4                                                                       \\ \hline
  \end{tabular}
  \parbox{0.9\linewidth}{\centering \small \textit{Blue cells \textcolor{lightblue}{\fsquare} = CodeLlama 7B model; purple cells \textcolor{lightpurple}{\fsquare} = CodeGemma 7B model; \\ yellow cells \textcolor{lightyellow}{\fsquare} = CodeLlama 7B with the content part of model input being masked with placeholders.}}
\end{table}

\begin{table}[h]
    \small
    \centering
    \caption{\label{tab:llm-fine-tuning}Comparison of fine-tuning times (in minutes) for CodeLlama 7B and CodeGemma 7B across multiple subjects.}
    \rowcolors{2}{gray!25}{white}
    \begin{tabular}{lrrrrr}
    \rowcolor{lightgrey} 
    \multicolumn{1}{c}{\cellcolor{lightgrey}Subject} & \multicolumn{1}{c}{\cellcolor{lightgrey}CodeGemma} & \multicolumn{1}{c}{\cellcolor{lightgrey}CodeLlama} & \multicolumn{1}{c}{\cellcolor{lightgrey}Custom 10k} & \multicolumn{1}{c}{\cellcolor{lightgrey}Custom 250k} & \multicolumn{1}{c}{\cellcolor{lightgrey}Custom 1M} \\
    Markdown - HTML                                     & 304.6                                                 & 296.0                                                 & 26.5                                                   & 407.4                                                   & 1359.1                                                \\
    HTML - Markdown                                     & 305.7                                                 & 297.6                                                 & 27.0                                                   & 407.2                                                   & 1354.3                                                \\
    SQL - KQL                                           & 517.9                                                 & 506.8                                                 & 18.1                                                   & 451.9                                                   & 1827.4                                                \\
    KQL - SQL                                           & 520.2                                                 & 508.2                                                 & 13.4                                                   & 330.8                                                   & 1322.3                                                \\
    PyExpres. - LaTeX                                   & 1248.8                                                & 1278.6                                                & 23.8                                                   & 619.9                                                   & 2534.4                                                \\
    LaTeX - PyExpres.                                   & 1244.4                                                & 1281.2                                                & 14.9                                                   & 375.3                                                   & 1488.2                                                \\
    LaTeX - AsciiMath                                   & 113.0                                                 & 119.9                                                 & 25.2                                                   & 642.2                                                   & 2626.6                                                \\
    AsciiMath - LaTeX                                   & 112.8                                                 & 120.2                                                 & 13.1                                                   & 315.8                                                   & 1318.7                                                \\
    MathML - LaTeX                                      & 261.6                                                 & 256.3                                                 & 28.0                                                   & 704.6                                                   & 2904.5                                                \\
    LaTeX - MathML                                      & 263.2                                                 & 258.5                                                 & 23.1                                                   & 570.5                                                   & 2314.2                                                \\
    AsciiMath - MathML                                  & 258.9                                                 & 253.8                                                 & 18.1                                                   & 446.8                                                   & 1823.4                                                \\
    MathML - AsciiMath                                  & 259.1                                                 & 254.6                                                 & 22.9                                                   & 579.5                                                   & 2374.0                                               
    \end{tabular}
\end{table}

After zero-shot prompting, we independently fine-tuned foundational models for conversion tasks based on every subject-to-conversion direction combination using 10,000 synthesized program input-output pairs to exclude the possibility of cross-subject interference and simplify the comparison with custom subject-specific models. 
We also reused the same instruction template from zero-shot prompting experiments during the model fine-tuning and inference phases.
We recorded significantly higher BLEU score values for all subjects, meaning that models now better understand the mapping between program input-output features, thus forming better syntactically and semantically structured predictions. 
However, Exact Match scores reflect that pre-trained LLMs cannot serve as in-place mocks for programs. 
Fine-tuning does not completely prevent models from injecting additional tokens, which is intolerable for programs expecting well-structured inputs.
We assume that iteratively refining incorrect model predictions using additional prompting rounds and fine-tuning on more samples could help fix hallucinations of LLMs in behavior-mocking and reverse engineering tasks. 
Once we additionally masked the content part of the model input with corresponding placeholders in the same way as we did in custom-trained models, we achieved better model performance in half of the subjects with \emph{CodeLlama} models. These cases are marked as outliers in \Cref{tab:llm-raw-tuned}.

We also compared the LLM fine-tuning time with training custom models from scratch on different training sets consisting of 10,000 (small), 250,000 (medium), and 1,000,000 (large) entries. While custom model training was performed on a widely used Nvidia RTX 4090 high-end gaming level card, fine-tuning of quantized LLMs required a more powerful data-center grade Nvidia A100 Tensor Core GPU.  Despite the dissimilarity in the computation capabilities of selected hardware accelerators, the results in \Cref{tab:llm-fine-tuning} indicate that LLM fine-tuning on a small dataset requires time similar to training custom models from scratch using a medium-size dataset with 250,000 entries.

In conclusion, despite our early results in the field of automated behavior extraction, we think that custom-trained models are more suitable for program behavior mocking and reverse engineering of evaluated subjects. 
We want to point out several core differences. 
Using LLMs sometimes requires processing data by third parties, which could be undesired or disallowed by terms and conditions, for example, because vendors host models themselves and only provide API access to end-users. 
Typically, proprietary models require significant computation resources, which are inaccessible to individual developers and small-sized companies.
Depending on usage patterns, the paywall could restrict API access to proprietary models and become costly for long-term use.
At the same time, our experiments with locally hosted, relatively small by current industry standards LLMs containing 7 billion parameters have demonstrated that such LLMs cannot replace programs in typical usage scenarios.
Even after fine-tuning, LLMs still struggle to replicate the behavior identical to the actual program.
In contrast, program-specific models learned by \Modelizer
\begin{itemize}
  \item achieve higher forward and inverse mocking accuracy;
  \item have significantly lower computation requirements; and
  \item allow for integration into existing deployment or testing scenarios.
\end{itemize} at a smaller cost. 
We foresee a big potential for the co-integration of lightweight program-specific models with large language models and large reasoning models.

\begin{result}
    Behavior mocking with LLMs is less efficient than training custom models because they are not task-specific, require more resources, and offer similar or lower performance.
\end{result}

\section{Limitations}
\label{sec:limitations}

In this section, we want to discuss the limitations of the \Modelizer framework and the current generation of sequence-to-sequence translation models. In particular, we want to address the following limitations:
\begin{description}

  \item [Stateless and Stateful Programs.] In the current study, we have explored programs that implement deterministic and stateless algorithms, including data format converters with recursive descent parsers. Such programs fit the nature and functionality of sequence-to-sequence models. In contrast, the modeling of stateful systems will require tracing and encoding additional program states as a part of the input data. Such a task implies further research and implementation of extensions to the framework to identify the meaningful fraction of the program's state representation that should be used to model its behavior and the development of efficient state encoding and tokenization algorithms. We assume that the ability of the \Modelizer framework to automatically validate and refine the learned models in combination with reinforcement learning algorithms~\cite{sutton2018reinforcement} will help to solve this problem. We plan to address this research topic in our future work.
  
  \item [Computational Complexity.] \Modelizer can not create models for all sorts of programs. The program logic or its individual capabilities can be too complex to be modeled by sequence-to-sequence models. The transformer models themselves have a quadratic computational complexity. For the input sequence length~$n$, hidden dimension size~$h$, and number of layers in the neural network~$L$, we get the following computational complexity:
    \[
    \text{Time Complexity:}\quad O\bigl(L \times (n^2d + nd^2)\bigr)
    \qquad
    \text{Space Complexity:}\quad O\bigl(L \times (n^2 + nd^2)\bigr)
    \]
  This means that modeling of extremely long input-output will be challenging and impractical. The models that are learned by \Modelizer are not intended to ubiquitously replace programs. We see \Modelizer as a tool that can help to understand the behavior of programs and assist in program testing and debugging.  

  % \item [Input Collisions.] Accurate prediction of program outputs requires the program under test (PUT) to be \emph{deterministic}, producing the same output for a given input every time. However, prediction becomes more complex when multiple inputs map to the same output. In such cases, using an intermediate representation may help, but it can lead to \emph{context loss}.
 
  \item[Context Window Size.] Similar to many Transformer-based systems, \Modelizer is sensitive to the length of the processed sequences. One of the factors that impact translation accuracy is the synthesis of training sequences. By default, the grammar-based fuzzers used in the implementation of our Generators tend to produce short sequences. To overcome this problem, we have introduced the sliding non-terminal expansion window, which partially resolves this issue. While the total token quantity per input and output sequences reached values similar to real-world samples, it does not guarantee a correlation for the input/output feature distributions between synthesized and real-world samples. At the same time, such a distribution can be learned from existing input samples~\cite{inputs-from-hell}. This approach introduces three requirements: such inputs have to be (1)~available before input generation, (2)~parsable with selected grammar, and (3)~the correct values have to be set for the minimal and maximal number of non-terminal expansions in the generator. In the current study, we intentionally have not followed this approach to check \Modelizer's capabilities in learning behavior models of arbitrary systems given only the formal specification of such systems.

  \item[Program-Specific Input Output Formats.] Certain subjects, specifically the output data types like KQL, required the implementation of custom parsers due to the unavailability of third-party parsers at the time this publication was prepared. The implementation of such parsers is a time-consuming and error-prone task. The third-party pre-trained tokenizers suboptimally partition the corresponding sequences by unnecessarily splitting atomic tokens into sub-tokens similar to the constructions that are found in natural language sequences, misinterpreting certain characters like punctuation symbols, which cover part of the format syntax. Partially, such an issue can be mitigated by defining custom post-processing routines that will refine the outcome of the tokenization phase, but, such a list of regular expressions and string replacement rules can quickly become very long, create dependencies and conflicts in the order or application, what makes them hard to maintain. Such measures do not simplify the tokenization process but rather move the complexity from the tokenization phase to the post-processing phase. While the custom subject-specific parser-tokenizers are robust and produce high-quality tokens, we do not think that automated model learning problems can benefit from this approach in the long run, specifically for cases when complex system behavior has to be modelized by chaining the behavior models of sub-systems. We think that such a domain-specific tokenizer can be automatically created by wrapping existing tokenization strategies like byte-pair encoding, word-piece, or sentence-piece tokenization with a reinforcement learning loop. Alternatively, we are looking forward to solutions that would be able to mine the input-output grammars from the PUTs and automatically infer the parser-tokenizers from them.
  
  \item[Semantic Validity of Synthesized Inputs.] Given the correct grammar definition, our generators, which rely on grammar-based fuzzers, are very efficient in producing syntactically correct inputs. However, the PUT still can reject these inputs because grammar-based fuzzers cannot guarantee that the generated input will be \emph{semantically} valid and pass all parsing guards in the PUT. Thus, an additional refinement of the synthesized data might be required. Steinhöfel et al.~\cite{steinhoefel.zeller-22} have proposed a technique and a formal language that allows manual specification of the semantic constraints inferred from the system specification, domain knowledge, and human expertise. Their approach, which builds on a modern Z3 SMT solver, can automatically refine the semantically invalid inputs using encoded constraints. However, fuzzing with semantic constraints is a computationally expensive task, and it slows down the formation of training datasets on a large scale. One type of inputs that we synthesized, and which required such a refinement, were SQL queries with a JOIN condition. Using our SQL grammar, the GrammarFuzzer was able to produce SQL queries similar to the one given below:
  \begin{center}
    \verb|SELECT COLUMN FROM TABLE JOIN TABLE ON TABLE.COLUMN = TABLE.COLUMN|
  \end{center}
  Once PlaceholderProcessor updates all placeholders with unique identifiers, the query starts to look like this:
  \begin{center}
    \verb|SELECT COLUMN_1 FROM TABLE_1 JOIN TABLE_2 ON TABLE_3.COLUMN_2 = TABLE_4.COLUMN_3|
  \end{center}
  The refined query is semantically invalid. To overcome this issue, we implemented an additional string refinement step, which parses the synthesized input and corrects the identifiers. While such an additional refinement step requires a bit of manual effort, it adds an insignificant performance penalty to the input generation process for simple cases like this. Thus, the current implementation of the \Modelizer framework does not support fuzzing with constraint solving, but we plan to support this functionality in future releases.

\end{description}

Some of the identified limitations can be resolved in the future releases of \Modelizer.

\section{Related Work}
\label{sec:related-work}
The evolution of language models and their integration into existing projects and solutions have opened new horizons in automating routine tasks and processes~\cite{chen2021evaluating,10.1145/3520312.3534864}. 
The so-called general-purpose LLMs like Llama~2~\cite{touvron2023llama}, Llama~3~\cite{llama3modelcard} or T5~\cite{2020t5} have been adapted to process code-related tasks by fine-tuning on a huge corpus of publicly available source codes and documentation (Code Llama~\cite{roziere2024code} or CodeT5~\cite{wang-etal-2021-codet5}). 
The latter are pre-equipped with a deep understanding of programming syntax, idioms, and semantic nuances. 
Such optimized code-specific models can be further fine-tuned on code generation and understanding downstream tasks.
Using such a transfer learning approach, these models can reduce the computation time and data quality demands compared to training a new model from scratch while enhancing the model's ability to generalize across different programs. 
However, LLMs generally require significantly more computational resources, leading to higher operational costs and energy consumption, thus limiting their application. 
While our Learner implementation based on a lightweight Transformer model with scaled~dot-product~attention~\cite{NIPS2017_3f5ee243} mechanism is suitable and sufficient for program behavior modeling and can be trained on a wide range of existing hardware, recent LLMs require access to data-center-grade hardware accelerators.
Still, LLMs remain the mainstream of modern software engineering, and, below, we discuss the most relevant applications of neural models in program behavior modeling and the challenges they face.

\subsection{Behavior Mocking} There have already been previous attempts to perform behavior mocking by training sequence-to-sequence neural networks. TransCoder~\cite{10.5555/3495724.3497454} represents two sequence-to-sequence models for cross-language translations of code written in C++, Java, and Python with additional support for back translations. 
Armengol-Estape et al.~\cite{armengolestape2022learning} train transformer models to replicate the behavior of the GCC compiler. 
The training data is obtained by selecting C functions from the Anghabench benchmark and compiling them to x86 assembly code. 
The learned models can correctly compile the C code in less than half of the cases, making replacing compilers with neural machine Translation models invisible in practice.
Zaremba et al., in their ``Learning To Execute'' publication~\cite{zaremba2015learning}, try to predict the execution results of short Python programs performing simple arithmetic operations and echo printing by training an LSTM~\cite{10.1162/neco.1997.9.8.1735} model of a Python interpreter on a set of synthesized programs. 
These works have shown that the current generation of neural network architectures cannot correctly learn the full behavior of such complex programs as compilers or interpreters. While learned models cannot fully replace complex programs, our approach has demonstrated promising results in modeling linear or linearithmic programs. 
In contrast to previous approaches that rely on sequence-to-sequence models to achieve behavior mocking, Shen. et al. and their Konure~\cite{rinard-active-learning} system relies on the active learning approach to mine the behavior model of applications that interact with relational databases. 
Konure synthesizes inputs with database content simultaneously, monitors the traffic exchange between the database and the application, and inspects the results of the database queries to continuously refine the application behavior hypothesis. 
Using the refined application model, Konure can generate the Python code for database querying that is semantically equivalent to the original application code, which might be written in a different programming language. 
Similar to Konure, our approach learns the program models from the input-output pairs. 
While Konure is designed purely to model the interaction between an application and a database, our approach does not have such a limitation and can be used to either model the behavior of the whole system or its components and individual functionality.

\subsection{Reversing Computation} 
Neural machine translation techniques have shown promising results in low-level code decompilation tasks. Such neural decompilers are implemented as sequence-to-sequence~\cite{8330222, Liang:2021aa} or sequence-to-tree~\cite{NEURIPS2019_093b60fd, 10.1145/3564625.3567998} models. 
While existing works focus on converting a program's binary representation to high-level code, our framework lets users reverse computation direction (predict the program input used to produce a given output) by training a model of the inverse program. 
This is done by simply swapping training data pairs, i.e., using program output as model input and program input as output. 
Even when an exact inverse program may not exist, our approach allows users to estimate the result of an inverse computation. 
Approaches by Katz et al. in the TRAFIX framework~\cite{katz2019neural} and Fu et al. in the N-Bref framework~\cite{fu2021nbref} suggest training a neural decompiler from generated data, which we find closest to our work. 
In contrast, our framework supports behavior modeling of arbitrary programs, learning both direct and inverse models of the selected PUT. 
It also generalizes the learning process from mapping finite input-output pairs to extracting common input properties and constraints.

\subsection{Data and Model Poisoning} 
Language models trained on crowd-sourced data suffer from poisoning problems. 
Pearce et al.~\cite{asleep_at_the_keyboard} studied the impact of dataset poisoning on language model predictions' quality. 
Code-generation models still tend to generate buggy or malicious code snippets because they are trained on very similar corpora of publicly available documents. 
Schuster et al.~\cite{you_autocomplete_me} demonstrated two types of attacks on generative models: data poisoning and model poisoning. 
In the first scenario, only a few specially crafted samples added to the training corpus are enough to make code generation models produce malicious snippets. 
Alternatively, such models can be attacked by manipulating the training parameters (weights) or fine-tuning them on a few malicious samples. 
Both attack strategies result in a deviation of generated sequences, and existing defensive measures against dataset poisoning cannot mitigate the problem completely~\cite{you_see_what_i_want_you_to_see, He2023ControllingLL}. 
Due to possible manipulation of model behavior, the model outcome cannot be trusted, and their application in mission-critical systems remains questionable. 
In contrast to other approaches relying on crowd-sourced training data, our framework allows full control over data collection and model training processes by specifying input synthesis rules, testing generated samples with the PUT, and filtering or sanitizing faulty, undesired, and dangerous samples.

\subsection{Pre-trained and Unsupervised Tokenizers}
Instead of implementing tokenizers from scratch, one can potentially rely on existing \emph{generic} pre-trained tokenizers. 
For example, the \emph{HuggingFace} repository~\cite{wolf-etal-2020-transformers}  provides access to many community-developed natural language tokenizers through the HuggingFace API. 
Such tokenizers cannot properly split the program data for several reasons. 
First, they ignore the data-specific or computation-specific features since they cannot distinguish structure from content, which impacts the tokenization granularity. 
Second, some tokenizers, like Llama3~\cite{llama3modelcard} tokenizer, modify the input by prepending or appending additional characters to the produced tokens. 
This can break the syntactical correctness of program data and impact the quality of learned models.

To control the tokenization process while maintaining low development effort for end users of \Modelizer, we provided support to train the Google SentencePiece~\cite{kudo-richardson-2018-sentencepiece} unsupervised tokenizer for the given program program data. 
SentencePiece learns a subword vocabulary from raw input text using a specified vocabulary size; it treats the text as a continuous stream of characters. 
Once the training is complete, the tokenizer uses this learned vocabulary to break new input data into token sequences. 
By setting the vocabulary size during training, one can control both the lengths of the resulting token sequences and the number of trainable parameters in the model.
Similar to other existing unsupervised tokenizers, SentencePiece also ignores the structural representation of data at the tokenization time. 
To reduce the size of token vocabularies in our experiments, we used custom tokenizers adapted for the subjects we were experimenting with.

\subsection{Domain-specific Adaptations} 
Because such domain-specific models were fine-tuned only on the program source codes, they cannot capture the program behavior patterns without further fine-tuning. 
Our evaluation shows that LLMs pre-trained on code do not behave as drop-in replacements for existing programs. 
While humans can tolerate a certain degree of noise and errors in the LLM output, other programs cannot interpret cannot interpret them without failure. 
Additionally, their operation is significantly more costly than interactions with custom-trained models with fewer parameters and is incomparable in terms of cost-efficiency with original programs. 
Fine-tuning could be an expensive process as well due to the sizes of models and specific hardware requirements. 
Zheng et al. and their LlamaFactory~cite{zheng2024llamafactory} framework provide estimated memory requirements implied in fine-tuning open-source LLMs. For example, the small versions of commonly used LLMs, like Llama2, Llama3, and Mixtral, with 7-to-8 billion parameters, require more than 60GB of memory for full-precision fine-tuning. 
Even though models are open-source, it is mainly impossible to fine-tune them locally without access to data-center-grade hardware. 
To mitigate these memory requirements problems, one can use a post-training model weights quantization\cite{quant1, quant2} technique that reduces the precision of model weights, thus requiring a lower number of bits to store them. 
Also, the low-rank adaptors\cite{hu2021loralowrankadaptationlarge} technique can further optimize computation and storage efficiency by introducing low-rank matrices into specific layers of LLMs like the attention mechanism to capture task-specific information by updating only these matrices during fine-tuning.
However, model weight quantization can tamper with the prediction quality due to quantization errors, which may significantly impact performance in learning uncommon behavior patterns. 
Similarly, LoRA adapters might limit the model's expressiveness and adaptability by constraining updates only to low-rank matrices.
Instead, our framework can learn ``lightweight'' program behavior models consisting of only several million parameters and can be trained on commodity hardware at a low cost.

\subsection{Neural Program Induction} 
We find that the neural program induction and neural program synthesis fields are also closely related to our research. 
While neural program induction models~\cite{10.5555/2969239.2969261, 10.5555/3294771.3294969, DBLP:journals/corr/NeelakantanLS15, graves2014neural, pmlr-v48-zaremba16, DBLP:journals/corr/KurachAS15} try to train a sequence-to-sequence generative model that learns data patterns from input-output examples, neural program synthesis~\cite{Vijayakumar2018NeuralGuidedDS, 10.5555/3305381.3305484, shin2019synthetic} use neural networks for predicting algorithms from corresponding input-output samples. 
Existing studies cannot capture the behavior of complex algorithms and programs because they had been relying on already outdated neural network architectures; they could not generalize to arbitrary complex inputs because the corresponding models had been trained on a limited number of training samples, which, for certain studies are in human-generated; require the additional manual modeling and encoding of the algorithm or program in a domain-specific language that has limited functionality. 
Recent advances in sequence-to-sequence prediction algorithms (Transformer architecture and its variations) make it possible to automatically infer complex data patterns from input-output pairs. 
Moreover, our approach can also automatically learn the inverse patterns, which allows for the revision of certain programs' computation results at a low cost.

\subsection{AI-driven Testing} 
AI-driven, or specifically LLM-driven software testing, is becoming more and more popular. 
We picked a few representative works from myriads of publications in this field to highlight common use case scenarios.
For example, the TOGA~\cite{10.1145/3510003.3510141} project explores neural test oracle generation problem by fine-tuning the CodeBERT~\cite{feng-etal-2020-codeBERT} model on automated inference of exceptional and assertion oracles from developer-written tests. 
Haluptzok et al.~\cite{haluptzok2023language} synthesize and solve Python programming puzzles using language models. 
The authors of DeepAnalyze~\cite{10.1145/3510003.3512759} learned a sequence labeling model to automatically mark memory frames in a stack trace that caused the crash of an application. 
The LIBRO~\cite{10.1109/ICSE48619.2023.00194} framework uses the Codex Large Language Model to generate and rank bug-reproducing test cases from bug reports.
Tufano et al.~\cite{8811910} train a neural machine translation model to automatically replicate in 36\% of cases the changes between pull requests in code repositories like bug fixes or code refactoring.
The DeepFix~\cite{Gupta_Pal_Kanade_Shevade_2017} approach randomly mutates correct programs to introduce errors. It uses these data pairs to train the neural machine translation model for automated program repair, which succeeds in fixing buggy programs in 27\% of cases.
The Break-It-Fix-It~\cite{pmlr-v139-yasunaga21a} approach achieves high accuracy in code repair tasks by chaining two generative neural machine translation models for synthesizing naturally looking code errors and training a fixer model from synthesized pairs.
Godefroid et al., in their ``Learn\&Fuzz'' publication~\cite{8115618}, learn a generative model for PDF structures by training a recurrent neural network from a set of PDF objects. The input sequence is represented by a sequence of characters in PDF objects, while the output sequence is obtained by shifting the corresponding input sequence by one character.
While each work mentioned follows its unique approach, they all require a significant amount of human effort to solve problems. 
In contrast, \Modelizer tries to automate the process of data collection (which is one of the most time-consuming parts of model preparation) and model training. 
The minimalistic set of requirements, in particular the ability to run a Python interpreter locally, simplifies the integration of the learned models into the existing projects and testing tools.

\subsection{AI-generated Synthetic Data} 
The capabilities of LLMs revealed several approaches of their application for synthetic data generation~\cite{li-etal-2023-synthetic}. 
They can generate human-like text or dialogues given the prompt query, augment existing datasets by producing variations of the original data, and produce structured representations of existing data like JSON or CSV. 
Such AI-generated synthetic data is mostly used for training classifiers or question-answering systems. 
For example, general-purpose LLMs like OpenAI ChatGPT or Google Bard have shown promising results~\cite{synthetic-data-llm-results} in generating synthetic data in a medical text production task. 
However, in certain tasks, like hateful speech detection and classification, the inclusion of syntactically generated samples to the training set does not improve the performance of classifiers~\cite{kruschwitz-schmidhuber-2024-llm}. 
Still, it is an active and promising field of research, and industry giants like Nvidia invest in the development of LLMs, like Nemotron-4~\cite{nemotron}, specifically designed for synthetic data generation. 
While synthetic data generation with LLMs implies specific hardware requirements to run models locally or leads to a privacy compromise while processing the data with a third-party cloud provider at an additional service cost, our approach, which relies on synthesizing data with grammar fuzzers, only requires a system with the ability to run a Python interpreter locally.

\subsection{Data Generation by Mutating Inputs} 
We also want to address the alternatives to grammar-based and AI-based input generation strategies. 
Potentially, an input set for the given PUT can be automatically constructed with the help of mutation-based fuzz-testing~\cite{10.1145/96267.96279}. 
A mutation fuzzer, like AFL++~\cite{10.5555/3488877.3488887}, can accept one input for a PUT as an initial seed and produce infinitely many new inputs by applying pre-defined mutation strategies, like bit flips, addition, or subtraction of integers, on different parts of the given seed. 
Such a technique produces two types of mutants. 
The first type of mutants has a broken syntactical or semantic structure. 
These mutants will either crash the PUT if it contains a bug in its implementation, or the input parser guards will reject such input, terminating program execution. 
The collected input-output pairs will only cover the behavior of the input parser rather than the whole program. In most cases, the second type of mutants, which would have managed to bypass the parser guards, will contain insignificant differences when compared to an initial input sample. 
Thus, testing the program with this kind of input will unlikely produce a diverse set of outputs. While fuzz testing is helpful for bug discovery in input parser implementations, we do not think the mutation-based input generation technique is applicable to automate program behavior modeling.

\subsection{Mining Input Grammars}
\label{rel:mining} 
In certain cases, input grammars can also be automatically mined from a PUT. 
The \emph{Mimid} approach by Gopinath et al.~\cite{10.1145/3368089.3409679} uses program instrumentation with string wrappers to identify control flow nodes accessing relevant portions of a program input. 
Kulkarni et al. presented the \emph{Arvada} algorithm~\cite{ARVADA2021} that can automatically learn a highly recursive grammar from a given program using positive examples and a Boolean-valued oracle. 
The \emph{GDBMiner}~\cite{gdbminer-eisele} tool by Eisele et al. can also automatically extract input grammars from binaries and executables written in any compiled language on any system architecture that can execute a GDB debugger. 
Finally, \emph{symbolic parsing}~\cite{stalagmite} by Bettscheider et al.\ promises to statically extract accurate input grammars from parser code, without requiring any sample inputs.
Currently, if input formats are not publicly available or are not recoverable from the PUT, the user of our framework must specify them as a grammar.
Once an input grammar is available, \Modelizer can generate inputs and learn the program behavior.

\section{Conclusion and Future Work}
\label{sec:conclusion}

With \Modelizer, we provide a principled approach to turning regular programs into models.
This opens up the manifold capabilities of machine learning models for several Software Engineering tasks such as mocking, testing, monitoring, and debugging.
Program models learned from synthetic inputs effectively represent real-world program execution scenarios by learning basic patterns from program input-output relations.
While the lack of knowledge of the end data feature distribution could limit accurate model training from scratch, few-shot fine-tuning on real-world examples efficiently enhances prediction accuracy.
The most promising feature of the learned translation models is their ability to predict inputs for given outputs, not only reversing executions but also offering new opportunities for test input generation.
We see \Modelizer as a pioneer in \emph{project-specific learning}---rather than trying to generate expensive LLMs that may or may not evolve into a one-size-fits-all, \Modelizer demonstrates that it is possible to learn small models for nontrivial programs that can successfully replicate their behavior---and their reverse.

Besides testing \Modelizer with more complex programs and larger model sizes, our future work on \Modelizer will focus on further applications, as well as addressing some of its current limitations.
Specifically:

\begin{description}
  \item[Targeted input generation.] One of the most promising applications of \Modelizer is its use for test input generation. 
  By treating \emph{traces} or \emph{coverage} as \emph{output} of the PUT (and supplying an output grammar that can parse trace logs), \Modelizer can learn which input features correlate with which traces or coverage features; and can then again predict which input would be required to achieve a particular trace or coverage.
  These inputs could then be tested automatically, and if the trace or coverage deviates, such feedback could be used to fine-tune the model.

  \item[Monitoring.]  Once \Modelizer has extracted a model from the reference program, one can also use it for \emph{monitoring} revisions of PUT by continuously comparing the program behavior against the extracted model. 
  This could be used to detect and prevent behavior not seen during testing.

  \item[Debugging.] By learning correlations between collected crash data (core dumps and stack traces) and input features, \Modelizer could predict inputs and input features likely to have caused a given program state.

  \item[Smarter test generation.] In our evaluation, we repeatedly found evidence that \Modelizer's performance greatly depends on the quality and diversity of synthesized training data and, thus, on the quality of the test generator. 
  In our future work, we will experiment with smarter black-box test generators, covering grammar features and combinations thereof~\cite{10.1109/ASE.2019.00027}, but also smarter \emph{grey-box fuzzing,} using code coverage in the PUT as guidance~\cite{pham2019smart,aschermann2019nautilus}.

  \item[Stateful programs.] At this point, the models extracted by \Modelizer do not consider internal state, as, say, a server might maintain. 
  Let us assume the PUT starts in a known initial state, and the input to the PUT is a sequence of commands. 
  Then, having \Modelizer learn the effects of command sequences should enable it to detect that certain outputs depend on specific subsequences of commands (implying a specific state). 
  Such reasoning will likely require larger models; we plan to evaluate \Modelizer on more complex and stateful programs.

  \item[Feature distribution.] The \emph{feature distribution} in the synthesized samples does not always correspond to real-world counterparts.
  Our attempt to add the sliding window strategy to the number of selected non-terminals and manually adjust probabilities of expansion rules only partially solves this problem.
  Without having the precise input specification in advance, it is not possible to initialize the generator to produce samples that fully cover the real-world input space.
  This problem can be resolved by first mining corresponding expansion probabilities from a few real-world samples and using them to fine-tune the model on additionally synthesized samples~\cite{inputs-from-hell}.
  Such samples will serve as guidance for the generator to focus on the uncovered areas of the input space.

  \item[Leveraging program feedback.] We estimate that the model's performance can be further improved by utilizing the feedback from the PUT when available.
  In contrast to fundamental LLMs, \Modelizer framework allows relatively cheap continuous re-training of the learned models.
  If the specifications of the target system have insignificantly changed, only affecting the system behavior but not extending the population of the input-output vocabularies with new elements, the learned models can be fine-tuned with a few samples collected from system executions.
  In \Cref{sec:deployment}, we have represented a scenario when the PUT serves as an oracle to validate the model's predictions. 
  Once the model prediction does not align with execution results, the collected feedback can be passed to the model as additional training data to adjust the model's parameters. 
  The model has to be repeatedly re-evaluated and fine-tuned until its predictions match the PUT's execution results. 
  We plan to study the capabilities of \Modelizer and respective learned models to serve as an oracle for automatic detecting of the behavior deviation cases between various revisions and implementations of the same protocol. 
  We also hope to explore the possibility of forging feedback-driven tokenizers, which will be able to automatically adjust the tokenization granularity and possibly mine the input-output grammars from black-box systems. 
  They will rely not only on the feedback from the PUT but also leverage the feedback from the model evaluation.

  % \item[1-bit models.] Lastly, we are excited by the idea of replacing 16-bit parameter weights with a 1-bit representation~\cite{ma2024era}, significantly reducing the memory requirements and thus allowing training more cost-effective models. We plan to integrate this feature into \Modelizer in future updates.
\end{description}
\Modelizer and all experimental data, including interactive plots, are available at
\begin{center}
  \url{https://github.com/2ral/Modelizer}
\end{center}

\noindent
\textbf{Acknowledgments.}
We appreciate the thorough review and critical insights provided by our colleagues, which significantly improved the quality of this manuscript. \\
This work is funded by the European Union (ERC S3, 101093186). Views and opinions expressed are, however, those of the author(s) only and do not necessarily reflect those of the European Union or the European Research Council. Neither the European Union nor the granting authority can be held responsible for them.

\bibliographystyle{ACM-Reference-Format}
\bibliography{modelizer}

\end{document}